\documentclass[showpacs,pra,preprint,amsmath,amssymb,aps]{revtex4-1}
\usepackage{graphicx}
\usepackage{bm}
\usepackage[colorlinks=true,pdfstartview=FitV,linkcolor=red,citecolor=blue,urlcolor=blue]{hyperref}

\begin{document}

\preprint{BNL-, RBRC-}

\title{Chiral matrix model of the semi-Quark Gluon Plasma in QCD}

\author{Robert D. Pisarski}
\email{pisarski@bnl.gov}
\affiliation{Department of Physics, Brookhaven National Laboratory,
Upton, NY 11973}
\affiliation{RIKEN/BNL, Brookhaven National Laboratory,
Upton, NY 11973}
\author{Vladimir V. Skokov}%
\email{vskokov@bnl.gov}
\affiliation{RIKEN/BNL, Brookhaven National Laboratory,
Upton, NY 11973}

\begin{abstract}
Previously, 
a matrix model of the region near the transition temperature,
in the ``semi''-Quark Gluon Plasma, was developed for the theory
of $SU(3)$ gluons without quarks.
In this paper we develop a a chiral matrix model applicable to QCD
by including dynamical quarks with $2+1$ flavors.
This requires adding a nonet of scalar fields, with both parities, 
and coupling these to quarks through a Yukawa coupling, $y$.  
Treating the scalar fields in mean field
approximation, the effective Lagrangian is computed by 
integrating out quarks to one
loop order.  As is standard, the potential for the scalar fields 
is chosen to be symmetric under the flavor symmetry of 
$SU(3)_L \times SU(3)_R \times Z(3)_A$, 
except for a term linear in the current quark mass, $m_{qk}$.  
In addition, at a nonzero temperature $T$
it is necessary to add a new term, $\sim m_{qk}\, T^2$.  
The parameters of the gluon part of the matrix model
%, including
%especially the deconfining transition temperature $T_d = 270$~MeV, 
 are identical to
that for the pure glue theory without quarks.
The parameters in the chiral matrix model are fixed
by the values, at zero temperature, of the pion decay constant 
and the masses of the pions, kaons, $\eta$, and $\eta'$.
The temperature for the chiral crossover at $T_{\chi} = 155$~MeV
is determined by adjusting the Yukawa coupling
$y$.  We find reasonable agreement with the results of 
numerical simulations on the lattice for the pressure and related
quantities.  In the
chiral limit, besides the divergence in the chiral 
susceptibility there is also a milder divergence in the
susceptibility between the Polyakov loop and the chiral order 
parameter, with critical exponent
$\beta - 1$.  We compute derivatives with respect to a
quark chemical potential to determine the susceptibilities
for baryon number, the $\chi_{2n}$.  
Especially sensitive tests are provided by
$\chi_4 - \chi_2$ and by $\chi_6$, which changes in sign about $T_\chi$.
The behavior of the susceptibilities in the chiral matrix model 
strongly suggests that as the temperature increases from $T_\chi$,
that the transition to deconfinement 
is significantly quicker than indicated by the measurements
of the (renormalized) Polyakov loop on the lattice.
\end{abstract}
\maketitle

\section{\label{sec:intro} Introduction}

Our understanding of the behavior of the collisions of heavy nuclei at 
ultra-relativistic energies rests upon the bedrock provided by 
numerical simulations of lattice QCD.
At present, for 
QCD with $2+1$ light flavors, these simulations
provide us with results, near the continuum limit, for the behavior of QCD
 in thermodynamic equilibrium
\cite{boyd_equation_1995, *umeda_fixed_2009, *borsanyi_precision_2012, cheng_transition_2006, *cheng_qcd_2008, *bazavov_equation_2009, *cheng_equation_2010, *bazavov_chiral_2012, bazavov_polyakov_2013, bazavov_polyakov_2016, buchoff_qcd_2014, bhattacharya_qcd_2014,  bazavov_equation_2014, ding_thermodynamics_2015, cheng_baryon_2009, *bazavov_fluctuations_2012, *bazavov_freeze-out_2012, aoki_qcd_2006, *fodor_phase_2009, *aoki_qcd_2009, *borsanyi_qcd_2010, *borsanyi_is_2010, *durr_lattice_2011, *endrodi_qcd_2011, borsanyi_full_2014, borsanyi_fluctuations_2012, *bellwied_is_2013, *borsanyi_freeze-out_2013, *borsanyi_freeze-out_2014, *bellwied_fluctuations_2015, ratti_lattice_2016}.  
Most notably, that there is a chiral
crossover at a temperature of $T_\chi \sim 155 \pm 9$~MeV.  

While this understanding is essential, there are many quantities 
of experimental interest which are much more difficult to obtain 
from numerical simulations of lattice QCD.
This includes all quantities 
which enter when QCD is out of but near thermal equilibrium, 
such as transport coefficients, the production of dileptons and photons, 
and energy loss.

For this reason, it is most useful to have phenomenological 
models which would allow us to estimate such quantities.  
Lattice simulations demonstrate that in equilibrium,
a non-interacting gas of hadrons works well up to rather high temperatures, 
about $\sim 130$~MeV
\cite{boyd_equation_1995, *umeda_fixed_2009, *borsanyi_precision_2012, cheng_transition_2006, *cheng_qcd_2008, *bazavov_equation_2009, *cheng_equation_2010, *bazavov_chiral_2012, bhattacharya_qcd_2014,  bazavov_equation_2014, ding_thermodynamics_2015, cheng_baryon_2009, *bazavov_fluctuations_2012, *bazavov_freeze-out_2012, aoki_qcd_2006, *fodor_phase_2009, *aoki_qcd_2009, *borsanyi_qcd_2010, *borsanyi_is_2010, *durr_lattice_2011, *endrodi_qcd_2011, borsanyi_full_2014, borsanyi_fluctuations_2012, *bellwied_is_2013, *borsanyi_freeze-out_2013, *borsanyi_freeze-out_2014, *bellwied_fluctuations_2015, ratti_lattice_2016}.  
Similarly, resummations of perturbation theory, such
as using Hard Thermal Loops (HTL's) at Next to- Next to- Leading order (NNLO), 
work down to about $\sim 300$ or $\sim 400$~MeV
\cite{andersen_three-loop_2010, *andersen_gluon_2010, *andersen_nnlo_2011, *andersen_three-loop_2011,*haque_two-loop_2013, *mogliacci_equation_2013, *haque_three-loop_2014}.
What is difficult to treat is the region between $\sim 130$ 
and $\sim 300-400$~MeV, which has been termed the ``sQGP'', 
or strong Quark-Gluon Plasma.  This name was suggested by T. D. Lee,
because analysis of heavy experiments appears to show that the
ratio of the shear viscosity to the entropy density, $\eta/s$,
is very small.
For QCD, in perturbation theory $\eta/s \sim 1/g^4$, and so a small
value of $\eta/s$ suggests that the QCD coupling constant, $g$, is large.

There is another way of obtaining a small value of $\eta/s$
without assuming strong coupling
\cite{pisarski_quark_2000, *dumitru_degrees_2002, *dumitru_two-point_2002, *scavenius_k_2002, *dumitru_deconfining_2004, *dumitru_deconfinement_2005, *dumitru_dense_2005, *oswald_beta-functions_2006, pisarski_effective_2006}.
At high temperature, the quarks and gluons are deconfined, and their density
can be estimated perturbatively.  At low temperatures, confinement implies that
the density of particles with color charge vanishes as
$T \rightarrow 0$.  Numerical simulations demonstrate that even with 
dynamical quarks, the density of color charge, 
as measured by the expectation value of the Polyakov
loop, is rather small at $T_\chi$, with $\langle \ell \rangle \sim 0.1$.
This presumes that the Polyakov loop is normalized so that its
expectation value approaches one at infinite temperature,
$\langle \ell \rangle \rightarrow 1$ as $T \rightarrow \infty$.

Because of the decrease in the density of color charge,
the region about $T_\chi$ can be termed
not as a strong, but as a ``semi''-QGP.
In this view, the dominant physics is assumed
to be the {\it partial} deconfinement of color charge,
analogous to partial ionization in Abelian plasmas
\cite{hidaka_suppression_2008, *hidaka_hard_2009, *hidaka_zero_2009, *hidaka_small_2010}.  

This partial deconfinement can be modeled in a matrix model
of the semi-QGP.
In such a matrix model, both the shear viscosity and
the entropy density decrease as the density of color charges decreases.  
It is not obvious, but calculation shows that the shear viscosity 
vanishes quicker than
the entropy density, so that the ratio $\eta/s \sim \langle \ell \rangle^2$
\cite{hidaka_suppression_2008, *hidaka_hard_2009, *hidaka_zero_2009, *hidaka_small_2010}.  
Thus in a matrix model, it is possible to obtain a small shear viscosity
not because of strong coupling, but because the density of color charge
is small.

A matrix model of the semi-QGP has been developed for the pure
gauge theory 
\cite{dumitru_eigenvalue_2008, *smith_effective_2013, dumitru_how_2011, dumitru_effective_2012, sasaki_effective_2012, dumitru_two-loop_2014}.
The fundamental variables are the eigenvalues of the thermal Wilson line,
and it is based upon the relationship between deconfinement and
the spontaneous breaking of the global $Z(N_c)$ symmetry of a $SU(N_c)$ gauge
theory.  This model
is soluble in the limit for a large number of colors,
and exhibits a novel ``critical first order'' phase transition
\cite{pisarski_gross-witten-wadia_2012, *lin_zero_2013}.
With heavy quarks, it has been used
to compute the critical endpoint for deconfinement
\cite{kashiwa_critical_2012}
and properties of the Roberge-Weiss
transition \cite{kashiwa_roberge-weiss_2013}.
The production of dileptons and photons has also been computed
\cite{gale_production_2015, *hidaka_dilepton_2015, *satow_chiral_2015};
the suppression of photon production in the semi-QGP may help to understand
the experimentally measured azimuthal anisotropy of photons.
In a matrix model, collisional energy loss 
behaves like the shear viscosity, and 
is suppressed as the density of color charges decreases
\cite{lin_collisional_2014}.  

In this paper we develop a chiral matrix model by including light,
dynamical quarks, as is relevant for QCD with $2+1$ light flavors.
Our basic assumption is the following.  The global $Z(3)$ symmetry
of a pure $SU(3)$ gauge theory is broken by the presence of dynamical
quarks, and generate a nonzero expectation value for
the Polyakov loop at nonzero temperature, 
$\langle \ell \rangle \neq 0$ when $T \neq 0$.
As noted above, however, this expectation value is remarkably small at
the chiral transition, with $\langle \ell \rangle \sim 0.1$.  Thus
in QCD, the breaking of the global $Z(3)$ symmetry by
dynamical quarks is surprisingly {\it weak} near $T_\chi$.
This is a nontrivial result of the lattice: it is related to the fact
that in the pure gauge theory, the deconfining phase transition occurs
at $T_d \sim 270$~MeV, which is much higher than 
$T_\chi \sim 155$~MeV.  We do not 
presume that this holds for arbitrary numbers of colors and flavors.
In QCD, though, it suggests that treating the global $Z(3)$ symmetry
breaking as small, and the matrix degrees of freedom as ``relevant'', is a
reasonable approximation.

Other than that, while the technical
details are involved, the basic physics is simple.
We start with a standard chiral Lagrangian for the nonet of light
pseudo-Goldstone mesons: pions, kaons, $\eta$, and the $\eta'$.
Because we wish to analyze the chirally symmetric phase, we 
add a nonet of mesons with positive parity,
given by the sigma meson and its associated partners
\cite{t_hooft_how_1986, donoghue_dynamics_1992, lenaghan_chiral_2000, roeder_chiral_2003, *janowski_glueball_2011, *parganlija_meson_2013, stiele_phase_2016, jaffe_multiquark_1977, *black_mechanism_2000, *close_scalar_2002, *jaffe_diquarks_2003, *maiani_new_2004, *pelaez_light_2004, *t_hooft_theory_2008, *pelaez_controversy_2015}.
The field for the mesons,
$\Phi$, couples to itself through a Lagrangian which includes 
terms which are invariant under the 
flavor symmetry of $SU(3)_L \times SU(3)_R \times U_A(1)$.

For the meson field $\Phi$ we take a linear sigma model, as then
it is easy to treat the chirally symmetric phase (this is possible,
but more awkward, with a nonlinear sigma model).
We include a chirally symmetric Yukawa coupling between $\Phi$ and 
the quarks, with a Yukawa coupling constant $y$.  
The quarks are integrated out to one loop order,
while the meson fields are treated in the mean field approximation,
neglecting their fluctuations entirely.
Dropping mesonic fluctuations is clearly a drastic approximation,
but should be sufficient for an initial study of the matrix model.

To make the pions and kaons massive, we add a term which is linear 
in the current quark mass, $m_{qk}$.  We demonstrate that in order 
for the constituent mass of
the quarks to approach the current quark mass at high temperature, 
it is also necessary to add an additional term 
$\sim m_{qk}$: this new term vanishes at zero temperature,
but dominates at high temperature.
This new term has not arisen previously, because
typically linear sigma models do not include fluctuations of the quarks.

The meson potential includes chirally symmetric terms for $\Phi$ at quadratic,
cubic, and quartic order.
For three flavors, the cubic term represents the effect of the axial anomaly.
The parameters of the model are
fixed by comparing to the meson masses at zero temperature, for the masses of
the pion, kaon, $\eta$, and $\eta'$, and the pion decay constant.  
This fitting is typical of models at zero temperature.
The quartic term includes a novel logarithmic term from the
fluctations of the quarks, but this does not markedly change the parameters of
the potential for $\Phi$.

The chiral matrix model can be considered as a generalization of Polyakov loop
models, as first proposed by Fukushima
\cite{fukushima_chiral_2004, *fukushima_phase_2011, *fukushima_phase_2013, sasaki_susceptibilities_2007, skokov_vacuum_2010, skokov_meson_2010, *morita_role_2011, islam_vector_2015, ishii_determination_2016, miyahara_equation_2016};
see also 
\cite{schaefer_higher-order_2012, *schaefer_qcd_2012, *chen_chemical_2015, *berrehrah_quark_2015, *tawfik_su3_2015, *tawfik_polyakov_2015}.
In a Polyakov loop model, the gauge fields are integrated out  to obtain an effective model of the
Polyakov loop and hadrons.  Because of this, except for one special case 
(dilepton production at leading order~\cite{islam_vector_2015}), 
Polyakov loop models can only be used to study processes in, and not near, 
equilibrium. 
In a matrix model, though, as $A_0$, is not integrated out
it is straightforward to compute processes neat equilibrium by analytic
continuation. This includes many quantities of experimental relevance, 
especially transport coefficients such as shear and bulk viscosities. 

There is another difference between the two models.  In a Polyakov loop
model, all thermodynamic functions are functions of the ratio $T/T_c$,
where $T_c$ is the critical temperature.  In a pure gauge theory,
$T_c$ is the temperature for the deconfining phase transition, $T_d$.
With dynamical quarks, $T_c$ is that for the restoration of chiral
symmetry, $T_\chi$.  

In contrast, in our chiral matrix model we take the gluon potential
to be {\it identical} to that of the pure gauge theory, keeping
the parameter $T_d = 270$~MeV.  The Yukawa coupling 
$y$ is then tuned to obtain a chiral crossover temperature
$T_\chi = 155$~MeV.  
We stress that in our model, $T_d$ is {\it not} the temperature for
deconfinement in QCD: it is just a parameter of the gluon
part of the effective, nonperturbative potential for $A_0$.  
Since dynamical quarks explicitly break the global $Z(3)$ symmetry
of the pure gauge theory, there is no precise definition of a
deconfining temperature in QCD.  One approximate measure
is provided by susceptibilities involving the Polyakov loop, as
considered in Sec. (\ref{sec:divergent}).  These indicate that deconfinement
occurs close to $T_\chi$, Fig. (\ref{fig:susc}).  

There are other models in which transport coefficients can be computed.
These include Polyakov quark meson models improved by using
the functional renormalization group
\cite{braun_nature_2010, herbst_phase_2011, *fister_confinement_2013, *herbst_phase_2013, *haas_gluon_2014, *herbst_thermodynamics_2014, *mitter_chiral_2015, herbst_confinement_2015, fu_relevance_2015, fu_correlating_2015}.

As a byproduct
we make some observations about linear sigma models.
For the special limit of three degenerate but massive flavors, 
in a general linear sigma model, we show that at zero temperature
the difference of the masses squared 
of the singlet and octet states $0^-$ states equals the difference
of the masses squared between the octet and singlet states for the $0^+$,
Eq. (\ref{eq:chiral_identity}).  This is identical to the same relation
for two degenerate, massive flavors \cite{t_hooft_how_1986}.

To fix the parameters of the chiral matrix model, we only use properties
of the $0^-$ nonet, not the $0^+$ nonet.  
This is fortunate, because the lightest
$0^+$ nonet may be formed not from a quark antiquark pair, but is a tetraquark,
composed of a diquark and diantiquark pair
\cite{jaffe_multiquark_1977, *black_mechanism_2000, *close_scalar_2002, *jaffe_diquarks_2003, *maiani_new_2004, *pelaez_light_2004, *t_hooft_theory_2008, *pelaez_controversy_2015}.

In this paper we do not consider a
nonzero quark density, $\mu$.  (We do consider derivatives of the
pressure with respect to $\mu$, but these are then always evaluated
at $\mu = 0$.)  Because at $\mu = 0$ lattice simulations indicate that
$T_\chi \ll T_d$, as one moves out in the plane of temperature 
and chemical potential, a quarkyonic phase in which $T_\chi < T_d$ 
when $\mu \neq 0$
\cite{mclerran_phases_2007, *andronic_hadron_2010, *kojo_quarkyonic_2010, *kojo_interweaving_2012} is very natural in a chiral matrix model.

\section{Simple example of a chiral matrix model}
\label{sec:toy_model}

%\subsection{Chiral symmetry breaking at nonzero temperature}
\label{sec:nonzero_T_chiral_bkng}

Before diving into all of the technicalities associated with
the chiral matrix model for $2+1$ flavors, it is useful to illustrate some
general ideas in the context of a simple toy model.
We take a
single flavor of a  Dirac fermion, interacting with a sigma field $\sigma$
through the Lagrangian
\begin{equation}
{\cal L}
=
\overline{\psi} 
\left(
\, \not \! \partial \, + \, y \, \sigma
\right) 
\psi 
+ \frac{m_\sigma^2}{2} \, \sigma^2 + \frac{\lambda}{4} \, \sigma^4 
\; .
\label{lagrangian_toy_model}
\end{equation}
To demonstrate our points we can even neglect the coupling to the gauge
field, although of course it is the coupling to gluons which drives
chiral symmetry breaking.  We neglect the kinetic term for the
$\sigma$ field, since that will not enter into our analysis, which is
entirely at the level of a mean field approximation for $\sigma$.

Notice that we include both the Lagrangian for the fermion $\psi$ as
well as for the scalar field $\sigma$.  Usually in sigma models, one
assumes that the quarks are integrated out, with their interactions
subsumed into those of the mesons.  We cannot do that, because we
need to include the effects of the quarks on the matrix model, as
we show in the next Section.  Consequently, we also include a
Yukawa coupling $y$ between the fermion $\psi$ and $\sigma$.

This Lagrangian is invariant under a discrete chiral symmetry of 
$Z(2)$,
\begin{equation}
\psi \rightarrow \gamma_5 \; \psi \;\;\; , \;\;\;
\sigma \rightarrow - \, \sigma \; .
\label{symmetry_toy_model}
\end{equation}
We take a Euclidean metric, where each Dirac matrix $\gamma^\mu$ satisfies
$(\gamma^\mu)^2 = +1$, and 
$\gamma_5 = \gamma_0 \gamma_1 \gamma_2 \gamma_3$, so
$\gamma_5^2 = 1$.

Integrating out the fermion gives the effective potential
\begin{equation}
{\cal V}^{{\rm eff}}_\sigma
= + \frac{m_\sigma^2}{2} \, \sigma^2 + \frac{\lambda}{4} \, \sigma^4 
- \frac{1}{V}\, {\rm tr} \, {\rm log} 
\left( 
\not \! \partial + \, y \, \sigma
\right)
\; ,
\label{eff_lag_toy}
\end{equation}
where $V$ is the volume of spacetime.

We thus need to compute the fermion determinant in the
background field of the $\sigma$ field, which 
in mean field approximation we take to be constant.
For ease of notation, we write
\begin{equation}
m_f = y \, \sigma \; .
\end{equation}
Taking two derivatives with respect to $m_f^2$,
\begin{equation}
- \; \frac{\partial^2}{(\partial m_f^2)^2} \; 
{\rm tr} \, \log( 
\; \slash \!\!\! \partial + m_f) 
= + 2 \; {\rm tr} \; \frac{1}{(K^2 + m_f^2)^2} \; .
\end{equation}
where
$\partial_\mu = - i K^\mu$.
Here the trace 
is the integral over the momentum $K$ in $4 - 2\epsilon$ dimensions,
\begin{equation}
{\rm tr} = \widetilde{M}^{2 \epsilon} 
\int \frac{d^{4 - 2 \epsilon}K}{(2 \pi)^{4 - 2 \epsilon}} \; .
\label{eq:trace_zeroT}
\end{equation}
A renormalization mass scale $\widetilde{M}$ is introduced so that the trace
has dimensions of mass$^4$.
The result is
\begin{equation}
{\rm tr} \; \frac{1}{(K^2 + m_f^2)^2} 
= \; + \; \frac{1}{16 \pi^2}
\left( \frac{1}{\epsilon} 
+ \log\left( \frac{\widetilde{M}^2}{m_f^2} \right)
+ \log(4 \pi) - \gamma \right) \; ,
\label{zeroTselfsame}
\end{equation}
where $\gamma \sim 0.577$ is the Euler-Mascheroni constant.
Integrating with respect to $m_f^2$, 
\begin{equation}
- \, \frac{1}{V} \, {\rm tr} \, \log( 
\; \slash \!\!\! \partial + m_f) 
=
+ \frac{m_f^4}{16 \pi^2}
\left( 
\frac{1}{\epsilon}
+ \log
\left( 
\frac{\widetilde{M}^2}{m_f^2} 
\right)
+ \log(4 \pi) - \gamma + \frac{3}{2} 
\right) \; .
\end{equation}
Defining
\begin{equation}
\log(M^2)
= \log \widetilde{M}^2 + \log(4 \pi) - \gamma + \frac{3}{2} \; , 
\end{equation}
we find
\begin{equation}
- \, \frac{1}{V} \, {\rm tr} \, \log( 
\; \slash \!\!\! \partial + m_f) 
= 
+ \frac{m_f^4}{16 \pi^2}
\left( 
\frac{1}{\epsilon}
+ \log
\left( 
\frac{M^2}{m_f^2} 
\right) 
\right) \; .
\label{eq:final_zero_temp}
\end{equation}

The integral in Eq. (\ref{zeroTselfsame})
is logarithmically divergent, $\sim d^{4-2\epsilon}K/(K^2 + m_f^2)^2$.  
The divergence in the ultraviolet
produces the usual factor of $1/\epsilon$ in $4 - 2 \epsilon$ dimensions.
Similarly, there is a logarithmic infrared divergence, 
cut off by the mass $m_f$.

We add a counterterm $\sim 1/\epsilon$ to the effective Lagrangian so
that the sum with the one loop fermion determinant is finite.  
We thus obtain a renormalized effective Lagrangian,
\begin{equation}
{\cal V}_{\sigma}^{{\rm eff, ren}}
= + \frac{m_\sigma^2}{2} \, \sigma^2 + 
\frac{1}{4} \, 
\left(
\lambda + \, \frac{y^4}{4 \, \pi^2} 
\log 
\left(
\frac{M^2}{y^2 \, \sigma^2}
\right)
\right)
\sigma^4 
\; .
\label{intgd_eff_lag_toy}
\end{equation}
This is resembles the standard effective Lagrangian, except that it
is no longer purely a polynomial in $\sigma$, but also has 
a term $\sim - \, y^4 \, \sigma^4 \log(\sigma^2)$.

While this logarithmic term changes the effective Lagrangian, it does not
really cause any particular difficulty.  As usual we tune
the scalar mass squared $m_\sigma^2$ to be negative at zero temperature,
so that $\sigma$ develops a vacuum expectation value
$\langle \sigma \rangle \neq 0$, and 
the fermion acquires a 
constituent mass $m_f = y \langle \sigma \rangle$.
Because the chiral
symmetry is discrete there are no (pseudo-) Goldstone bosons, but
for the points we wish to make here this is irrelevant.

There is one feature which we
must note.  The sign of the logarithmic term in the effective Lagrangian,
$\sim - \, y^4 \, \sigma^4 \, \log(\sigma^2)$,
is {\it negative}.
This means that the quartic term is positive for small values of $\sigma$,
so to obtain chiral 
symmetry breaking, we must tune $m_\sigma^2$ to be negative.
That is no problem, but it also implies that for
large values of $\sigma$, the potential is unbounded from
below, as the logarithmic term $\sim - \, y^4 \, \sigma^4 \, \log(\sigma^2)$
inevitably wins over $\sim + \, \lambda \, \sigma^4$.

It is useful to contrast this to 
the Gross-Neveu model in
$1+1$ spacetime dimensions \cite{gross_dynamical_1974}.  In this model
there 
is a potential term $\sigma^2$, and from the one loop fermion determinant,
a term $\sim + \, \sigma^2 \, \log(\sigma^2)$.  
Because the sign of logarithmic term is {\it positive}, 
the potential is unstable at small $\sigma$, which implies that
there is chiral symmetry breaking for any value of the coupling constant.
Conversely, the total 
potential is stable at large values of 
$\sigma$.  This is opposite what happens in our
effective model in $3+1$ dimensions.

The reason for this difference is clear: the Gross-Neveu model is
asymptotically free \cite{gross_dynamical_1974}, while our model is
infrared free.  As such, we do not expect our theory to be
well behaved at arbitrarily high momenta, which as an effective model
is hardly surprising.  It does imply that we need to check that
we do not obtain results in a regime where there is instability, which we
do.  For the chiral matrix model which is applicable to QCD,
this is easy to satisfy, because $\lambda$ is rather large, $y$ relatively
small, and we never probe large $\sigma$.
We comment
that a similar instability at large $\sigma$ is present in 
renormalization group optimized perturbation theory
\cite{kneur_alpha_s_2013, *kneur_chiral_2015, *kneur_scale_2015, *kneur_renormalization_2015}.

The restoration of chiral symmetry at nonzero temperature is 
straightforward.  In the imaginary time formalism, 
the four momenta $K^\mu = (k_0,\vec{k})$, $k = |\vec{k}|$,
where for a fermion the energy $k_0 = (2 n + 1) \pi T$ for integral
``$n$''.  The trace is
\begin{equation}
{\rm tr} = T \sum_{n=-\infty}^{+\infty}
\widetilde{M}^{2 \epsilon} \int 
\frac{d^{3 - 2 \epsilon}k}{(2 \pi)^{3 - 2 \epsilon}} \; .
\label{eq:trace_T}
\end{equation}
Computing the fermion determinant to one loop order with
$m_f = y \, \sigma \ll T$,
\begin{equation}
- \, \frac{1}{V} \, {\rm tr} \, \log( 
\; \slash \!\!\! \partial + m_f) 
\approx
\frac{1}{12}\; y^2 \, T^2 \, \sigma^2
+ \frac{y^4}{16 \pi^2} \, \sigma^4 
\left( 
\frac{1}{\epsilon}
+ \log
\left( 
\frac{M^2}{T^2} 
\right) 
\right) + \ldots 
\label{expand_ferm_det_small_m}
\end{equation}
From the term quadratic in $\sigma$, we see that there is a second
order chiral phase transition at a temperature
\begin{equation}
T_\chi^2 = - 12 \; \frac{m_\sigma^2}{y^2} \; ,
\label{Tchi_toy_model}
\end{equation}
which is standard.

What is also noteworthy are the subleading terms in the fermion determinant.
At zero temperature we saw that there is a logarithmic term from an
infrared divergence, $\sim \sigma^4 \, \log(\sigma^2)$.  
Eq. (\ref{expand_ferm_det_small_m}) shows that the logarithm of $\sigma$
does not occur at nonzero temperature when $y \sigma \ll T$.  This
is not surprising: for fermions, the energy $k_0$ is always an odd
multiple of $\pi T$.  Thus the energy itself cuts off the infrared divergence,
and the $\log(y^2 \, \sigma^2)$ is replaced by $\log(T)$.

The disappearance of the $\log(\sigma^2)$ at nonzero temperature 
is important to include in our analysis.  It implies that if, as we
show is convenient, we divide the integral into two pieces,
one from $T = 0$, and the other from $T \neq 0$, that the 
$- \sigma^4 \log(y^2 \, \sigma^2)$ in the piece at $T=0$ must cancel against
a similar term, $+ \sigma^4 \log(y^2 \, \sigma^2)$, from the piece
at $T \neq 0$ \cite{skokov_vacuum_2010}.

We conclude our discussion of the toy model by considering the terms
which must be added to describe the explicit breaking of chiral symmetry.
The usual term is just
\begin{equation}
{\cal V}_h = - \, h \, \sigma \; .
\label{h_toy}
\end{equation}
This is perfectly adequate at zero temperature.  
Consider the limit at high temperature, though, where
the effective Lagrangian, including the fermion determinant, is
\begin{equation}
{\cal V}_{\sigma}^{{\rm eff, ren}}
\approx - \, h \, \sigma + \frac{1}{12} \; y^2 \, T^2 \, \sigma^2 + \ldots 
\;\;\; , \;\;\; T \rightarrow \infty \; .
\label{high_temp_toy}
\end{equation}
where the terms of higher order in $\sigma$ do not matter.  Then at
high temperature, 
\begin{equation}
m_f = y \langle \sigma \rangle
\rightarrow  \, \frac{6 \, h}{y \, T^2}  \;\;\; , \;\;\;
T \rightarrow \infty \; ,
\label{high_temp_toy_just_h}
\end{equation}
and the effective fermion mass, $m_f$,
{\it vanishes} as $T \rightarrow \infty$.

For the light quarks in QCD, though, we know that while the constituent
quark mass is much smaller at high temperature than at $T=0$, 
as $T \rightarrow \infty$ it does not vanish, but should asymptote to
the {\it current} quark mass.  In terms of the
original Lagrangian in Eq. (\ref{lagrangian_toy_model}),
we need to require that
\begin{equation}
m_f = y \, \langle \sigma \rangle \rightarrow
m_0 \;\;\; , \;\;\; T \rightarrow \infty \; ,
\label{correct_sigma_highT}
\end{equation}
where $m_0$ is the analogy of the current quark mass in our toy model.

The obvious guess is just to put the current quark mass in the fermion
Lagrangian in the first place, and so start with a modified Lagrangian,
\begin{equation}
{\cal L}_{mod}
= \overline{\psi} 
\left(
\, \not \! \partial \, + m_0 + \, y \, \sigma
\right) 
\psi 
- h \, \sigma 
+ \frac{m_\sigma^2}{2} \, \sigma^2 + \frac{\lambda}{4} \, \sigma^4  \; .
\label{modified_fermion_lag}
\end{equation}
However, at high temperature the effective Lagrangian just becomes
\begin{equation}
{\cal V}_{\sigma}^{{\rm mod, ren}}
\approx - \, h \, \sigma + \frac{1}{12} \; (m_0 + y \, \sigma)^2 
\, T^2 + \ldots \; .
\label{high_temp_modified_toy}
\end{equation}
With this modification we have $\langle \sigma \rangle = - m_0/y$, 
which looks fine.  However, it is clear that in
Eq. (\ref{modified_fermion_lag}), the total effective fermion mass
is $m_f = m_0 + y \langle \sigma \rangle$, so the total effective fermion
mass still {\it vanishes} like
$\sim 1/T^2$ as $T \rightarrow \infty$. 

This problem has not arisen previously because typically
the quarks are integrated out to give an effective chiral model.
In a chiral matrix model, though, we need to keep the quarks as
fundamental degrees of freedom, and so we need
$\sigma$ to approach a small but nonzero value, proportional to the
current quark mass.

In the symmetry breaking term of Eq. (\ref{high_temp_toy})
we assume that $h \sim m_0$.  One solution is then simply to add a new term
which only contributes at nonzero temperature,
\begin{equation}
{\cal V}_{m_0}^T
= - \; \frac{y}{6} \; m_0 \; T^2 \sigma
\; .
\label{new_sym_bkg_T}
\end{equation}
Consequently, at high temperature the effective Lagrangian is now
\begin{equation}
{\cal V}_{\sigma}^{{\rm eff, ren}}
\approx - \, h \, \sigma 
- \; \frac{y}{6} \; m_0 \; T^2 \sigma
+ \frac{1}{12} \; y^2 \, T^2 \, \sigma^2 + \ldots 
\;\;\; , \;\;\; T \rightarrow \infty \; .
\label{high_temp_toy_improved}
\end{equation}
At high temperature the first term $\sim h$ can be neglected.
In this way, the effective fermion mass is just the Yukawa coupling
times the expectation value of $\sigma$, and so 
by construction we obtain the desired behavior, 
\begin{equation}
m_f = y \, \langle \sigma \rangle \rightarrow 
m_0 \;\;\; , \;\;\; T \rightarrow \infty
\; .
\label{highT_toy_sigma}
\end{equation}
That is, we add an additional
term to the effective Lagrangian to ensure that we obtain the requisite
breaking of the chiral symmetry at high temperature, as we did by
adding a term $\sim h \, \sigma$ at zero temperature.

While admittedly inelegant, this is typically the way effective
models are constructed.  
In fact we take a term which is
analogous but not identical to Eq. (\ref{new_sym_bkg_T}), so that
the effective mass is close to the current quark mass even at relatively
low temperatures.  We defer a discussion of the detailed form of the
new symmetry breaking term until Sec. (\ref{sec:sym_bkng_nonzeroT}).

The toy model in this section displays all of the essential physics
in the chiral matrix model which we develop in the following for QCD.
There is one last point which is worth emphasizing.  
In the chiral limit, where $m_0 = h = 0$, we would expect
a chiral transition of second order.  The concern is whether a spurious
first order transition is induced by integrating over quark
fluctuations.  For instance, if the fluctuations are over a bosonic field,
then the energy $k_0$ is an even multiple of $\pi T$, and
there is a mode with zero energy.  Integrating over that mode
generates a cubic term $\sim - (\sigma^2)^{3/2}$, which
drives the transition first order \cite{carrington_effective_1992}.
In our model, however, we integrate over a fermion field, 
where the energy $k_0$ is an odd multiple of $\pi T$, and there
is no mode with zero energy.  Thus the fermion determinant
is well behaved for small $\sigma$, 
Eq. (\ref{expand_ferm_det_small_m}), and the transition is of second
order.  Depending upon the universality class, there can be a first
order transition from fluctuations in the would-be critical
fields \cite{pisarski_remarks_1984}, but at least the model does not generate
one when it should not.

\section{ Matrix model with massless quarks}
\label{sec:matrix_model_massless_quarks}

\subsection{Matrix model for $SU(3)$ gluons without quarks}
\label{sec:matrix_pure_glue}

Following Refs. \cite{dumitru_how_2011, dumitru_effective_2012}, we define
the parameters of a matrix model for a $SU(3)$ theory without quarks.
The basic idea is to incorporate partial confinement in the semi-QGP
through a background gauge field for the timelike component of the
gauge field, $A_0$.  We take the simplest possible ansatz, and
neglect the formation of domains.  Instead, we assume that the
background $A_0$ field is constant in space.  
By a global gauge rotation, we can assume that this field is a diagonal
matrix, and so take the background field to be 
\begin{equation}
A^{bk}_0 = \frac{2 \pi T}{3 \, g} \; 
\left(
q \; \lambda_3 +
r \; \lambda_8
\right) \; ;
\label{eq:ansatz}
\end{equation}
$\lambda_3$ and $\lambda_8$ are proportional to the analogous
Gell-Mann matrices 
\begin{equation}
\lambda_3 = 
\left( 
\begin{array}{ccc}
1 & 0 & 0 \\
0 & -1 & 0 \\
0 & 0 & 0
\end{array}
\right)
\;\;\; ; \;\;\;
\lambda_8 = 
\left( 
\begin{array}{ccc}
1 & 0 & 0 \\
0 & 1 & 0 \\
0 & 0 & -2
\end{array}
\right)
\; .
\label{eq:def_cartan}
\end{equation}

From the background field we can compute the Wilson line in the
direction of imaginary time, $\tau$:
\begin{equation}
{\bf L}(A_0) = {\cal P} \; \exp 
\left( 
i g \int^{1/T}_0 \; A_0 \; d \tau 
\right)
\; ,
\label{define_Wilson_line}
\end{equation}
with ${\cal P}$ path ordering.  
Under a gauge transformation $\Omega$, 
${\bf L} \rightarrow \Omega^\dagger \; {\bf L} \; \Omega$, so
the thermal Wilson line is gauge dependent.  The trace
of powers of ${\bf L}$ are gauge invariant; more generally,
the gauge invariant quantities are 
the eigenvalues of the Wilson line.

For three colors there are two independent eigenvalues,
related to the variables $q$ and $r$.  
As only the exponentials enter into the Wilson line, these are
then periodic variables.  (Mathematically,
this periodicity is related to the Weyl chamber.)  We note that at one
loop order the eigenvalues of the thermal Wilson line
are directly given by $q$ and $r$, but 
beyond one loop order, there is a finite, gauge and field dependent
shift in these variables
\cite{bhattacharya_interface_1991, *bhattacharya_zn_1992, korthals_altes_constrained_1994, dumitru_two-loop_2014}.  

This periodicity can be
understood from the Polyakov loop,
as the trace of the Wilson line in the
background field of Eq. (\ref{eq:ansatz}):
\begin{equation}
\ell_{bk} = 
\frac{1}{3} \; {\rm tr} \; 
{\bf L}(A_0^{bk})
= \frac{{\rm e}^{2 \pi i r/3}}{3} 
\left( 
{\rm e}^{- 2 \pi i \, r} + 2 \cos\left(\frac{2 \pi }{3} \, q\right) 
\right) \; .
\label{eq:loop_gen_qr}
\end{equation}
In the perturbative vacuum, $\ell_{bk} = 1$.  

When $r = 0$, the Polyakov
loop is real; the confined vacuum 
in the pure gauge theory corresponds to $q = 1$, with $\ell_{bk} = 0$.
We can always assume that 
the Polyakov loop is real.  Thus one goes from the perturbative
vacuum at high temperature, to the confining vacuum
at low temperatures, by varying $q$ along a path with $r = 0$.

Rotations in $Z(3)$ correspond to $r \neq 0$: for example, 
$q = 0$ and $r = \pm 1$ gives $\ell_{bk} = \exp(\pm 2 \pi i/3)$,
so these represent $Z(3)$ rotations of the perturbative vacuum.
The interface tension between different $Z(3)$ can be computed
semiclassically, by varying $r$ along
a path with $q = 0$ 
\cite{bhattacharya_interface_1991, *bhattacharya_zn_1992};
near $T_d$ in the semi-QGP, one moves from $r=0$ to $r=1$ along
a path where both $q$ and $r$ vary
\cite{dumitru_how_2011, dumitru_effective_2012}.

Since the background field is a constant, diagonal matrix, the gluon field
strength tensor vanishes, and all $q$ are equivalent.  This degeneracy
is lifted at one loop order.  As typical of background field computations,
one takes
\begin{equation}
A_\mu = A_\mu^{bk} + A_\mu^{qu} \;  .
\label{eq:bkgd_field}
\end{equation}
and expands to quadratic order in the quantum fluctuations, $A_\mu^{qu}$.
This is best done in background field gauge
\cite{bhattacharya_interface_1991, *bhattacharya_zn_1992, korthals_altes_constrained_1994, dumitru_two-loop_2014}.  

For three colors the result is
\begin{equation}
{\cal V}^{gl}_{pert}(q,r)
= \frac{1}{V} \; {\rm tr} \; {\rm log} 
\left(
- D_{bk}^2 
\right)
= \pi^2 \; T^4 \; 
\left(
- \; \frac{ 8}{45} + \frac{4}{3} \; {\cal V}_4(q,r) 
\right) \; .
\label{perturbative_gluon_potential}
\end{equation}
The first term is minus the pressure of eight massless gluons.  The
second term is the potential 
\begin{equation}
{\cal V}_4(q,r)
= \left|\frac{2 q}{3}\right|^2 
\left( 1 - \left|\frac{2 q}{3}\right| \right)^2
+  \left| \frac{q}{3} + r \right|^2 
\left( 1 - \left| \frac{q}{3} + r \right|\right)^2 
+  \left| \frac{q}{3} - r \right|^2 
\left( 1 - \left| \frac{q}{3} - r \right|\right)^2  \; .
\end{equation}
In this and {\it all} further expressions, 
each absolute value is defined modulo one:
\begin{equation}
|x| \equiv |x|_{modulo \; 1} \; .
\label{eq:define_modulo}
\end{equation}
This arises because in thermal sums over integers ``$n$'', 
$D_0^{bk} = i \, 2 \pi T (n + x)$, and clearly any integral shift
in ``$x$'' can be compensated by one in ``$n$''.

When $r=0$,
\begin{equation}
{\cal V}^{gl}_{pert}(q,0) = \frac{8 \pi^2}{45} \;  T^4 \;
\left(
- \, 1 + \; 5 \, q^2 
\left( 
1 - \, \frac{10}{9} \, q + \, \frac{1}{3} \, q^2
\right) 
\right)
\; .
\label{pert_q_gluon_pot}
\end{equation}
Since ${\cal V}^{gl}_{pert}(1,0) > {\cal V}^{gl}_{pert}(0,0)$,
the pressure in the confined vacuum is less than that of the perturbative
vacuum, and so disfavored.

To obtain an effective theory for the confined vacuum, by hand we add
a term to drive the transition to confinement:
\begin{equation}
{\cal V}_{non}^{gl}(q,r) = 
 \; \frac{4 \pi^2}{3} \; T^2 \, T_d^2 \; \left(
- \; \frac{1}{5} \; c_1 \; {\cal V}_2(q,r)
- \;  c_2 \; {\cal V}_4(q,r) + \frac{2}{15} \; c_3 \right) \; ,
\label{nonpert_gluon_pot}
\end{equation}
where
\begin{equation}
{\cal V}_2(q,r) 
= \left|
\frac{2 q}{3}
\right| 
\left( 
1 - 
\left|
\frac{2 q}{3}
\right| 
\right)
+  
\left| 
\frac{q}{3} + r 
\right|
\left( 
1 - 
\left| 
\frac{q}{3} + r 
\right|
\right)
+  
\left| 
\frac{q}{3} - r 
\right|
\left( 
1 - 
\left| 
\frac{q}{3} - r 
\right|
\right) 
\; ;
\end{equation}
again, each absolute value is defined modulo one.
When $r=0$,
\begin{equation}
{\cal V}_{non}^{gl}(q,0) = 
\frac{8 \pi^2}{45} \; T^2 \; T_d^2 \; 
\left( 
- 2 \; c_1 \; q 
\left( 
1 - \frac{q}{2} 
\right)
- 5 \; c_2 \; 
q^2 
\left( 
1 - \frac{10}{9}\, q + \frac{q^2}{3} 
\right) 
+ c_3
\right) 
\; .
\label{eq:non_pert_gluon_pot}
\end{equation}

The nonperturbative terms are assumed to be proportional to $T^2$
because of the following.  Numerical simulations of lattice $SU(3)$
gauge theories find that the leading correction to the leading $\sim T^4$ term
in the pressure is $\sim T^2$
\cite{boyd_equation_1995, *umeda_fixed_2009, *borsanyi_precision_2012}.
This was first noticed by Meisinger, Miller, and Ogilvie
\cite{meisinger_phenomenological_2002, *meisinger_complete_2002},
and then by one of us 
\cite{pisarski_effective_2006}.
This is a {\it generic} property  of pure gauge theories, and holds
for $SU(N_c)$ gauge theories from $N_c = 2 \rightarrow 8$
\cite{panero_thermodynamics_2009, *datta_continuum_2010}.
In $2+1$ dimensions, where the ideal gas term is $\sim T^3$, again
the leading correction is $\sim T^2$ when $N_c = 2 \rightarrow 6$
\cite{caselle_thermodynamics_2011}.  In both cases, if one divides
the pressure by the number of perturbative gluons, $= N_c^2 -1 $, one
finds a universal curve, independent of $N_c$, for $T > 1.1 \; T_d$ 
(closer to $T_d$, differences in the order of the transition enter).

The results of these lattice simulations in pure $SU(N_c)$ gauge
theories strongly suggests 
that massless strings, with a free energy $\sim T^2$, persist in the
{\it de}confined phase.  
Strings can be either closed or open.  In the confined phase,
both are color singlets, with a free energy $\sim N_c^0$.  For
open strings, this implies that the color charge at one end of the
string matches the color charge at the other.  In the deconfined phase,
however, near $T_d$ lattice simulations show that the free energy of
the deconfined strings, $\sim T^2$, has a free energy which is 
$\sim N_c^2-1$.  This must then be due to open strings where the
color charges at each end do {\it not} match.

Returning to the matrix model for $SU(3)$,
the three parameters $c_1$, $c_2$, and $c_3$ 
are reduced to one parameter by imposing two conditions.
The first is that the transition occurs at $T_d$.
For the second, we approximate the small, but nonzero
\cite{meyer_high-precision_2009}, pressure in the confined phase
by zero.  These two equations give
\begin{equation}
c_1 = \frac{50}{27} (1 - c_2) \;\;\; , \;\;\;
c_3 = \frac{47 - 20 \, c_2}{27} \; ,
\end{equation}
Eqs. (77) and (78) of Ref. \cite{dumitru_effective_2012}.
The single remaining parameter, $c_2$,  is then adjusted to agree with
the results from lattice simulations for $(e-3p)/T^4$.  The best fit gives
\begin{equation}
c_1 = 0.315       \;\;\; ; \;\;\;
c_2 = 0.830 \;\;\; ; \;\;\;
c_3 = 1.13 \; .
\end{equation}

We remark that besides terms $\sim T^2$, it is also natural to add
terms $b \sim T^0$, which represent a nonzero MIT ``bag'' constant
\cite{dumitru_effective_2012}.  We do not include such a term for
the following reason.  From lattice simulations, 
in QCD the chiral crossover takes place at a temperature 
$T_\chi \ll T_d$.  Consider the interaction measure, defined as
$\Delta = (e-3p)/T^4$, where $e$ is the energy density, and $p$ the
pressure, each at a temperature $T$.  Clearly, terms $\sim T^2 \, T_d^2$
contribute to the interaction measure $\Delta \sim T_d^2/T^2$, while
a bag constant gives $\Delta \sim b/T^4$.  In the pure gauge theory,
where only temperatures $T \geq T_d$ enter, a better fit
is found with $b \neq 0$ \cite{dumitru_effective_2012}.  With
dynamical quarks, however, as the model is pushed to much lower
temperatures $\sim T_\chi$, we find that at such relatively low
temperatures, that a nonzero bag constant uniformly is difficult to
incorporate into the model.  

The parameters of the model are chosen to agree with the pressure
obtained from the lattice \cite{dumitru_effective_2012}.  The results
for the 't Hooft loop agree well with the lattice, but there
is sharp disagreement for the Polyakov loop, as that
in the matrix model approaches unity much quicker than on the lattice.
Consequently, in Sec. (\ref{sec:alternate}) we consider an alternate
model: while involving many more parameters, the value of the Polaykov
loop is in agreement with the lattice.  We then use this model to compute
susceptibilities in QCD.

\subsection{Adding massless quarks to the matrix model}
\label{sec:quark_potential}

The Lagrangian for massless quarks is
\begin{equation}
{\cal L}^{qk} = \overline{\psi}
\left( \, \not \!\! D \, 
+  \, \mu \, \gamma^0 \, \right)\psi \; ,
\end{equation}
with $D_\mu = \partial_\mu - i g A_\mu$ the covariant derivative
in the fundamental representation, and $\mu$ is the quark chemical
potential.  In the background field of Eq. (\ref{eq:bkgd_field}), 
for a single massless quark flavor, to
one loop order quarks generate the potential
\cite{altes_potential_2000}
\begin{equation}
{\cal V}^{qk}_{pert}(q,r,\widetilde{\mu}) 
= - \; \frac{1}{V} \; 2 \; {\rm tr} \, 
\log
\left(
\, \not \!\! D \, 
+  \, \mu \, \gamma^0 \,
\right)^2
= 
\pi^2 \,
T^4 
\left( 
- \, \frac{2}{15} 
+ \, \frac{4}{3} \; 
{\cal V}^{qk}_4(q,r,\widetilde{\mu})
\right) 
\; ,
\end{equation}
where 
\begin{equation}
\widetilde{\mu} = \frac{\mu}{2 \pi T}
\end{equation}
and
\begin{eqnarray}
{\cal V}^{qk}_4(q,r,\widetilde{\mu})
&=& 
\left|
\frac{q + r}{3} + \frac{1}{2} + i \, \widetilde{\mu} 
\right|^2 
\left( 
1 - \left|\frac{q+r}{3}
+ \frac{1}{2} + i \, \widetilde{\mu} 
\right| 
\right)^2
\nonumber \\
&+&  
\left| \frac{-q + r}{3} + \frac{1}{2} + i \, \widetilde{\mu} 
\right|^2 
\left( 
1 - \left| \frac{-q + r }{3} + \frac{1}{2} + i \, \widetilde{\mu} 
\right|
\right)^2 
\nonumber \\
&+&  
\left|  \frac{ - 2 r}{3} + \frac{1}{2} + i \, \widetilde{\mu} 
\right|^2 
\left( 
1 - 
\left| 
\frac{- 2 r}{3} + \frac{1}{2} + i \, \widetilde{\mu} 
\right|
\right)^2  
\; .
\label{eq:qk_potential_general}
\end{eqnarray}
At a temperature $T$,
bosons satisfy periodic boundary conditions in imaginary time,
and fermions, antiperiodic; the factor of $1/2$ in the above
is because the energy is $2 n \pi T$ for bosons, and $(2n + 1) \pi T$
for fermions, with ``$n$'' an integer.

There are subtleties which arise when the quark chemical potential
is nonzero.  To understand these, first consider the case in which
the chemical potential is purely {\it imaginary}.  As noted before,
a $Z(3)$ transformation of the perturbative vacuum is given by
$q=0$ and $r=1$, with the Polyakov loop $\ell = \exp(2\pi i/3)$.
Inspection of the quark potential in Eq. (\ref{eq:qk_potential_general})
shows that when $r=1$, we can compensate this by choosing
$i \widetilde{\mu} = - 1/3$.  This is obvious for the first two 
terms, where $r/3 + i \widetilde{\mu}$ enters.  For the last term,
which involves $|- 2r/3 + 1/2 + i \widetilde{\mu}|$, this
occurs because the absolute value is defined modulo one,
Eq. (\ref{eq:define_modulo}).  

This is an illustration of the Roberge-Weiss phenomena 
\cite{roberge_gauge_1986, kashiwa_roberge-weiss_2013, aarts_introductory_2015}.
While the theory with dynamical quarks does not respect a global
$Z(3)$ symmetry, it does exhibit a symmetry under shifts by an
imaginary chemical potential.  As this is related to $Z(3)$, in $SU(3)$
the corresponding generator is $\lambda_8 = {\rm diag}(1,1,-2)$,
Eq. (\ref{eq:def_cartan}).  For a $SU(N)$ gauge
theory, the corresponding generator is that related to $Z(N)$ transformations,
which is $\lambda_N = {\rm diag}({\bf 1}_{N-1}, -(N-1))$.

Thus nonzero, real values of $r$ naturally involve imaginary $\mu$.  
We bring up this point because it also helps understand the converse,
which is that for {\it real} values of the chemical potential $\mu$,
the stationary point involves values of $r$ which are {\it imaginary}.

Remember that a chemical potential biases particles over antiparticles.
The loop, as the propagator of an infinitely heavy test quark,
tends to enter effective Lagrangians as ${\rm e}^{-\mu/T} \ell$; the antiloop,
as ${\rm e}^{\mu/T} \ell^*$ \cite{dumitru_dense_2005}.
Thus when $\mu \neq 0$, the expectation values of both the loop
and the antiloop are real, but {\it unequal}.

For this to be true in a matrix model, at any stationary point where
$q \neq 0$, $r$ must be imaginary,
\begin{equation}
r = i {\cal R} \; .
\label{eq:imag_rR}
\end{equation}
For this background field, from Eq. (\ref{eq:loop_gen_qr}) the loop is
\begin{equation}
\ell_{bk} = 
\frac{{\rm e}^{- 2 \pi {\cal R}/3}}{3} 
\left( 
{\rm e}^{2 \pi \, {\cal R}} + 2 \cos\left(\frac{2 \pi }{3} \, q\right) 
\right) \; ,
\label{eq:loopR}
\end{equation}
while the antiloop is given by
\begin{equation}
\ell_{bk}^* = 
\frac{{\rm e}^{2 \pi {\cal R}/3}}{3} 
\left( 
{\rm e}^{ - 2 \pi \, {\cal R}} + 2 \cos\left(\frac{2 \pi }{3} \, q\right) 
\right) \; .
\label{eq:loopRconj}
\end{equation}
Hence imaginary $r$ generates different values for the loop and the antiloop.

In Sec. (\ref{sec:flavor_susp}) we shall need to use the fact that
the stationary point when $\mu \neq 0$ involves imaginary
values of $r = i {\cal R}$.  For now we conclude this discussion
by making one comment about periodicity of the potential.
In previous expressions for the potential, the absolute value is
defined modulo one, Eq. (\ref{eq:define_modulo}).  One then needs
to understand how to extend this definition when the argument is complex.
The correct prescription is
to take the absolute value, modulo one, only for the {\it real} part of the
argument, leaving the imaginary part unaffected \cite{altes_potential_2000}:
\begin{equation}
|x + i y| \equiv |x|_{modulo \; 1} + i y \; ,
\label{eq:define_modulo_imag}
\end{equation}
As before, this is natural in considering the sum over thermal energies
which arises in the trace.

When $r = \mu = 0$,
\begin{equation}
{\cal V}^{qk}_{pert}(q,0,0) = 
\pi^2 \, T^4 \left(
- \; \frac{7}{60} + \frac{4}{27} \; q^2 
- \, \frac{8}{243} \; q^4 \right) \; .
\end{equation}

In the following, we make the simplest possible assumption, 
which is that we only need to 
add the perturbative potential for quarks in $q$ and $r$.
Doing so, we find a very good fit to the pressure and other
thermodynamic quantities.  
That is, unlike the gluonic part of the theory, at least
from the pressure we see 
no evidence to indicate that it is necessary to add a nonperturbative
potential in $q$ from the quarks.

We note, however, that in Sec. (\ref{sec:alternate}), we consider
alternate models where different potentials are used.  We show that
they lead to strong disagreements with either the pressure or
quark susceptibilities.

\section{Chiral matrix model for three flavors}
\label{sec:meson_potential}

\subsection{Philosophy of an effective model, with and without quarks}

For a $SU(N_c)$ gauge theory without quarks, the matrix model of
Refs. \cite{dumitru_how_2011, dumitru_effective_2012} is clearly
applicable only at temperatures above the deconfining transition
temperature.  This is because even for two colors, the pressure in
the confined phase is very small 
(for three colors, see Ref. \cite{meyer_high-precision_2009}).
This is evident by considering large $N_c$, where the pressure 
of deconfined gluons in the
deconfined phase is $\sim N_c^2$, while that of confined
glueballs in the confined phase is $\sim N_c^0$.  

This is not true with dynamical quarks.  To make the argument
precise, assume that we have $N_f$ flavors of massless quarks.
If the chiral symmetry is spontaneously broken at zero temperature,
then the low temperature has a pressure which is $\sim N_f^2 - 1$
from the Goldstone bosons, plus other contributions from confined
hadrons.  At high temperature, deconfined quarks contribute
$\sim N_f \, N_c$ to the pressure, while the gluons contribute
$\sim N_c^2$.  

Thus for three colors and three flavors, it is not 
obvious that the pressure is small at low temperatures,
and becomes large at high temperature.  Nevertheless, numerical
simulations on the lattice find that for
$2+1$ flavors and three colors, at a chiral crossover
temperature of $T_\chi \sim 155$~MeV,
the pressure is rather small.  

Similarly, consider the order parameter for deconfinement in the
$SU(N_c)$ gauge theory without quarks, which is the expectation value
of the Polyakov loop.  This is a strict order parameter because there
is a global $Z(N_c)$ symmetry which is restored in the confined phase,
and spontaneously broken in the deconfined phase.  
Dynamical quarks do not respect this $Z(N_c)$ symmetry, and so
the Polyakov loop is no longer a strict order parameter.
This is seen in lattice QCD, where the expectation value of the Polyakov
loop is nonzero at {\it all} temperatures $T > 0$.  Nevertheless,
as for the pressure, the expectation value of the Polyakov loop is
surprisingly small in  QCD at $T_\chi$, 
$\langle \ell \rangle \sim 0.1$.

As with so much else, this is important input from lattice QCD.  There
is no reason to believe that this remains true as $N_f$ and $N_c$ change;
in particular, as $N_f$ increases for three colors.

This is surely related to the fact that lattice QCD
finds that $T_\chi = 155$~MeV is {\it much}
less than the deconfining transition temperature in the $SU(3)$
gauge theory without quarks, $T_d = 270$~MeV.  Thus adding
dynamical quarks inexorably requires us to push the matrix model
to much lower temperatures than in the pure glue theory.

Further, in our effective theory we do not presume to be able to
develop a model by which we can derive chiral symmetry breaking from
first principles.  Rather, as described at the beginning of the
Introduction, Sec. (\ref{sec:intro}),
we merely wish to develop an effective theory which
can be used to extrapolate results from lattice QCD in equilibrium
to quantities near equilibrium.

To do so, unsurprisingly it is necessary to explicitly introduce
degrees of freedom to represent the spontaneous breaking of chiral
symmetry, through a field $\Phi$.  
What is not so obvious is that we find that it is also
necessary to introduce parameters for a potential for $\Phi$, which
we describe shortly.  In principle, we might ask that lattice
QCD determine these parameters directly, say at a temperature near
but below $T_\chi$.  For example, at a temperature $\sim 130$~MeV, where
the hadronic resonance gas first appears to break down.

While possible, in practice determining such couplings from lattice QCD is a
rather daunting task.  Instead, since the hadronic resonance gas does appear
to work at temperatures surprisingly close to $T_\chi$, we require
that our effective chiral model describe the mass of the (pseudo-)Goldstone
bosons in QCD all the way down to {\it zero} temperature.

While clearly a drastic assumption, it is a first step towards a more
complete effective theory.  With these caveats aside, we turn to the
detailed construction of our chiral matrix model.

\subsection{Linear sigma model}
\label{sec:std_meson_pot}

One thing which we certainly do need to add with dynamical quarks
are effective degrees of freedom to model the restoration of chiral
symmetry.   We do this by introducing a scalar field $\Phi$,
and an associated linear sigma model 
\cite{donoghue_dynamics_1992}.
To be definite, in this work we follow 
the conventions of Ref. 
\cite{lenaghan_chiral_2000}; for related work, see Refs. 
\cite{roeder_chiral_2003, *janowski_glueball_2011, *parganlija_meson_2013, stiele_phase_2016, jaffe_multiquark_1977, *black_mechanism_2000, *close_scalar_2002, *jaffe_diquarks_2003, *maiani_new_2004, *pelaez_light_2004, *t_hooft_theory_2008, pelaez_controversy_2015}.

We only treat the three lightest flavors of quarks in QCD,
up, down, and strange.
In the chiral limit, classically there is a global flavor symmetry of
$G_f^{cl} = SU(3)_L \times SU(3)_R \times U(1)_A$, where the $U(1)_A$
axial flavor
symmetry is broken quantum mechanically by the axial anomaly to a discrete
$Z(3)_A$ symmetry,
$G_f^{qu} = SU(3)_L \times SU(3)_R \times Z(3)_A$. 

For three flavors the $\Phi$ field is a complex nonet,
\begin{equation}
\Phi = (\sigma^A + i \pi^A) \; t^A \;\;\; , \;\;\;
{\rm tr} \left( t^A t^B \right) = \frac{1}{2} \, \delta^{A B} \; .
\end{equation}
The flavor indices $A = 0,1\ldots8$, where $t^0 = {\bold 1}/\sqrt{6}$,
and $t^1\ldots t^8$ are the usual Gell-Mann matrices.  

For particle nomenclature, we follow 
that of the Particle Data Group \cite{olive_review_2014}.
The field $\Phi$ includes a nonet with spin-parity
$J^P = 0^-$: $\pi^{1\ldots 3}$ are pions,
$\pi^{4 \ldots 7}$ are kaons, while $\pi^8$ and $\pi^0$ mix to form
the observed $\eta$ and $\eta'$ mesons.
The nonet with $J^0 = 0^+$ includes the following particles.
First, there is an isotriplet,  
$\sigma^{1 \ldots 3}$, which could be the isotriplet $a_0(980)$.
Second, there are its associated strange mesons, 
$\sigma^{4 \ldots 7}$.
This state may be the $K_0^*$; there are candidate states at
both $800$ and $1430$~MeV \cite{olive_review_2014}.
Lastly, analogous to the $\eta$ and the $\eta'$ there are
isoscalar and iso-octet states, which are commonly referred to as
the $f_0$ and the $\sigma$.  Experimentally, the candidates for
these states are $f_0(1500)$ and  $\sigma(500)$.

Under global flavor rotations, 
\begin{equation}
\psi_{L,R} \equiv {\cal P}_{L,R} \; \psi \;\;\; ; \;\;\; 
\psi_{L,R} \rightarrow \; {\rm e}^{\pm \, i \, \alpha/2} 
\; U_{L,R} \; \psi_{L,R}
\;\;\; ; \;\;\; 
\Phi \rightarrow {\rm e}^{- \, i \alpha} \; U_{R} \; \Phi \; U_{L}^\dagger \; ;
\label{eq:flavor_rotations}
\end{equation}
where 
\begin{equation}
{\cal P}_{L,R} = \frac{1 \pm \gamma_5}{2}
\end{equation}
are the chiral projectors,
${\rm e}^{\pm i \alpha/2}$ represent axial $U(1)_A$ rotations,
and $U_{L}$ and $U_R$ rotations for the chiral symmetries of
$SU(3)_L$ and $SU(3)_R$, respectively.
Hence the $\Phi$ field then transforms as
${\bf \overline{3}} \times {\bf 3}$ under $SU(3)_L \times SU(3)_R$.

Coupling quarks to $\Phi$ in a chirally invariant manner, 
the quark Lagrangian becomes
\begin{equation}
{\cal L}^{qk}_{\Phi} = 
\overline{\psi} 
\left(
\, \not \!\! D \; + 
 \mu \, \gamma^0
+ \; y 
\left(
\Phi \, {\cal P}_L
+ \Phi^\dagger \, {\cal P}_R 
\right)
\right) 
\psi \; ,
\label{eq:quark_lagrangian_with_phi}
\end{equation}
where $y$ is a Yukawa coupling between the quarks and the $\Phi$ field.
Note that by construction the theory is invariant under both the
$SU(3)_L \times SU(3)_R$ and $U(1)_A$ chiral symmetries.

To model chiral symmetry breaking we assume a potential for $\Phi$ which
will produce a constituent quark mass in the low temperature phase.  
In the chiral limit, this potential must respect the flavor
symmetry, $G_f^{qu}$.  Including
terms up to quartic order, the most general potential is
\begin{equation}
{\cal V}_\Phi 
= \; m^2 \; {\rm tr} 
\left(
\Phi^\dagger  \Phi 
\right)
- c_A 
\left( 
\det \Phi + \det\Phi^\dagger
\right)
+ \; \lambda  \; {\rm tr} 
\left(
\Phi^+ \Phi 
\right)^2 
+ \; \lambda_V  \; 
\left(
{\rm tr} 
\left(
\Phi^+ \Phi 
\right)
\right)^2  \; .
\label{eq:phi_potential}
\end{equation}
All terms are manifestly invariant under
$SU(3)_L \times SU(3)_R$.  
As they are formed from combinations of $\Phi^\dagger \Phi$,
they are also invariant under the axial $U(1)_A$ symmetry.
The cubic determinantal term is only invariant when the axial phase
$\alpha = 2 \pi \, j/3$, where $j = 0, 1, 2$, 
which is a discrete symmetry of axial $Z(3)_A$
\cite{goldberg_chiral_1983, pisarski_remarks_1984}.
We define $\Phi$ to have axial charge one.

The last quartic term, 
$\sim ({\rm tr}(\Phi^\dagger \Phi))^2$, is
invariant under a larger flavor symmetry of $O(18)$.
This term is suppressed when the number of colors, $N_c$, is large,
with the coupling constant $\lambda_V \sim 1/N_c$
\cite{coleman_chiral-symmetry_1980}.  
Phenomenologically, this coupling is very small:
Ref. \cite{lenaghan_chiral_2000} finds $\lambda_V \approx 1.4$, while
$\lambda \approx 46$, so $\lambda_V \ll  \lambda$. 
Thus we neglect $\lambda_V$ in our analysis.

We also add a term to break the chiral symmetry,
\begin{equation}
{\cal V}^{0}_{H} = 
- \; {\rm tr} 
\left(
H 
\left(
\Phi^\dagger  + \Phi 
\right)
\right) \; .
\label{eq:zero_temp_mass_term}
\end{equation}
The background field $H$ is proportional to the
current quark masses, $m_{qk}$.  We shall assume isospin degeneracy
between the up and down quarks, and so take
\begin{equation}
H = \; {\rm diag}(h_u,h_u,h_s) \; ,
\label{definition_H}
\end{equation}
where $h_{u,s} \sim m_{u,s}$, with $m_u = m_d$ and $m_s$ the current
quark masses.  The superscript in ${\cal V}^{0}_{H}$
denotes that this symmetry breaking term is at zero temperature;
in Sec. (\ref{sec:nonzero_T_chiral_bkng}), we show that an additional
term is required at nonzero temperature,  ${\cal V}^{T}_{H}$.

\subsection{Logarithmic terms for $2+1$ flavors}

The novel term is ultraviolet finite, $\sim m^4 \log(m^2)$.  
Generalizing to three flavors of quarks this becomes
\begin{equation}
{\cal L}^\psi(m_i) = 
+ \sum_{i = 1}^{3} \frac{3 \, m_i^4}{16 \pi^2}
\left( 
\frac{1}{\epsilon}
+ \log
\left( 
\frac{M^2}{m_i^2} 
\right) 
\right) \; ,
\label{eq:zero_temp_several_flavors}
\end{equation}
where ``$i$'' is the flavor index, and the overall factor of
three is from color.

We wish to generalize Eq. (\ref{eq:zero_temp_several_flavors})
to a form which is manifestly chirally symmetric.  To do this,
we simply need to recognize that a mass corresponds to
an expectation value for the diagonal components of
$\Phi$,
\begin{equation}
m_i = y \; 
\langle 
\Phi_{i i} 
\rangle \; .
\end{equation}
Hence the expression for several flavors is just the sum over flavors
of each term
\begin{equation}
{\cal V}_{T=0}(m_i) = 
+ \sum_{i = 1}^{N_f} \frac{3 \, m_i^4}{16 \pi^2}
\left( 
\frac{1}{\epsilon}
+ \log
\left( 
\frac{M^2}{m_i^2} 
\right) 
\right) \; .
\label{eq:zero_temp_sum_over_flavors}
\end{equation}
It is then evident that for arbitrary $\Phi$, we need to add a counterterm
\begin{equation}
{\cal V}^{ct}_{\Phi} = 
- \; \frac{3 \, y^4}{16 \pi^2} \; \frac{1}{\epsilon} 
\; {\rm tr} \; 
\left(
\Phi^\dagger \Phi
\right)^2 
\; ,
\label{eq:counterterm}
\end{equation}
which is standard.

However, this computation shows that it is also
necessary to include in the effective Lagrangian a novel term,
\begin{equation}
{\cal V}^{{\rm log}}_{\Phi} = 
\; \frac{3 \, y^4}{16 \pi^2} \; 
\; {\rm tr} \; 
\left[
\left(
\Phi^\dagger \Phi
\right)^2 
\log
\left(
\frac{M^2}{\Phi^\dagger \Phi}
\right)
\right]
\; ,
\label{eq:new_counterterm}
\end{equation}
where the trace is only over flavor indices.
This term does not arise in the usual analysis of effective Lagrangians,
which assumes that all terms are polynomials in $\Phi$.  We cannot
avoid introducing such a term, since it will be induced by
integrating over the quarks.  The necessity of introducing such a
term was noted by Stiele and Schaffner-Bielich 
\cite{stiele_phase_2016}.

We comment that if one were to compute in our model beyond one loop
order, that many other logarithmic terms will obviously be introduced.  
These include
\begin{equation}
{\rm tr}
\left(
\Phi^\dagger \Phi
\right)^2
\; 
{\rm tr}
\log
\left(
\Phi^\dagger \Phi
\right)
\;\;\; ; \;\;\;
\left(
{\rm tr}
\; \Phi^\dagger \Phi
\right)^2
\; 
{\rm tr}
\log
\left(
\Phi^\dagger \Phi
\right)
\; ,
\end{equation}
and so on.  Since they involve two traces over flavor, they
are suppressed by $\sim 1/N_c$ 
\cite{coleman_chiral-symmetry_1980}.

\subsection{Sigma model at zero temperature: masses}
\label{sec:masses_sigma_model}

In this section we determine the parameters of the linear sigma
model by fitting to the spectrum of the light Goldstone bosons in QCD.
Because of the novel term in Eq. (\ref{eq:new_counterterm}), with
a term which involves the logarithm of $\Phi$, this is similar,
but not identical, to the analysis where only polynomials in $\Phi$
are included:
\begin{eqnarray}
{\cal V}_\Phi^{tot}
= 
{\cal V}^{0}_{H}
+ {\cal V}_\Phi
+ {\cal V}^{{\rm log}}_{\Phi} 
&=& 
- \; {\rm tr} 
\left(
H 
\left(
\Phi^\dagger  + \Phi 
\right)
\right)
+ \,  m^2 \; {\rm tr} 
\left(
\Phi^\dagger  \Phi 
\right)
\nonumber \\
&-& c_A 
\left( 
{\rm det}\; \Phi + {\rm det} \;\Phi^\dagger
\right)
+ \; {\rm tr} 
\left[
\left(
\Phi^+ \Phi 
\right)^2 
\left(
\lambda
+ \kappa \; 
\log 
\left( 
\frac{M^2}{\Phi^\dagger \Phi}
\right)
\right)
\right]  .
\label{eq:chiral_lagrangian}
\end{eqnarray}
For ease of notation we redefine
\begin{equation}
\kappa = \frac{3 \, y^4}{16 \pi^2} \; .
\end{equation}

We assume a nonzero expectation value for $\Phi$,
\begin{equation}
\langle \Phi \rangle  = t^0 \langle \Phi_0 \rangle 
+ t^8 \langle \Phi_8 \rangle= 
 \begin{pmatrix}
  \Sigma_u & 0 & 0 \\
  0 & \Sigma_u  & 0 \\
  0 & 0 & \Sigma_s 
 \end{pmatrix} \; .
\label{Eq:Phi}
\end{equation}
Since we treat the high temperature phase, we find it convenient to use
the flavor diagonal expectation values, $\Sigma_u$ and $\Sigma_s$, which
are related to the $SU(3)_f$ values by
\begin{eqnarray}
\Sigma_u &=& \frac{1}{\sqrt{6}} \left( 
\langle \Phi_0\rangle+ \frac{1}{\sqrt{2}} \;
\langle \Phi_8 \rangle \right) \; , \\ 
\Sigma_s &=& \frac{1}{\sqrt{6}} \left( 
\langle \Phi_0\rangle - \, \sqrt{2} \; \langle \Phi_8 \rangle \right) 
\; .
\label{Eq:sl_ss}
\end{eqnarray}
At zero temperature, where the effects of the axial anomaly, $c_A \neq 0$,
are large, then it is natural to use eigenstates of $SU(3)_f$ flavor.
At high temperature, however, the mass eigenstates are more natural
in a flavor diagonal basis.  It is for this reason that we use
both the $SU(3)_f$ expectation values $\Phi_{0,8}$ and
the flavor diagonal $\Sigma_{u,s}$.

We define
\begin{equation}
\Phi = \langle \Phi \rangle + \delta \Phi 
\; ,
\label{define_expansion}
\end{equation}
and expand the potential in the fluctuations, $\delta \Phi$.

Expanding to linear order in $\delta \Phi$ gives the equations of motion,
\begin{eqnarray}
\frac{h_u}{\Sigma_u}
&=& m^2 - c_A \, \Sigma_s + 2 \, \lambda \, \Sigma_u^2
+ \kappa \, \Sigma_u^2 
\left(
-\, 1 + 2 \, \log
\left(
\frac{M^2}{\Sigma_u^2}
\right)
\right) \; ,
\label{eq:eq_motion_u}
\\
\frac{h_s}{\Sigma_s}
&=& m^2 - c_A \, \frac{\Sigma_u^2}{\Sigma_s} + 2 \, \lambda \, \Sigma_s^2
+ \kappa \, \Sigma_s^2 
\left(
-\, 1 + 2 \, \log
\left(
\frac{M^2}{\Sigma_s^2}
\right)
\right)
\; .
\label{eq:eq_motion_s}
\end{eqnarray}
For the meson masses at zero temperature, by using the equations of motion
we can eliminate all factors
of $\log(M^2/\Sigma^2)$ for $h_u$ and $h_s$, and thus eliminate any
dependence upon the renormalization mass scale $M$.  This agrees
with the expectation that physical quantities are independent of
$M$.

The mass squared for the pion can be derived directly by simply expanding
the effective Lagrangian to quadratic order in the pion field,
\begin{equation}
m_\pi^2
= m^2 - c_A \; \Sigma_s + 2 \lambda \; \Sigma_u^2
+ \kappa \; \Sigma_u^2 
\left(
- 1 + 2 \log
\left(
\frac{M^2}{\Sigma_u^2}
\right)
\right) \; .
\label{eq:pion_mass}
\end{equation}

For the kaon, it is necessary to be a bit more careful.  
This is due to the presence of $\log (\Phi^\dagger \Phi)$ in the potential,
and because the expectation value 
$\langle \Phi^\dagger \rangle \langle \Phi \rangle$, while diagonal,
is not proportional to the unit matrix.
However, it is simply necessary
to compute the logarithm of $\Phi^\dagger \Phi$ to quadratic order in
the kaon field and then expand, giving
\begin{eqnarray}
m_K^2
&=& m^2 - c_A \; \Sigma_u + 
2 \lambda \; 
\left(
\Sigma_u^2 - \Sigma_u \; \Sigma_s + \Sigma_s^2
\right)
\nonumber \\
&+& \kappa \; 
\left[
- \, \Sigma_u^2 + \Sigma_s \; \Sigma_u - \, \Sigma_s^2  
+ \frac{2}{\Sigma_u + \Sigma_s}
\left(
\Sigma_u^3
\log
\left(
\frac{M^2}{\Sigma_u^2}
\right)
+
\Sigma_s^3
\log
\left(
\frac{M^2}{\Sigma_s^2}
\right)
\right)
\right] \; .
\label{eq:kaon_mass}
\end{eqnarray}

Using the equations of motion,
Eqs. (\ref{eq:eq_motion_u}) and (\ref{eq:eq_motion_s}),
we find that the masses of the pion and kaon reduce to
\begin{equation}
m_\pi^2 = \frac{h_u}{\Sigma_u} \;\; ; \;\;
m_K^2 = \frac{h_u+h_s}{\Sigma_u+\Sigma_s} \; .
\label{eq:reduced_pion_kaon_mass}
\end{equation}

The results in Eq. (\ref{eq:reduced_pion_kaon_mass})
are familiar from chiral perturbation theory \cite{donoghue_dynamics_1992}.
In the present case, by introducing the background fields 
$h_u$ and $h_s$ we have eliminated
the ungainly dependence upon the logarithms of $\Sigma_u$ and $\Sigma_s$
in Eqs. (\ref{eq:pion_mass}) and (\ref{eq:kaon_mass}).
This is true generally, and helps explain
why there is a rather mild dependence 
upon the logarithmic coupling $\kappa$.

The masses for the $\eta$ and $\eta'$ mesons is complicated by their mixing,
because $h_u \neq h_s$.  We find
\begin{eqnarray}
(m^\pi_{00})^2 
&=& m^2 + \frac{2}{3} \, c_A \, 
\left(
2 \, \Sigma_u + \Sigma_s
\right)
+ \frac{2}{3} \, \lambda
\left(
2 \, \Sigma_u^2 + \Sigma_s^2
\right)
\nonumber \\
&+&
\frac{\kappa}{3}
\left(
- \, 2 \, \Sigma_u^2 - \, \Sigma_s^2
+ 4 \, \Sigma_u^2 \, \log \frac{M^2}{\Sigma_u^2}
+ 2 \, \Sigma_s^2 \, \log \frac{M^2}{\Sigma_s^2}
\right)
\; .
\label{eq:eta_mass_ugly}
\end{eqnarray}
\begin{eqnarray}
(m^\pi_{88})^2 
&=& m^2 + \frac{c_A}{3} 
\left(
- \, 4 \, \Sigma_u + \Sigma_s
\right)
+ \frac{2}{3} \, \lambda
\left(
\Sigma_u^2 + 2 \, \Sigma_s^2
\right)
\nonumber \\
&+&
\frac{\kappa}{3}
\left(
- \, \Sigma_u^2 - \, 2 \, \Sigma_s^2
+ 2 \, \Sigma_u^2 \, \log \frac{M^2}{\Sigma_u^2}
+ 4 \, \Sigma_s^2 \, \log \frac{M^2}{\Sigma_s^2}
\right)
\; .
\label{eq:etap_mass_ugly}
\end{eqnarray}
The sum of these masses squared is equal to that for the $\eta$
and $\eta'$,
\begin{equation}
m^2_\eta + m^2_{\eta'} = (m^\pi_{00})^2 + (m^\pi_{88})^2 
= 
\frac{h_u}{\Sigma_u} 
+ \frac{h_s}{\Sigma_s} 
+ c_A 
\left(
\frac{\Sigma_u^2}{\Sigma_s} + \, 2 \, \Sigma_s 
\right)
\; .
\label{eq:sum_eta_etap_masses}
\end{equation}
The difference of these masses is
\begin{equation}
(m^\pi_{00})^2 - (m^\pi_{88})^2 
= 
\frac{1}{3}
\left(
+ \, \frac{h_u}{\Sigma_u} 
- \, \frac{h_s}{\Sigma_s} 
+ c_A 
\left(
8 \, \Sigma_u - \, \frac{\Sigma_u^2}{\Sigma_s} 
+ \, 2 \, \Sigma_s 
\right)
\right)
\; .
\label{eq:diff_singlet_octet_pion}
\end{equation}
In addition, there is a mixing term between the singlet and octet states,
\begin{eqnarray}
(m^\pi_{08})^2
&=& \frac{2 \, \sqrt{2}}{3}
\left(
c_A 
\left(
- \, \Sigma_u + \Sigma_s 
\right)
+ 2 \, \lambda
\left(
\Sigma_u^2 - \, \Sigma_s^2
\right)
\right.
\nonumber \\
&+& 
\left. \kappa
\left(
- \, \Sigma_u^2 + \Sigma_s^2 
+ 2 \, \Sigma_u^2 \, \log \frac{M^2}{\Sigma_u^2}
- 2 \, \Sigma_s^2 \, \log \frac{M^2}{\Sigma_s^2}
\right)
\right)
\nonumber \\
&=& \frac{\sqrt{2}}{3}
\left(
+ \, \frac{h_u}{\Sigma_u} 
- \, \frac{h_s}{\Sigma_s} 
+ c_A 
\left(
- \, \Sigma_u - \, \frac{\Sigma_u^2}{\Sigma_s} 
+ \, 2 \, \Sigma_s 
\right)
\right)
\; .
\label{eq:mixing_singlet_octet_pion}
\end{eqnarray}
Using this, algebra shows
\begin{eqnarray}
\left(
m_{\eta'}^2-m^2_\eta
\right)^2 &=&
\left(
(m^\pi_{00})^2 - (m^\pi_{88})^2 
\right)^2 
+ 4 \, (m^\pi_{08})^4
\nonumber \\
&=&
\left(
+ \frac{h_u}{\Sigma_u} 
- \; \frac{h_s}{\Sigma_s} 
+ \,  c_A \, 
\left(
- \, \frac{\Sigma_u^2}{\Sigma_s } + 2 \, \Sigma_s
\right) 
\right)^2
+ \, 8 \, c_A^2 \, \Sigma_u^2 
\; .
\label{eq:diff_masses_eta_etap}
\end{eqnarray}

We next compute the masses of the scalar nonet, with $J^P = 0^+$.  
The analogies of the pion and kaon are the $a_0$ and $K_0^*$,
whose mass squared are
\begin{eqnarray}
m^2_{a_0} &=& m^2 + c_A \, \Sigma_s + 6 \, \lambda \, \Sigma_u^2
+ \kappa \, \Sigma_u^2 
\left( 
- 7 + 6 \log \frac{M^2}{\Sigma_u^2}
\right) 
\; ,
\label{eq:ugly_a0_mass}
\\
m^2_{K_{0}^{*}} &=& m^2 + c_A \, \Sigma_u
+ 2 \, \lambda
\left(
\Sigma_u^2 + \Sigma_u \, \Sigma_s + \Sigma_s^2
\right)
\nonumber
\\
&+& \kappa
\left(
- \left(
\Sigma_u^2 + \Sigma_u \, \Sigma_s + \Sigma_s^2
\right)
+ \frac{2}{\Sigma_s - \Sigma_u}
\left(
- \, \Sigma_u^2 \log \frac{M^2}{\Sigma_u^2}
+ \Sigma_s^2 \log \frac{M^2}{\Sigma_s^2}
\right)
\right)
\; .
\label{eq:ugly_kappa_mass}
\end{eqnarray}
It can be shown that these can be reduced to
\begin{eqnarray}
m^2_{a_0} &=& 3 \, m^2_\pi
- 2 \, m^2 + 4 \, c_A \, \Sigma_s - 4 \, \kappa \, \Sigma_u^2 \; ,
\label{eq:pretty_a0_mass}
\\
m^2_{K_0^*} &=&
\frac{h_s - h_u}{\Sigma_s - \Sigma_u} 
\label{eq:pretty_kappa_mass}
\; .
\end{eqnarray}
The mass of the $K_0^*$ looks like that of current
algebra \cite{donoghue_dynamics_1992}, but is not, because it involves
the ratio of differences, $h_s - h_u$ over $\Sigma_s - \Sigma_u$.

The final two mesons are the $\sigma$ and the $f_0$.  After some
computation, 
\begin{eqnarray}
(m^\sigma_{00})^2 
&=& m^2 - \frac{2}{3} \, c_A \, 
\left(
2 \, \Sigma_u + \Sigma_s
\right)
+ 2 \, \lambda
\left(
2 \, \Sigma_u^2 + \Sigma_s^2
\right)
\nonumber \\
&+&
\kappa
\left(
- \, \frac{14}{3} \, \Sigma_u^2 - \, \frac{7}{3} \, \Sigma_s^2
+ 4 \, \Sigma_u^2 \, \log \frac{M^2}{\Sigma_u^2}
+ 2 \, \Sigma_s^2 \, \log \frac{M^2}{\Sigma_s^2}
\right)
\; ,
\label{eq:sigma_mass}
\end{eqnarray}
\begin{eqnarray}
(m^\sigma_{88})^2 
&=& m^2 + \frac{c_A}{3} 
\left(
4 \, \Sigma_u - \Sigma_s
\right)
+ 2 \, \lambda
\left(
\Sigma_u^2 + 2 \, \Sigma_s^2
\right)
\nonumber \\
&+&
\kappa
\left(
- \, \frac{7}{3} \, \Sigma_u^2 - \, \frac{14}{3}  \, \Sigma_s^2
+ 2 \, \Sigma_u^2 \, \log \frac{M^2}{\Sigma_u^2}
+ 4 \, \Sigma_s^2 \, \log \frac{M^2}{\Sigma_s^2}
\right)
\; .
\label{eq:f0_mass}
\end{eqnarray}
The sum of these masses squared equals the sum of the
masses squared for the $\sigma$ and $f_0$,
\begin{eqnarray}
m_{\sigma}^2 + m_{f_{0}}^2 &=&
(m^\sigma_{00})^2 + (m^\sigma_{88})^2 
\nonumber \\
&=& 
3 \, \frac{h_u}{\Sigma_u} + 3 \, \frac{h_s}{\Sigma_s}
- 4 \, m^2 + c_A 
\left( 
3 \, \frac{\Sigma_u^2}{\Sigma_s} + 2 \, \Sigma_s 
\right)
- 4 \, \kappa 
\left(
\Sigma_u^2 + \Sigma_s^2 
\right)
\; .
\end{eqnarray}
The difference of these masses is
\begin{equation}
(m^\sigma_{00})^2 - (m^\sigma_{88})^2 
= 
+ \, \frac{h_u}{\Sigma_u}  - \, \frac{h_s}{\Sigma_s} 
+ \frac{c_A }{3}
\left(
- \, 8 \, \Sigma_u - \, 3 \, \frac{\Sigma_u^2}{\Sigma_s} 
+ \,  2 \, \Sigma_s
\right)
+ \, \frac{4}{3} \, \kappa 
\left(
- \, \Sigma_u^2 + \Sigma_s^2
\right)
\; .
\label{diff_sigma_mass_sq}
\end{equation}
The mixing between the two states is 
\begin{eqnarray}
(m^\sigma_{08})^2
&=& \frac{\sqrt{2}}{3}
\left(
c_A 
\left(
\Sigma_s - \Sigma_u
\right)
+ \, 6 \, \lambda
\left(
\Sigma_u^2 - \Sigma_s^2
\right)
\right.
\nonumber \\
&+& 
\left. \kappa
\left(
- 7 \, \Sigma_u^2 + 7 \, \Sigma_s^2 
+ \, 6 \, \Sigma_u^2 \, \log \frac{M^2}{\Sigma_u^2}
- 6 \, \Sigma_s^2 \, \log \frac{M^2}{\Sigma_s^2}
\right)
\right)
\nonumber \\
&=& \sqrt{2}
\left(
\frac{h_u}{\Sigma_u} 
- \; \frac{h_s}{\Sigma_s} 
+ \frac{c_A}{3}
\left(
\Sigma_u - 3 \, \frac{\Sigma_u^2}{\Sigma_s} + 2 \, \Sigma_s 
\right)
+ 
\frac{4}{3} \, \kappa
\left(
- \, \Sigma_u^2 + \, \Sigma_s^2 
\right)
\right)
\; .
\label{eq:mixing_singlet_sigma}
\end{eqnarray}
Using these expressions,
\begin{eqnarray}
\left(
m^2_{f_0}-m^2_\sigma
\right)^2 
&=&
\left(
(m^\sigma_{00})^2 - (m^\sigma_{88})^2 
\right)^2 
+ 4 \, (m^\sigma_{08})^4
\\
&=& 
9 \, \left(
\frac{h_u}{\Sigma_u} 
- \, \frac{h_s}{\Sigma_s} 
+ \, c_A \, 
\left(
- \, \frac{\Sigma_u^2}{\Sigma_s} + \, \frac{2}{3} \, \Sigma_s
\right)
+ \, \frac{4}{3} \, \kappa
\left(
- \, \Sigma_u^2 + \, \Sigma_s^2
\right)
\right)^2
+ \, 8 \, c_A^2 \, \Sigma_u^2 
\; .
\nonumber
\label{diff_sigma_f0_masses_crct}
\end{eqnarray}
This is a surprisingly elegant form, analogous to the expression for
the splitting between the masses for the $\eta$ and $\eta'$ in 
Eq. (\ref{eq:diff_singlet_octet_pion}).

We next turn to two applications of these results: in the chiral
limit, and to QCD.

\subsubsection{Masses in the chiral limit: the $\sigma$ meson and the axial anomaly}
\label{eq:sec_chiral_limit_masses}

In the limit of exact $SU(3)_f$
symmetry, $h_u = h_s = h$, and so $\Sigma_u = \Sigma_s = \Sigma$.
The two equations of motion in 
Eqs. (\ref{eq:eq_motion_u}) and (\ref{eq:eq_motion_s}) reduce to one,
and the masses become
\begin{eqnarray}
m_\pi^2 &=& m_K^2 = m_\eta^2 = \frac{h}{\Sigma}  \; ,
\label{eq:chiral_sym_pion}
\\
m_{\eta'}^2 &=& m_\pi^2 + \, 3 \, c_A \, \Sigma \; ,
\label{eq:chiral_sym_etap}
\\
m_{a_{0}}^2 &=& m_{K_{0}^{*}}^2 = m_{f_0}^2 =
3 \, m_\pi^2 - \, 2 \, m^2 + \, 4 \, c_A \, \Sigma - \, 4 \, \kappa \, \Sigma^2
\; ,
\label{eq:chiral_sym_a0}
\\
m_\sigma^2 &=& m_{a_{0}}^2  - \, 3 \, c_A \, \Sigma \; , 
\label{eq:chiral_sym_sigma}
\end{eqnarray}
All of these expressions can be derived directly from the corresponding
equations, except for the mass of the $K_0^*$, which takes some care.

As expected by the explicit $SU(3)_f$ symmetry, the pions, kaons, and the
$\eta$ form a degenerate octet.  The mass squared of the $\eta'$ is 
larger than that for this octet by an amount $= + \, 3 \, c_A \, \Sigma$.
This explains the negative sign of the term $\sim c_A$ in the chiral
Lagrangian of Eq. (\ref{eq:chiral_lagrangian}), because experiment
tells us that the $\eta'$ is heavy.

For the scalar mesons, again the $a_0$, $K_0^*$, and the $f_0$ 
form a degenerate octet.
This mass, Eq. (\ref{eq:chiral_sym_a0}), explicitly
involves the mass parameter of the chiral Lagrangian,
$m^2$ in Eq. (\ref{eq:chiral_lagrangian}).
Notice that we chose to include $m^2$ with a positive sign.  
As we show in the next section, this is because to fit the
observed hadronic spectrum with $c_A \neq 0$, 
$m^2>0$; this is also true when $\kappa = 0$ \cite{lenaghan_chiral_2000}.  
With $c_A = \kappa = 0$, though, then it is necessary to take $m^2 < 0$
so that the $a_0$ is heavy.

What is striking, however, is that if we chose $c_A$ to be positive,
so that the mass of the $\eta'$ is driven {\it up}, that
the mass of the $\sigma$ meson is driven {\it down}, by {\it exactly} the
same amount:
\begin{equation}
m^2_{\eta'} - m^2_\pi
= m^2_{a_0} - m^2_\sigma \;\;\; , \;\;\; h_u = h_s \; .
\label{eq:chiral_identity}
\end{equation}
The same relation was first derived by 't Hooft in a
linear sigma model with
two flavors \cite{t_hooft_how_1986}.

One motivation for including tetraquarks
\cite{jaffe_multiquark_1977, *black_mechanism_2000, *close_scalar_2002, *jaffe_diquarks_2003, *maiani_new_2004, *pelaez_light_2004, *t_hooft_theory_2008, *pelaez_controversy_2015}
is that they naturally give an ``inverted'' spectrum, where for
$0^+$ mesons, the isosinglet state is lighter than the octet.  
Eq. (\ref{eq:chiral_identity}) shows that this inverted spectrum
arises naturally in a linear sigma model for three flavors.  
It is also a clear demonstration that the axial anomaly is as important
for the $0^+$ mesons as it is for the $0^-$.

\subsubsection{Parameters of the chiral model in QCD}
\label{sec:fit_sigma_model}

We now use our results for the masses to derive the values
of the parameters of our chiral model in QCD.

In contrast to the standard linear sigma model, as treated in Ref. 
\cite{lenaghan_chiral_2000}, we have one more parameter, the
Yukawa coupling between the two scalar nonets and the quarks,
$y$.  We keep $y$ as a free parameter, and use this
to adjust the temperature for the chiral crossover.  

To determine the parameters, we take the known masses of the pseudoscalar
nonet, 
\begin{equation}
m_\pi = 140 \;\; , \;\;
m_K = 495 \;\; , \;\;
m_\eta = 540 \;\; , \;\;
m_{\eta'} = 960 \; ;
\label{input_masses}
\end{equation}
in this expression and henceforth, all mass dimensions are
assumed to be MeV.  

We take the value of the light quark condensate from its relation to the pion
decay constant, $f_\pi = 93.$~MeV \cite{lenaghan_chiral_2000}:
\begin{equation}
\Sigma_u = \frac{f_\pi}{2} = 46.0 \; .
\label{eq:value_sigmau}
\end{equation}
There is a similar relation for the strange quark condensate,
\begin{equation}
  \Sigma_s = f_K - \frac{f_\pi}{2} \; ,
\label{eq:value_sigmas_Kdecay}
\end{equation}
which was used in Ref. 
\cite{lenaghan_chiral_2000} to fix $\Sigma_s$.

Instead, we prefer to proceed as following. 
First we set the renormalization scale $M$ to $\Sigma_u$ in vacuum, i.e. 
$M= f_\pi/2$. Then, 
we take the four masses
in Eq. (\ref{input_masses}), and $\Sigma_u$ from 
Eq. (\ref{eq:value_sigmau}) as input, and use these to determine
$\Sigma_s$, the background fields $h_u$ and $h_s$, and the axial 
coupling $c_A$, from 
Eqs. (\ref{eq:reduced_pion_kaon_mass}), (\ref{eq:sum_eta_etap_masses}),
and (\ref{eq:diff_masses_eta_etap}).  The result is
\begin{equation}
\Sigma_s = 76.1 \;\; , \;\;
h_u = (96.6)^3 \;\; , \;\;
h_s = (305.)^3 \;\; , \;\;
c_A = 4560. \;\;\; .
\label{results_fixed_parameters}
\end{equation}

These values are all independent of the Yukawa coupling $y$.
The remaining two parameters of the linear sigma model $m^2$
and $\lambda$, can be determined from
the equations of motion in 
Eqs. (\ref{eq:eq_motion_u}) and (\ref{eq:eq_motion_s}),
\begin{equation}
m^2 = (538.)^2 - \, (11.3)^2 \, y^4 \;\; ; \;\;
\lambda = 18.3+\, 0.0396 \, y^4 \; ,
\label{eq:mass_coupling_fit}
\end{equation}
and do depend upon $y$.

These values agree approximately with those of a linear sigma model
without a logarithmic coupling, 
as studied by Lenaghan, Rischke, and Schaffner-Bielich (LRS) in
Ref. \cite{lenaghan_chiral_2000}.
Using Eqs. (\ref{Eq:sl_ss}), we find that they obtain
$h_u^{LRS} = (98)^3$, versus our $h_u = (96.6)^3$; their
$h_s^{LRS} = (299)^3$, versus our $h_s = (305)^3$;
their $c_A^{LRS} = 4808$, versus our $c_A = 4560$.  The differences
arise primarily not because of the differences in the
potential for $\Phi$, but because they fix
$\Sigma_s$ from the kaon decay constant, Eq. (\ref{eq:value_sigmas_Kdecay}).
In contrast, we determine $\Sigma_s$
from the $\eta$ and $\eta'$ masses,
Eq. (\ref{eq:diff_masses_eta_etap}).  Thus their 
$\Sigma_s^{LRS} = 66.8$, versus our $\Sigma_s = 76.1$.

The difference in $\Sigma_s$ affects the mass of the $K^*_0$, which
in both models is given by Eq. (\ref{eq:pretty_kappa_mass}).  
Using their value for the strange quark condensate,
Ref. \cite{lenaghan_chiral_2000} finds that the mass of the 
$K_0^*$ is $m^{LRS}_{K_{0}^{*}} = 1124$, while we find that 
$m_{K_{0}^{*}} = 957$.  

This leaves the masses of the rest of the $0^+$ nonet, the $a_0$,
$\sigma$, and $f_0$.  These masses explicitly depend upon the Yukawa
coupling $y$, which is determined by the temperature for the
chiral crossover, $T_\chi$.

\subsection{Symmetry breaking term at $T\neq 0$}
\label{sec:sym_bkng_nonzeroT}

In Sec. (\ref{sec:nonzero_T_chiral_bkng}) we argued that a new
symmetry breaking term needs to be added to ensure that the effective
fermion mass is nonzero in the limit of high temperature.  It
is necessary to fix this term in order to determine $T_\chi$.

One possible approach would be simply to take the analogy of Eq.
(\ref{high_temp_toy_improved}), taking a symmetry breaking which is
computed perturbatively, with the matrix variables $q = r = 0$.  
Since the temperature for the chiral crossover is so much lower than
the deconfining transition, however, this seems unduly naive.

In fact it is not difficult generalizing the term.  
Starting from Eq. (\ref{eq:quark_lagrangian_with_phi}), 
for a quark of mass ``$m$'', the quark contribution
to the effective potential is
\begin{equation}
{\cal V}^{qk}_{pert}
= - \; \frac{1}{V} \; {\rm tr} \, 
\log
\left(
\, \not \!\! D \, 
+  \, m
+ \,  \mu \, \gamma^0
+ \; y 
\left(
\Phi \, {\cal P}_L
+ \Phi^\dagger \, {\cal P}_R 
\right)
\right) 
\; .
\end{equation}
Now consider the derivative of this quantity with respect to $m$,
evaluated at $m = 0$, times the current quark mass $m_{qk}$:
\begin{equation}
- \, m_{qk} \; \frac{1}{V} \;
{\rm tr}
\; 
\frac{1}{
\not \!\! D \, 
+ \,  \mu \, \gamma^0
+ \; y 
\left(
\Phi \, {\cal P}_L
+ \Phi^\dagger \, {\cal P}_R 
\right)
} \; .
\label{eq:sym_brkng_chiral_sym}
\end{equation}
It is then obvious from the discussion in 
Sec. (\ref{sec:nonzero_T_chiral_bkng}) that adding this term will
accomplish our objective, to ensure that the constituent quark mass
approaches the current quark mass at high temperature.

Further, this term is linear in the symmetry breaking parameter 
$m_{qk}$, times a form which is manifestly chirally symmetric.
In fact the form in Eq. (\ref{eq:sym_brkng_chiral_sym}) is a bit awkward
for our purposes.  The computation of susceptibilities involves taking
derivatives with respect not just to $\sigma_0$ and $\sigma_8$, but
all components of $\Phi$.  While this can be done, the contribution from
the symmetry breaking term is {\it prima facie} small.  Thus we ease
our computational burden by taking the symmetry breaking term to be
\begin{equation}
{\cal V}_{h}^{T}
= - \; \frac{m_{qk} }{V} 
\left(
\;
\left.
{\rm tr}
\; 
\frac{1}{
\not \!\! D \, 
+ \,  \mu \, \gamma^0
+ \; y \, \sigma_{ii}
}
\right|_{T \neq 0}
-
\left.
{\rm tr}
\; 
\frac{1}{
\not \!\! D \, 
+ \,  \mu \, \gamma^0
+ \; y \, \sigma_{ii}
}
\right|_{T=0}
\right) \; .
\label{eq:sym_brkng_chiral_sym_approx}
\end{equation}
That is, we only take the real, diagonal components of $\Phi$ in the
symmetry breaking term.  Thus Eq. (\ref{eq:sym_brkng_chiral_sym_approx})
is not linear in $m_{qk}$, but implicitly involves terms which are
of higher order.

We do not view this as a serious drawback, but of course a more careful
study, which would not be trivial, would be most welcome.

We comment that it is absolutely necessary to use a symmetry breaking
term which involves the dynamically generated quark mass, through
the components of $\sigma$.  At first we tried a term which involves
only the form of symmetry breaking at high temperature, so that
the trace in Eq. (\ref{eq:sym_brkng_chiral_sym_approx}) is computed
for massless quarks.  This gives the correct behavior at high temperature,
but because $T_\chi \ll T_d$, as discussed previously, this greatly
affects the results near $T_\chi$.  This is manifestly unphysical:
near $T_\chi$ the quarks do have a dynamically generated mass, and this
mass suppressed the contribution of the temperature dependent symmetry
breaking term above.

\section{Chiral matrix model at nonzero temperature}
\label{sec:chiral_matrix_nonzero_T}

\subsection{Complete model}

With the symmetry breaking term in hand, we only need to 
put all of the pieces together.  In mean field approximation
for the $\Phi$ mesons, this is
\begin{equation}
{\cal V}_{eff}(q,r,\Sigma_f) 
= {\cal V}^{gl}(q,r) + {\cal V}_\Phi^{tot}(\Sigma_f) 
+ {\cal V}^{qk}(q,r,\Sigma_f) +
{\cal V}^{T}_{h}(q,r,\Sigma_f)  \; .
\label{Eq:Vgeneral}
\end{equation}
We assume isospin symmetry, so there are two quark condensates,
$\Sigma_u = \Sigma_d$ and $\Sigma_s$.

The gluon potential ${\cal V}^{gl}(q,r)$ is the sum of the perturbative
term in Eq. (\ref{perturbative_gluon_potential}) and the
nonperturbative term in Eq. (\ref{nonpert_gluon_pot}).  
As discussed previously, 
we do not change the value of the deconfining temperature,
$T_d$, in the nonperturbative part of the gluon potential.

The chiral potential ${\cal V}_\Phi^{tot}(\Sigma_f)$ is that of 
Eq. (\ref{eq:chiral_lagrangian}).  For the time being, we do not
incorporate any temperature dependence in the parameters of the chiral 
Lagrangian.
In the mean-field approximation, 
\begin{equation}
{\cal V}_\Phi^{tot}(\Sigma_f) = 
-\, 2\, h_u \, \Sigma _u -\, h_s \, \Sigma _s +
m^2 \left(2 \, \Sigma _u^2+\Sigma _s^2\right)
-2 \, c_A \, \Sigma _u^2 \, \Sigma _s
+ \lambda  \, \left(2 \, \Sigma _u^4 + \, \Sigma _s^4\right) \; .
\label{Eq:MF_V_Phi}
\end{equation}

The quark contribution is 
\begin{equation}
{\cal V}^{qk}(q,r,\Sigma_f)
=  \sum_{f=u,d,s} {\cal V}^{qk}_f
=  \sum_{f=u,d,s} 
\left(
- \frac{3}{8 \pi^2} \; y^4 \, \Sigma_f^4 
\ln\left(\frac{y^2\, \Sigma_f^2}{M^2}\right)
+   {\cal V}_f^{qk,T}(q,r,\Sigma_f)
\right)
\; .
\end{equation}
The first two terms are just the usual vacuum contributions from the quark
loop, Eqs. (\ref{eq:zero_temp_sum_over_flavors})
and (\ref{eq:new_counterterm}).  
We assume that the renormalization scale $M$ is chirally symmetric,
and so the same for light and strange quarks.

The thermal term is also straightforward, just the sum over free energies
for each quark flavor, at nonzero chemical potential $\mu$ and $q_a$,
\begin{equation}
{\cal V}_f^{qk,T}(q,r,\Sigma_f)  
= -\, 2 \, T \sum_{a=1}^{3}  \int \frac{d^3k}{(2\pi)^3} 
\left[ 
\ln\left(  1 + e^{-(E_f -\mu)/T + 2\pi i q_a/3  }\right)
+
\ln\left(  1 + e^{-(E_f +\mu)/T - 2\pi i q_a/3  }\right) \right] \; .
\label{Eq:Vq}
\end{equation}
The energy and mass of each quark is
\begin{equation}
E_f^2 = k^2 + m_f^2; \quad 
m_f =  y \, \Sigma_f \; .  
\label{Eq:masses}
\end{equation}
The sum over ``$a$'' is over colors, where from 
Eqs. (\ref{eq:ansatz}) and (\ref{eq:def_cartan}),
the holonomy parameters $q_a$ are
\begin{equation}
\vec{q} = \left( q + i \, {\cal R}, 
- \, q + i \, {\cal R}, -\, 2 \, i \, {\cal R}  \right) \; .
\label{Eq:qcomplex}
\end{equation}
As discussed previously, when $\mu \neq 0$, $r = i {\cal R}$ is imaginary.

Lastly, for the symmetry breaking term, explicitly the form of
Eq. (\ref{eq:sym_brkng_chiral_sym_approx}) becomes
\begin{equation}
{\cal V}^{T}_{h}(q,r,\Sigma_f) =  - \sum_{f=u,d,s} 
\Sigma_{f}^0 \; 
\frac{\partial   }{\partial \Sigma_f}
{\cal V}_f^{qk,T}(q,r,\Sigma_f)
\; .
\label{Eq:HT_SBT}
\end{equation}  

\subsection{Mass spectrum, $T=0$ and $T \neq 0$}
\label{sec:mass}

We have one free parameter left to determine in the model, the Yukawa
coupling $y$.  Then at any temperature, we have a set of three coupled
equations which determine the condensates $q$, $\Sigma_u$, and $\Sigma_s$.
The quark condensates are determined by the equations of motion.
Taking derivatives of Eq.~\eqref{Eq:MF_V_Phi} with respect to 
$\Sigma_{u,s}$ we get 
\begin{equation}
 \frac{\partial }{\partial \Sigma_u} {\cal V}_u^{qk}
-  \, h_u -  2 \, c_A \, \Sigma_u \, \Sigma_s 
+ 2 \, m^2 \, \Sigma_u + 4 \, \lambda \, \Sigma_u^3 \, 
- \Sigma_u^0 \, \frac{\partial^2 }{\partial \Sigma_u^2}{\cal V}_u^{qk, T}
= 0  \; ,
\label{Eq:eom_l}
\end{equation}
and 
\begin{equation}
 \frac{\partial }{\partial \Sigma_s} {\cal V}_s^{qk}
- \, h_s + \, 2 \, m^2 \, \Sigma_s - \, 2 \, c_A \, \Sigma_u^2 
+  4 \, \lambda \, \Sigma_s^3 \, 
- \Sigma_s^0\, \frac{\partial^2 }{\partial \Sigma_s^2}
{\cal V}_s^{qk, T} =0  \; .
\label{Eq:eom_s}
\end{equation}
The first term in each expression is the derivative of the quark
potential, ${\cal V}^{qk}_f$, for that flavor.  Next are the terms
from the potential for $\Phi$.  The last term is the derivative
of the mass term at nonzero temperature.
The derivative with respect to $q$ is similar, and determined
numerically.

\begin{figure}
\includegraphics[width=0.8\linewidth]{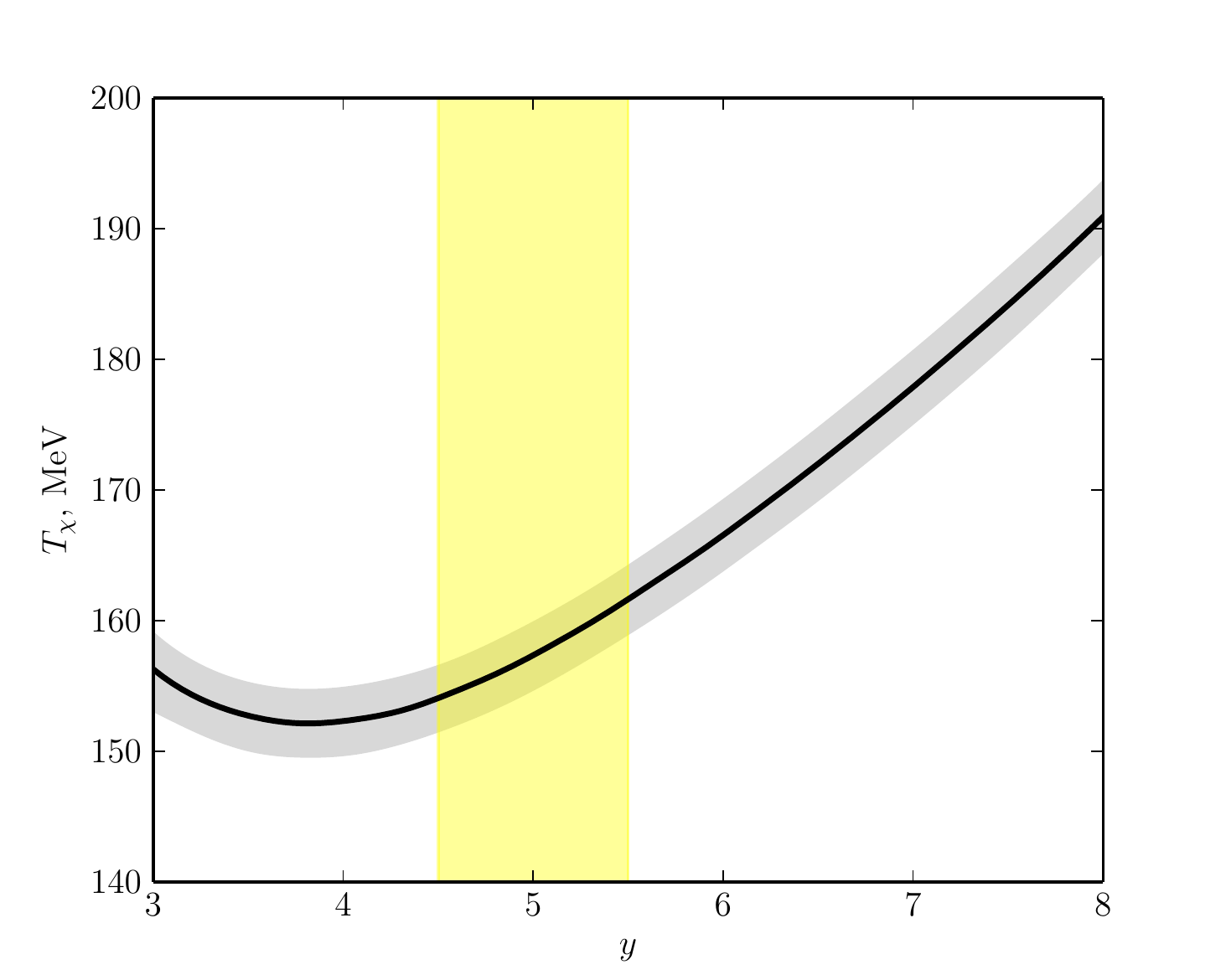}
\caption{The chiral crossover temperature $T_\chi$
as a function of the Yukawa coupling, $y$. 
In the horizontal shaded region $T_d$ varies from
$260$ to $280$ MeV, with the line $T_d=270$ MeV.
The vertical shaded region corresponds to $y: 4.5 \rightarrow 5.5$.
}
\label{fig:Tchi_on_y}
\end{figure}

To fix the Yukawa coupling, we fit to $T_\chi$, which we define
as the maximum in the derivative of the condensate
for the light quark, $|\partial \Sigma_u/\partial T|$.
the peak in the chiral susceptibility for light quarks.  This is
shown in Fig. (\ref{fig:Tchi_on_y}).  We consider varying the deconfining
temperature $T_d$ from $260$ to $280$ MeV, with the central line
corresponding to $270$ MeV.  The vertical shaded region demonstrates
varying $y$ from $4.5$ to $5$.

Given the range in the Yukawa coupling, we can then determine the
masses of the $0^+$ mesons at zero temperature.
In Table~(\ref{fig:mesons_masses}) we show the values of the 
$a_0$, $f_0$, and $\sigma$, for values of $y = 4.5$, $5$,
and $5.5$.  

\begin{figure}
\includegraphics[width=0.8\linewidth]{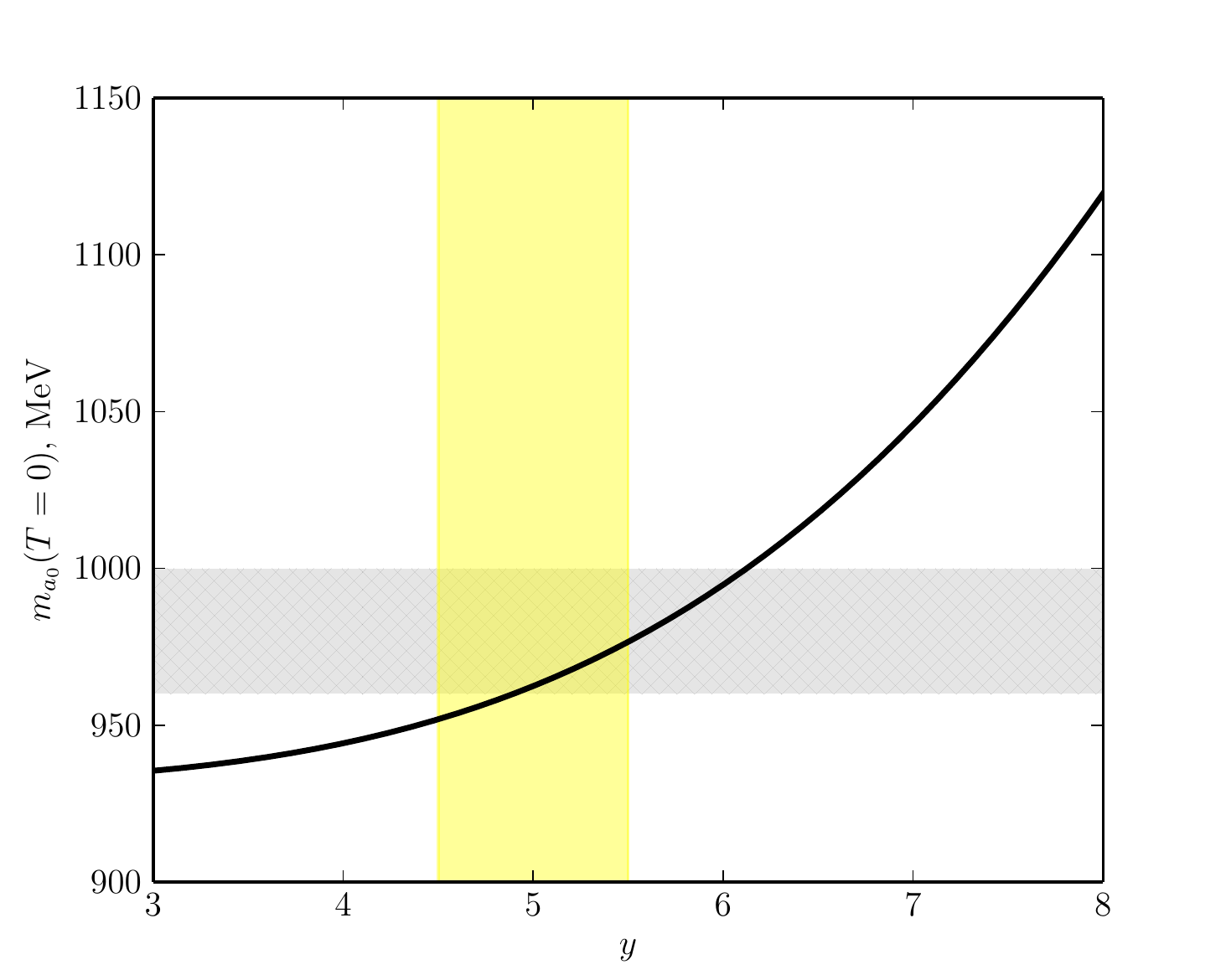}
\caption{
The mass of $a_0$ meson at zero temperature 
as a function of the Yukawa coupling, $y$. 
The horizontal shaded region corresponds to the experimental 
uncertainty in the $a_0$ mass. 
}
\label{fig:ma0_on_y}
\end{figure}

The variation of the mass of the $a_0$ at $T=0$, as a function of the Yukawa
coupling, is shown in Fig. (\ref{fig:ma0_on_y}).  

The mass of the $a_0$ in all cases is near the experimental
value of $980$~MeV, although low by $\sim 3\%$.  The mass of the $f_0$
is a bit below $1$~GeV, while the $\sigma$ is very low, $\sim 325$~MeV.
These values are typical of linear sigma models \cite{lenaghan_chiral_2000}.

\begin{table}
\begin{tabular}{  |c || c | c | c|    }
 \hline
$y$   & $m_{a_0}$ &  $m_{f_0}$ & $m_\sigma$ \\
 \hline
  4.5 & 952 &  982	& 309 \\
  \hline
  5   & 962 &  966 	& 328 \\
  \hline
  5.5 & 977 &  945 	& 348 \\
  \hline
\end{tabular}
\caption{Meson masses as functions of the Yukawa coupling.}
\label{fig:mesons_masses}
\end{table}

We choose the central value of ``$y = 5$''.  The properties of the
theory at $\mu = 0$ then follow directly.

\begin{figure}
\includegraphics[width=0.8\linewidth]{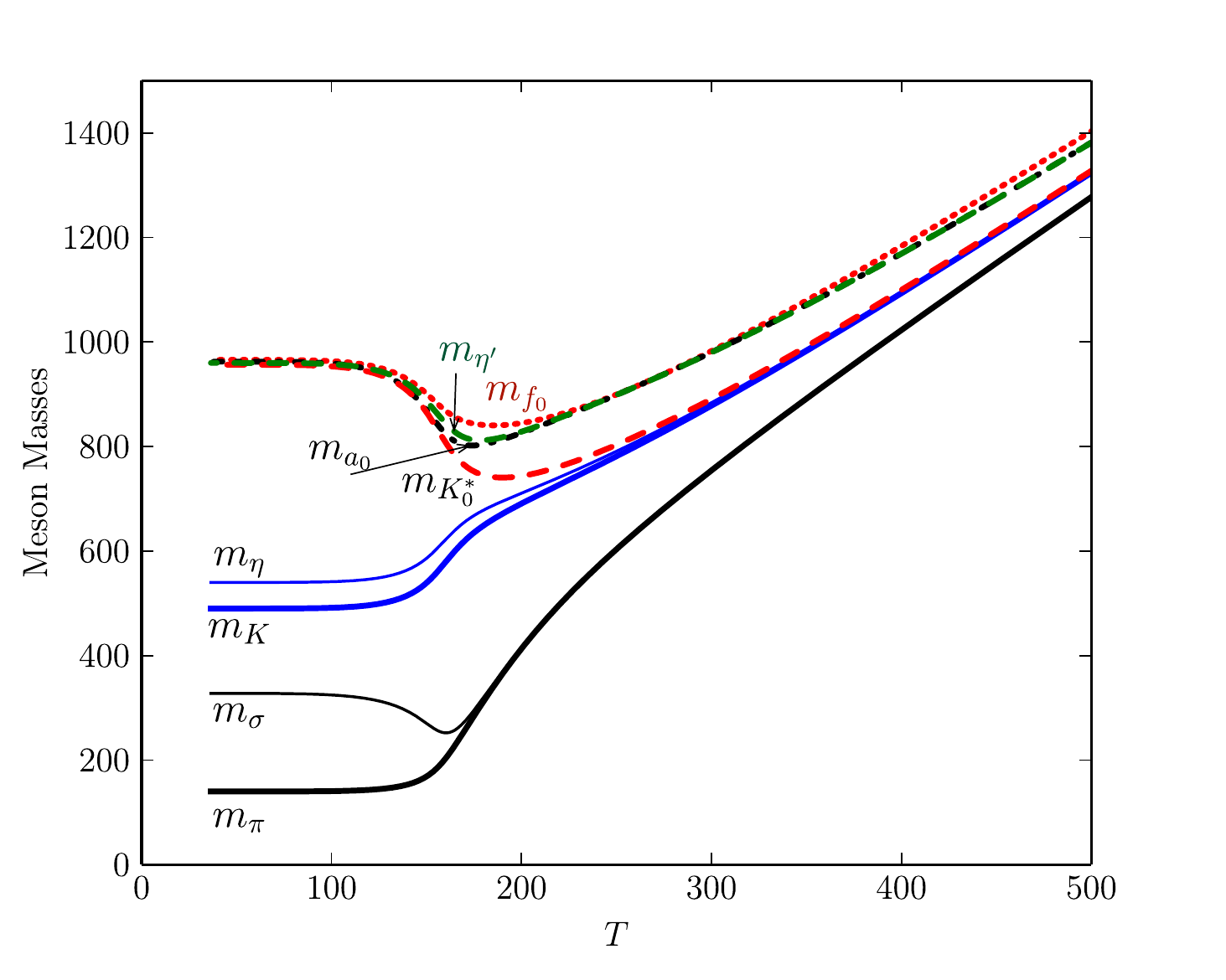}
\caption{Temperature dependence of the meson masses for $y=5$. }
\label{fig:masses}
\end{figure}

The temperature dependence of the meson masses at nonzero temperature
are shown in Fig. (\ref{fig:masses}).  Above $T \sim 200$~MeV, we find that
the following masses are degenerate: the $\pi$ and $\sigma$;
the $K$, $\eta$, and $K_0^*$; and the $a_0$, $f_0$, and $\eta'$.
This is expected for the restoration of the $SU(3)_L \times SU(3)_R$
chiral symmetry, with the small mass splittings due
to the residual symmetry breaking from $m_u = m_d \ll m_s \neq 0$.

Notice that the mass spectrum does not exhibit the
restoration of the axial $U(1)_A$ symmetry, as the $\eta'$ meson 
is heavier than the $\eta$ meson.  This is because we assume that
the coefficient $c_A$ is fixed, and does not vary with temperature.
This is clearly unphysical, 
as seen in lattice simulations \cite{buchoff_qcd_2014}, and
as we discuss in Sec. (\ref{sec:chiral_suscep}).

\subsection{Thermodynamics}
\label{sec:thermo}

\begin{figure}
\includegraphics[width=0.8\linewidth]{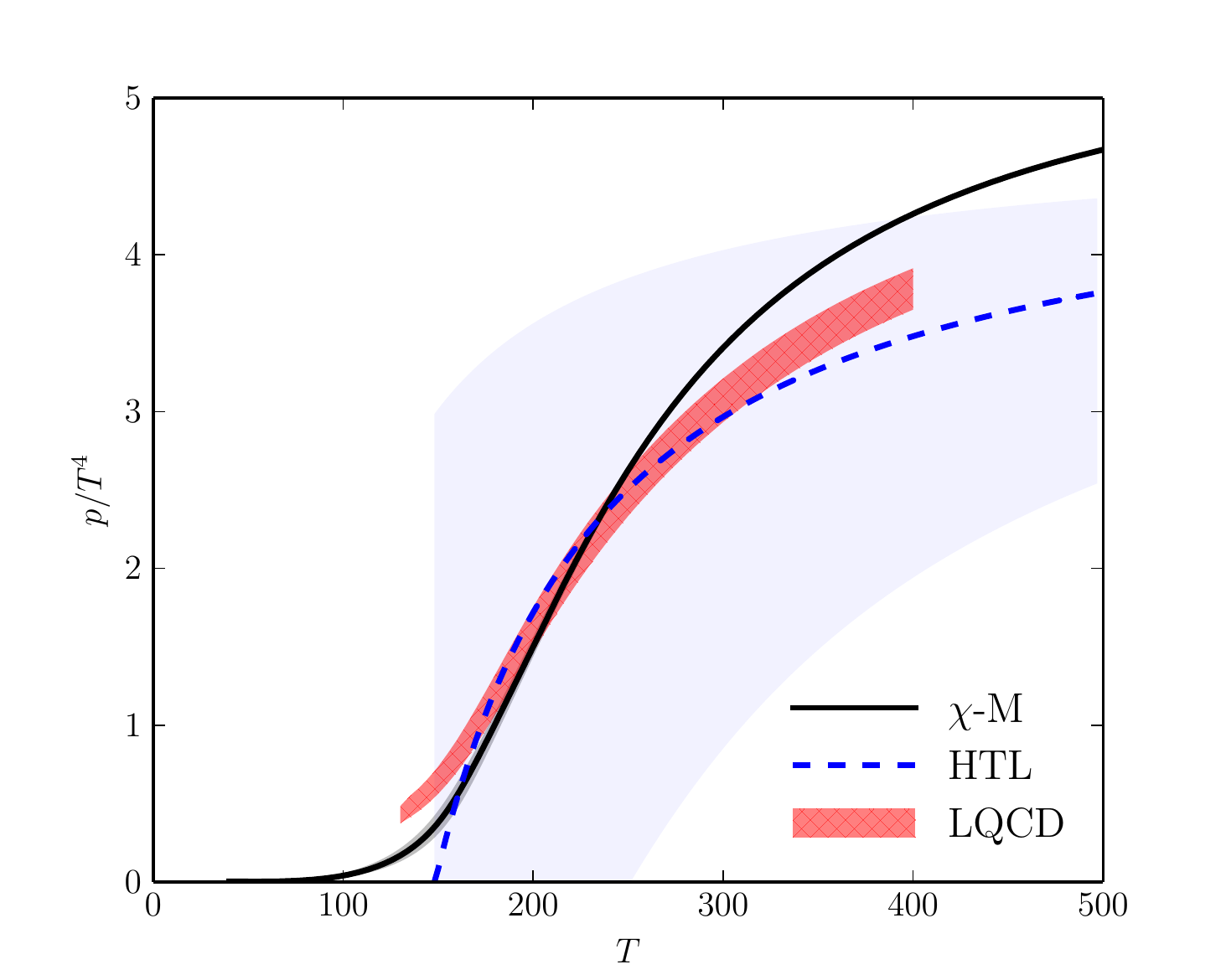}
\caption{The pressure as a function of temperature. The solid black
line is our chiral matrix model, $\chi$-M.  The shaded region about
this denotes the variation of the Yukawa coupling, $y:4.5\rightarrow 5.5$.
The red band are the results from lattice simulations
\cite{bazavov_equation_2014}.
The dashed blue line is that of NNLO HTL perturbation theory, with
the band changing the renormalization mass scale by a factor of two
\cite{andersen_three-loop_2010, *andersen_gluon_2010, *andersen_nnlo_2011, *andersen_three-loop_2011,*haque_two-loop_2013, *mogliacci_equation_2013, *haque_three-loop_2014}.
}
\label{fig:p_on_T}
\end{figure}

Turning to thermodynamics, the pressure is illustrated
in Fig. (\ref{fig:p_on_T}).  The agreement with the pressure is
reasonable, but not spectacular.  The pressure in the chiral matrix
model is too small at low temperature, below $T_\chi$.
This is because we do not include light hadrons
such as pions, kaons, {\it etc.} as dynamical degrees of freedom.

At high temperature, above $250$~MeV, the pressure in our model overshoots
that from the lattice data.  This is because we choose the parameters
in the gluon potential to be identical to those in the pure glue theory.
A better fit could be obtained if we allowed this potential to vary.

\begin{figure}
\includegraphics[width=0.8\linewidth]{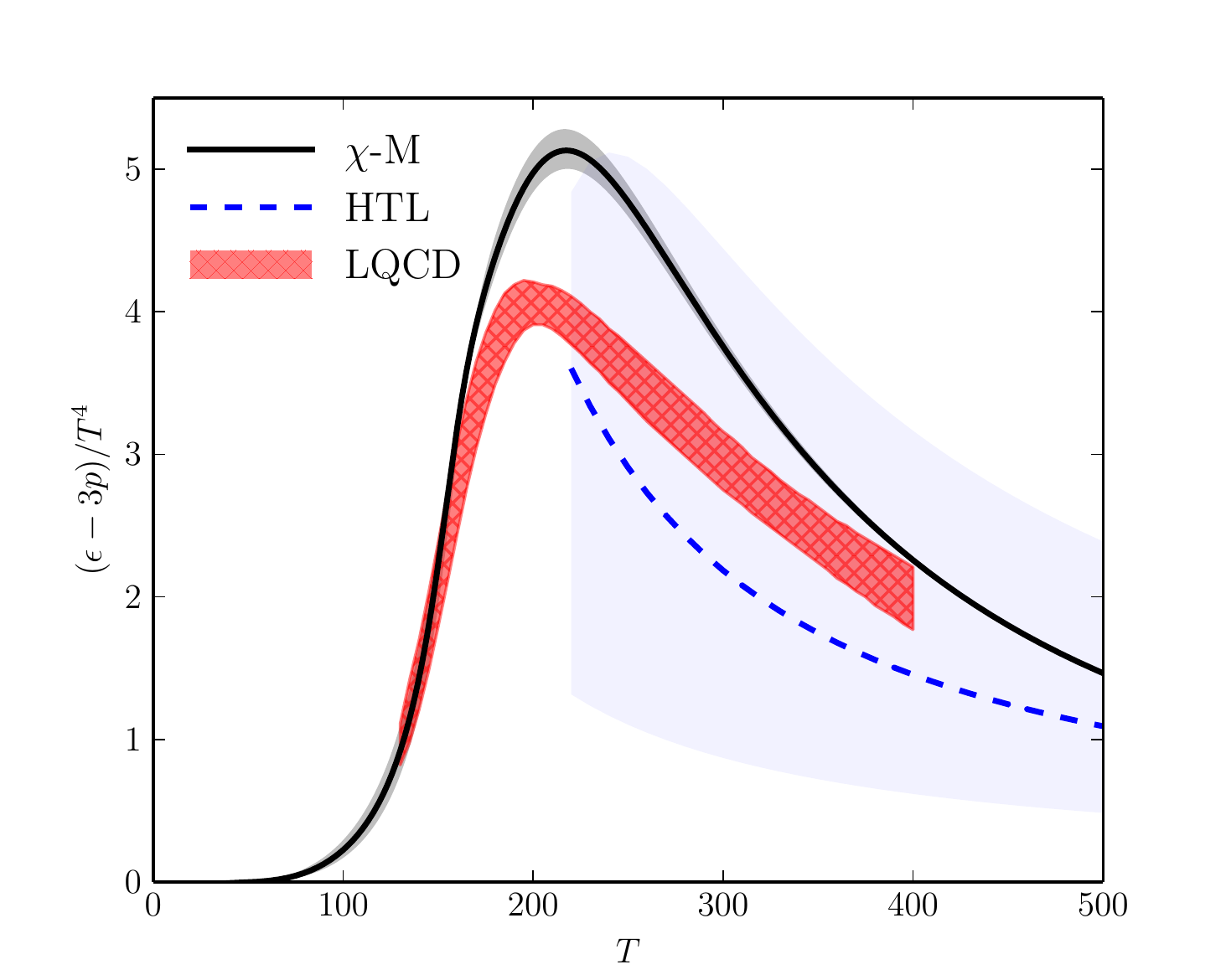}
\caption{
The interaction measure as a function of $T$. 
The black line denotes the chiral matrix model, $\chi$-M, with the shaded
band the variation in $y:4.5\rightarrow 5.5$.  
The red band are the results from lattice simulations
\cite{bazavov_equation_2014}.
The dashed blue line is that of NNLO HTL perturbation theory, with
the band changing the renormalization mass scale by a factor of two
\cite{andersen_three-loop_2010, *andersen_gluon_2010, *andersen_nnlo_2011, *andersen_three-loop_2011,*haque_two-loop_2013, *mogliacci_equation_2013, *haque_three-loop_2014}.
}
\label{fig:int_on_T}
\end{figure}

To see the discrepancy with the lattice results, in Fig. (\ref{fig:int_on_T})
we show the interaction measure, $(e-3p)/T^4$, where $e(T)$ is the energy
density.  This peak in the interaction measure is about $25\%$
too high: it is $\sim 5$, versus $\sim 4$ from the lattice.  Also,
the peak in the interaction measure is at $\sim 220$~MeV, versus
$\sim 200$~MeV from the lattice.

\subsection{Behavior of the order parameters}
\label{sec:order}

\begin{figure}
\includegraphics[width=0.8\linewidth]{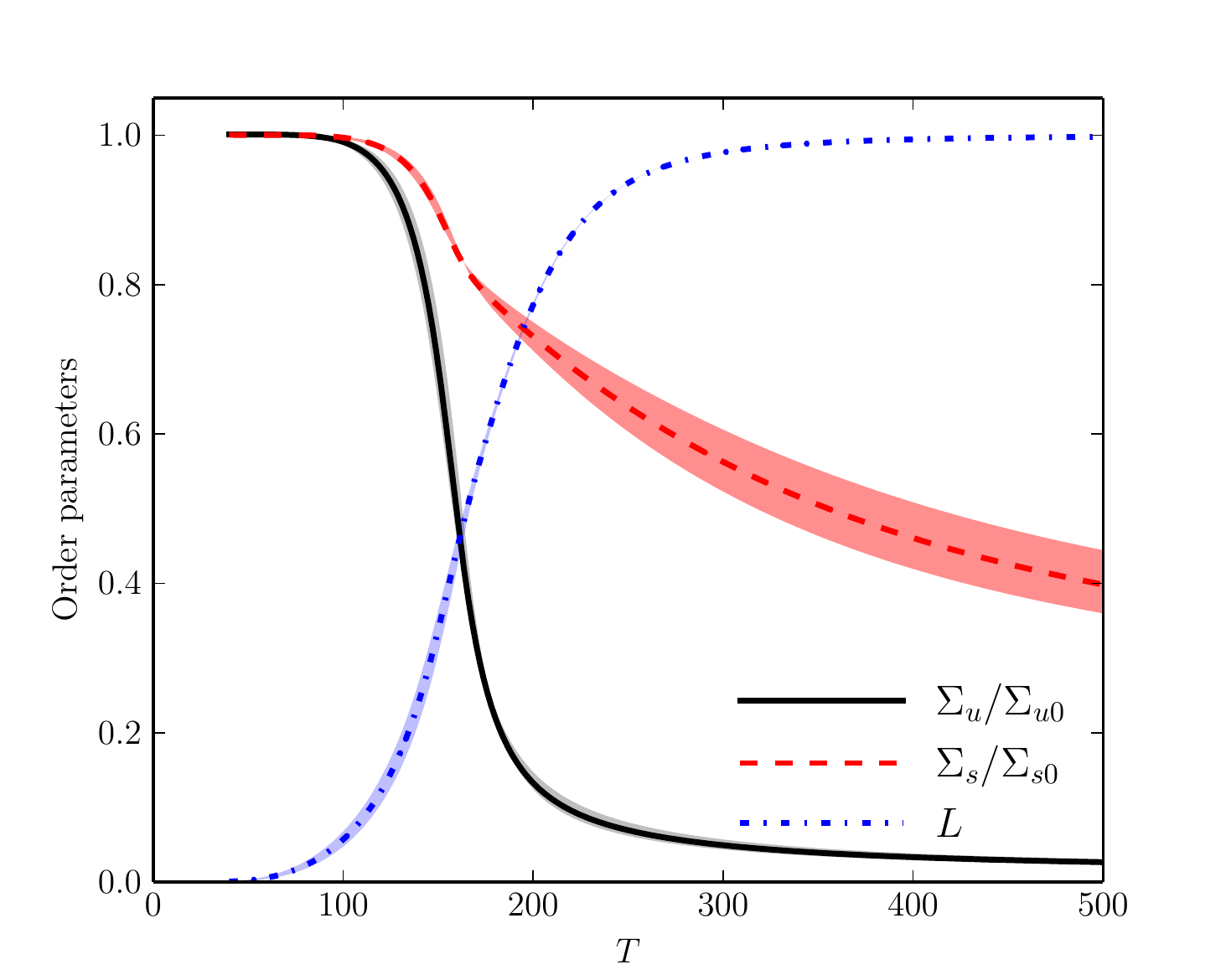}
\caption{
The chiral and deconfining order parameters as functions of $T$. 
The light and strange chiral condensates are normalized to their
value at zero temperature.  
The shaded regions correspond to varying the Yukawa
coupling $y$ in the range $y:4.5\rightarrow 5.5$.
}
\label{fig:op_on_T}
\end{figure}

How the order parameters change with temperature
is illustrated in Fig. (\ref{fig:op_on_T}).  We show the
Polyakov loop directly, while for the chiral order parameters, we
show the ratio of the condensate at $T\neq 0$ to that at $T=0$.
This figure shows that in
our matrix model there is an {\it extremely} close correlation
between the restoration of chiral symmetry, and deconfinement,
as the decline in the light quark condensate mimics the rise
in the Polyakov loop, for temperatures between $100$ and $300$ MeV.
To be more precise, one can compute the associated susceptibilities
for the order parameters.  We defer this to Sec. (\ref{sec:order}),
so that we can discuss at length which susceptibilities diverge in
the chiral limit.  As expected for a heavy quark, the 
strange quark condensate declines much slower than that for
the light quarks.

The chiral order parameters cannot be directly compared to those on
the lattice.  Even their mass dimensions are different: in our model
$\Sigma$ has dimensions of mass, while 
in QCD $\langle \overline{\psi} \psi \rangle$ has dimensions of mass$^3$.

Further, in QCD the quark condensate has a quadratic
ultraviolet divergence.  Analytically we can eliminate this
divergence by using dimensional regularization, but on the lattice, there
are terms $\sim 1/a^2$, where
$a$ is the lattice spacing.  In numerical simulations, this divergence
is eliminated by computing the difference between the condensates
between the light and heavy quarks, weighted by the quark mass difference:
\begin{equation}
\Delta_{u,s}^{lattice}(T) = 
\frac{\langle \overline{\psi} \psi \rangle_{u,T}
- (m_u/m_s) \langle \overline{\psi} \psi \rangle_{s,T}}
{\langle \overline{\psi} \psi \rangle_{u,0}
- (m_u/m_s) \langle \overline{\psi} \psi \rangle_{s,0}} \; .
\label{define_Delta}
\end{equation}
Here $m_u$ and $m_s$ are the current quark masses for the up and
strange quarks, and $\langle \overline{\psi} \psi \rangle$
the corresponding condensates.  

\begin{figure}
\includegraphics[width=0.8\linewidth]{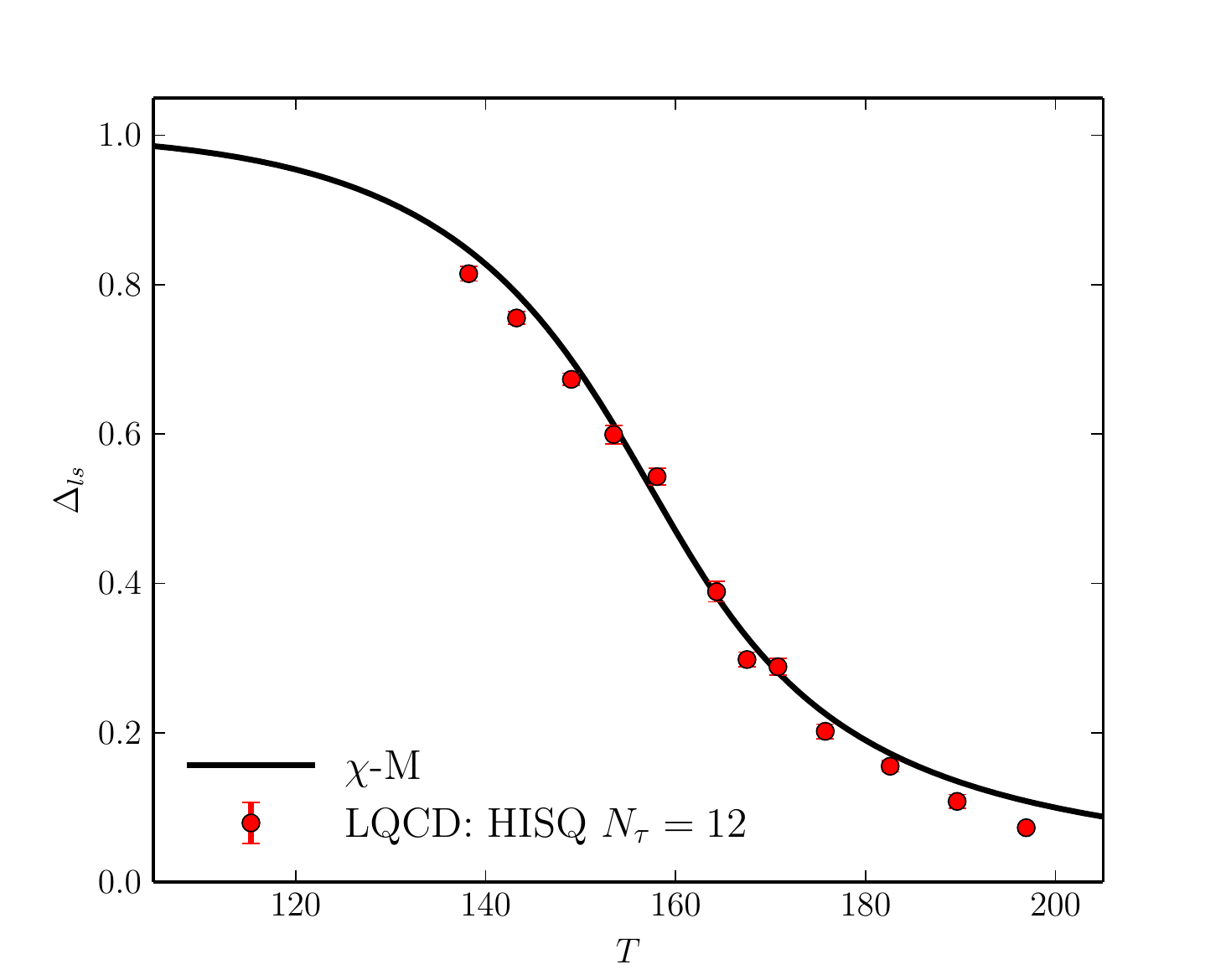}
\caption{
The subtracted chiral condensates, from the lattice
\cite{bazavov_chiral_2012},
Eq. (\ref{define_Delta}), and the matrix model, Eq. (\ref{model_Delta}).
}
\label{fig:chiral_cond}
\end{figure}

We then compute this ratio of condensates in our model, where 
the analogous quantity is 
\begin{equation}
\Delta_{u,s}^{\chi-M}(T) = 
\frac{\Sigma_u(T) - (h_u/h_s) \Sigma_s(T)}
{\Sigma_u(0) - (h_u/h_s) \Sigma_s(0)} \; .
\label{model_Delta}
\end{equation}
These two quantities are shown in Fig. (\ref{fig:chiral_cond}).  The close
agreement between the lattice results of 
Ref. \cite{bazavov_chiral_2012} and the matrix model is
satisfying.

\begin{figure}
\includegraphics[width=0.8\linewidth]{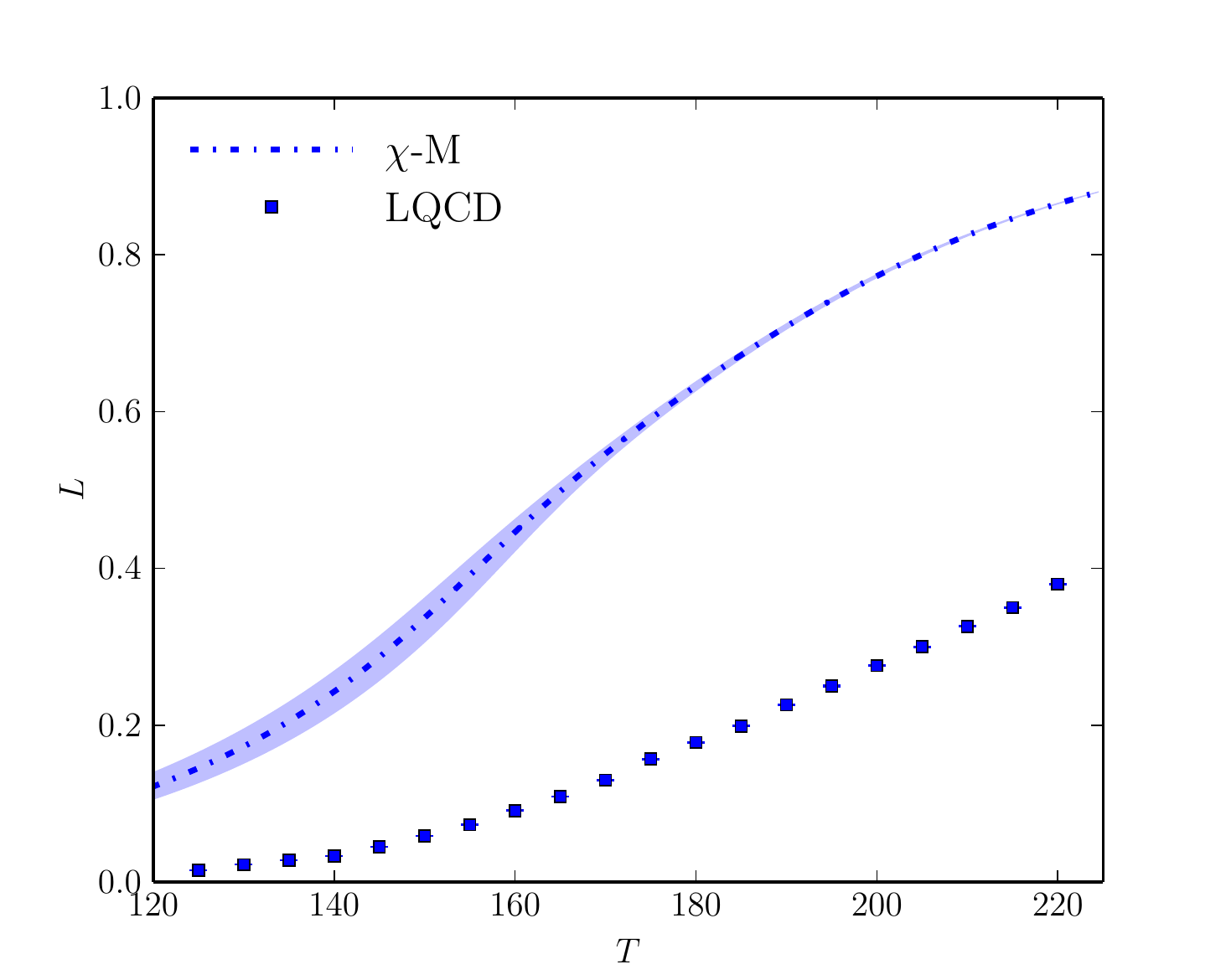}
\caption{
The Polyakov loop in the matrix model and from the lattice
\cite{bazavov_polyakov_2013, bazavov_polyakov_2016}.
The band in the matrix model corresponds to the variation in the Yukawa
coupling from $y: 4.5 \rightarrow 5.5$, as shown before.
}
\label{fig:loop_compare}
\end{figure}

In contrast, there is a {\it strong} difference in the value of Polyakov
loop in our model, and from the lattice 
\cite{bazavov_polyakov_2013, bazavov_polyakov_2016}.  
This is illustrated in Fig. (\ref{fig:loop_compare}).  The
Polyakov loop in the matrix model approaches unity {\it much} quicker
than measurements of the (renormalized) Polyakov loop on the lattice.

Given the good qualitative agreement between the susceptibilities
in the model and the lattice, this disagreement for the Polyakov
loop must be considered the outstanding puzzle of our model.  We note
that a similar disagreement was seen in the pure gauge theory
\cite{dumitru_how_2011, dumitru_effective_2012}.  For this reason,
in Sec. (\ref{sec:alternate}) we consider alternate models in which
we fit the Polyakov loop, more or less by hand.  We show that doing so
obviates any agreement for other quantities, such as the pressure
and susceptibilities.

%The behavior of the chiral order parameters is reasonably close to
%the results of numerical simulations on the lattice.  In contrast,
%the behavior of the Polyakov loop in the chiral matrix model
%is {\it very} different than indicated by measurements of the
%renormalized Polyakov loop, and
%approaches unity much faster than indicated by the lattice.
%The same was found in the matrix model of the pure glue theory
%\cite{dumitru_how_2011, dumitru_effective_2012}.
%We comment on this further 
%when we consider alternate models in Sec. (\ref{sec:alternate})
%and in the Conclusions, Sec. (\ref{sec:conclusions}).  

\subsection{Susceptibilities for the order parameters, and their
divergences in the chiral limit}
\label{sec:divergent}

\begin{figure}
\includegraphics[width=0.8\linewidth]{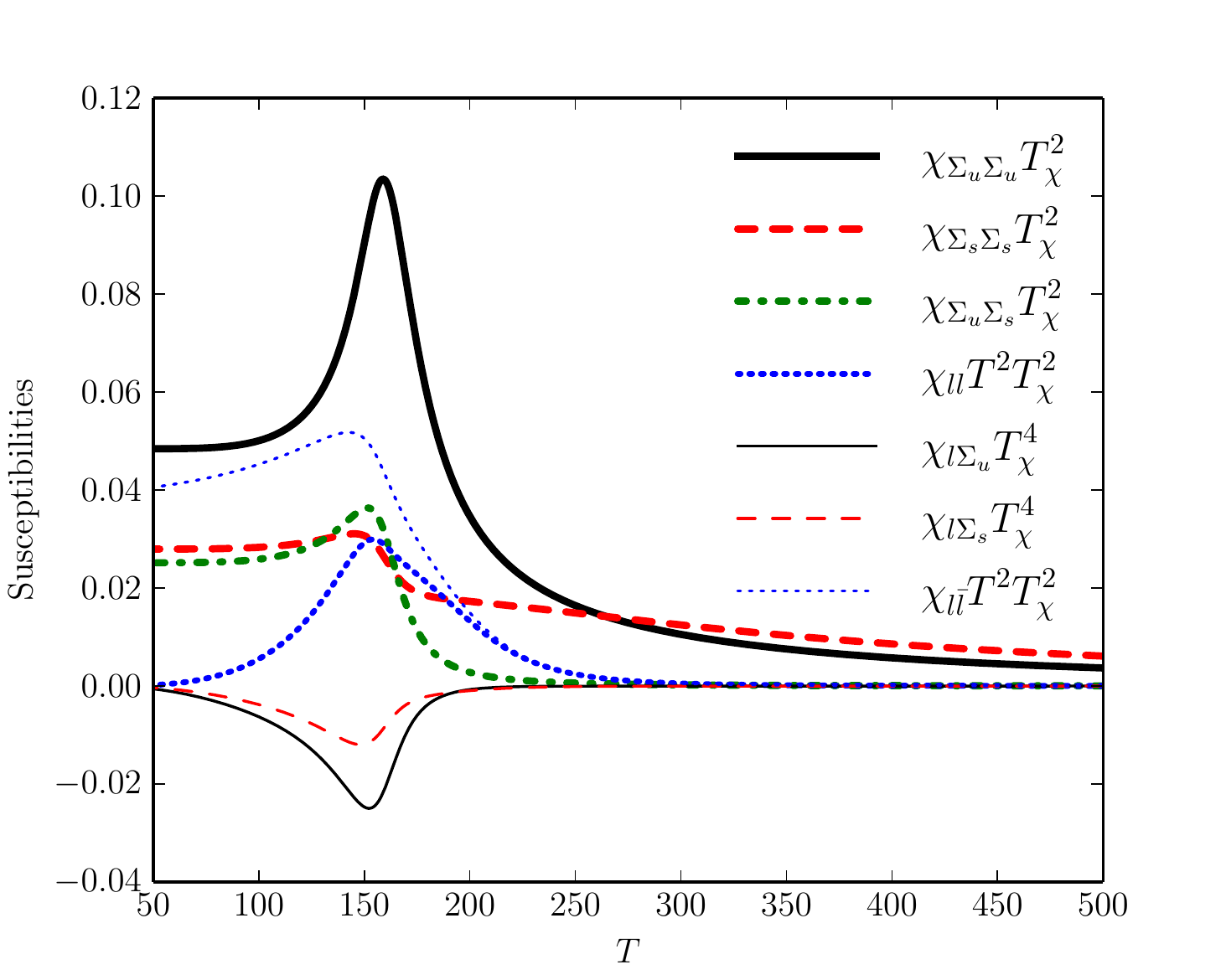}
\caption{
The susceptibilities for the chiral and deconfining
order parameters, as a function of the temperature $T$.
}
\label{fig:susc}
\end{figure}

To better understand how the chiral and deconfining order parameters
are related, it is useful to compute their associated
susceptibilities.  This is shown in 
Fig. (\ref{fig:susc}).  These are normalized to be dimensionless quantities
by multiplying by the relevant powers of $T_\chi$, except for those
for the loop-loop and loop-antiloop, where we use $T^2 T_\chi^2$.  

As expected, the largest peak is that for the light quark condensate,
$\Sigma_u-\Sigma_u$.  That for $\Sigma_u-\Sigma_s$ is less sharp, and
even more so for $\Sigma_s-\Sigma_s$.  This is unremarkable, demonstrating
that a heavy quark is farther from the chiral limit than light quarks.

The susceptibility for the loop correlations are broad.  For both
the loop-loop and loop-antiloop correlations, they peak about $T_\chi$,
with a wide width, due to their coupling to the light quark fields.

The susceptibilities of the loop-antiloop have been computed on the
lattice by Bazavov {\it et al.} \cite{bazavov_polyakov_2016}.  Their
results peak at a significantly higher temperature than we find in
the chiral matrix model, at $\sim 200$~MeV.  This presumably is due
to the fact that the lattice Polyakov loop is shifted to higher 
temperatures than in the chiral matrix model.  They did not investigate
the susceptibility between the loop and the chiral order parameter.

Returning to our results, 
after the $\Sigma_u-\Sigma_u$ correlation, the sharpest peak
is for that between the loop and the light quark condensate, $\Sigma_u-\ell$.
This is not an artifact.  In a Polyakov loop model, 
Sasaki, Friman, and Redlich \cite{sasaki_susceptibilities_2007} 
found that the $\Sigma$-loop correlation is divergent:
see Fig. (19) of Ref. \cite{sasaki_susceptibilities_2007}.

This is a general result for a chiral transition of second order.
To show this, we consider the interaction of the lowest mass dimension
between a chiral field $\Phi$ and the Polyakov loop $\ell$,
\begin{equation}
\left(
\ell + \ell^*
\right)
{\rm tr} \left( \Phi^\dagger \Phi \right)
\; .
\label{eq:model_sigma_q}
\end{equation}
This coupling respects all of the relevant symmetries of gauge invariance
and chiral symmetry. 
It is not invariant under the global color
symmetry of $Z(3)$, but since this symmetry of the pure gauge theory
is violated by the presence of dynamical quarks, it does arise.
In particular, such a coupling appears in our chiral matrix model.
In general, and in the chiral matrix model, 
there is an infinite series of Polyakov loops, in different
representations, which couple to ${\rm tr} \Phi^\dagger \Phi$.  
We shall argue that this does not alter our conclusions
about the critical behavior which follow.

Consider the mass matrix between the chiral field and the Polyakov loop.
We can concentrate on the field $\Sigma$ which is nonzero in the
phase with chiral symmetry breaking.  The mass squared matrix between $\Sigma$
and $\ell$ is
\begin{equation}
{\cal M}^2 =
\left(
\begin{array}{cc}
m^2 & \kappa \, \Sigma_f \\
\kappa \, \Sigma_f & \widetilde{m}^2 \\
\end{array}
\right) \; ,
\label{scaling_masses_second}
\end{equation}
where $\kappa$ is some constant, and $\widetilde{m}^2$ the mass for
the loop.  Assuming the chiral transition is of second order, 
\begin{equation}
m^2 \sim \delta t 
\;\;\; ; \;\;\;
\Sigma_f \sim \delta t^\beta
\;\;\; ; \;\;\;
\delta t \equiv \left| \frac{T-T_\chi}{T_\chi} \right| \; .
\end{equation}
That the mass of the $\Sigma$ field vanishes as the reduced temperature
$\delta t$ is standard.  Similarly, 
the expectation value of $\Sigma$ vanishes with critical
exponent $\beta$.  The mass of the Polyakov loop is
assumed to be nonzero at the chiral phase transition, since it is
not a critical field.

The susceptibilities are determined by the inverse of this matrix.
Consequently, for that between the loop and the condensate, we
obtain
\begin{equation}
\left. 
\frac{1}{{\cal M}^2}
\right|_{ \ell \, \Sigma}
\sim \delta t^{\beta - 1}
\; .
\label{scaling_loop_sigma}
\end{equation}
In this we assume that $\beta < 1/2$, which is true for the $O(4)$
universality class, which is what enters for two massless flavors
\cite{pisarski_remarks_1984}.

It is direct to show that Eq. (\ref{scaling_loop_sigma}) is
true in a chiral matrix model.  In such a model the coupling is
not between the loop $\ell$ and the scalar field, but between $q$
and $\Phi$.  What matters is that in the phase with $\Sigma_f \neq 0$,
there is a coupling between $q$ and $\Phi$ which is $\sim \Sigma_f$.
This factor can be understood as follows.  The loop diagram between
a $q$ field and the $\Sigma$ is proportional to
\begin{equation}
{\rm tr} \; \gamma^0 \; \lambda_3 \;
\frac{1}{(\slash \!\!\!\! D^{bk} + m_f)^2}
\end{equation}
The factor of $\gamma^0$ is from the coupling to $q$, while the
coupling of a quark antiquark to $\Sigma$ is proportional to unity.
This diagram is nonzero only if the Dirac trace is over two Dirac matrices,
so one of the propagators must bring in a factor of the
quark mass, $m_f \sim y \Sigma_f$.
The mixed susceptibility between the loop and $q$ then behaves
as $\sim y \Sigma_f/m^2 \sim 1/\delta t^{1/2}$.  This is the expected
behavior in mean field theory, where $\beta = 1/2$.

Viewed in a general context of second order phase transitions, it is
not surprising that the coupling between a critical field $\Phi$, and
a noncritical field, $\ell$, gives a weak but divergent susceptibility
for the off-diagonal susceptibility between $\Phi$ and $\ell$.  
Indeed, assuming
that the expectation value of the loop is nonzero at $T_\chi$, even
$Z(3)$ symmetric operators such as $|\ell|^2 {\rm tr} \Phi^\dagger \Phi$
would produce a divergent susceptibility.  However, they would be smaller
by powers of the expectation value of the loop, which is small in QCD
at $T_\chi$.  

\subsection{Chiral susceptibilities and $U(1)_A$}
\label{sec:chiral_suscep}

In Fig. (\ref{fig:masses}) we showed the meson masses as a function of
temperature.  As discussed at the end of Sec. (\ref{sec:mass}), it
still exhibits a violation of the axial $U(1)_A$, with the mass of
the $\eta'$ meson heavier than that of the $\eta$ meson.

This splitting is controlled by the coefficient $c_A$ in the effective
Lagrangian.  Dynamically, at high temperature $c_A$ should decrease
with temperature, as instanton fluctuations are suppressed by the Debye
mass \cite{gross_qcd_1981}.  

To study the restoration of the axial $U(1)_A$ symmetry,
numerical simulations have studied chiral susceptibilities
which are sensitive to this breaking \cite{buchoff_qcd_2014}.
In a chirally symmetric phase, the susceptibilities for the 
$\sigma$ and $\pi$ are equal, as are those for the $\eta'$ and the $a_0$.
This degeneracy is demonstrated by the meson masses in 
Fig. (\ref{fig:masses}).  That the $\pi$ and $a_0$ masses are unequal
is manifestly due to $c_A \neq 0$.  Neglecting the temperature
dependent symmetry breaking term, this is clear from the expressions
for these masses in Eqs. (\ref{eq:pion_mass}) and (\ref{eq:ugly_a0_mass}):
with $\Sigma_u \approx 0$, $m_\pi^2 = m^2 - c_A \, \Sigma_s$ and
$m_{a_0}^2 = m^2 + c_A \, \Sigma_s$.  

Numerical simulations find that while the $\pi$ and $a_0$ susceptibilities
differ at $T_\chi \sim 155$~MeV, 
they are essentially equal by $T_{U(1)_A} \sim 200$~MeV.
At zero temperature there is a close relationship between the spontaneous
breaking of chiral symmetry and anomalous amplitudes, such as for
$\pi^0 \rightarrow \gamma \gamma$.  Naively this suggests
that $T_{U(1)_A} \approx T_\chi$.
However, at nonzero temperature
Lorentz invariance is lost, and this relationship is much more involved
\cite{pisarski_how_1997}.  Consequently, the
two temperatures $T_{U(1)_A}$ and $T_\chi$ can differ.  The lattice
shows that $T_{U(1)_A} > T_\chi$; for other numbers of flavors and colors,
to us it seems possible that $T_{U(1)_A} < T_\chi$.

One might hope to compute the $\pi$ and $a_0$ susceptibilities in the matrix
model, to fix the temperature dependence of $c_A$. 
This was done in Ref. \cite{ishii_determination_2016} in a Polyakov
Nambu-Jona-Lasino model.

The difficulty is that while our chiral matrix model can be used to compute
many quantities, it cannot be used to compute all.   
Consider the quark operator with pion
quantum numbers, $J_5^a = \overline{\psi} \tau^a \gamma_5 \psi$.
The chiral susceptibility for the pion is
dominated by single pion exchange, 
$\sim \langle 0 | J_\pi | \pi \rangle 1/m_\pi^2 
\langle \pi | J_\pi | 0 \rangle$. 

The form factors are determined by partially conserved axial current.  
The axial current satisfies $\partial_\mu J_\mu^{5,a} = 2 m_{qk} J_5^a$,
where $J_\mu^{5,a} = \overline{\psi} \tau^a \gamma_\mu \gamma_5 \psi$,
and $m_{qk}$ is the current quark mass.
Since $\langle 0 | J_\mu^{5,a} | \pi \rangle \sim P^\mu f_\pi$,
using $P^2 = m_\pi^2 = m_{qk} \langle \overline{\psi}\psi\rangle/f_\pi^2$,
we find that 
$\langle \pi | J_\pi | 0 \rangle \sim 
\langle \overline{\psi}\psi\rangle/f_\pi$.  

In QCD, the expectation value is  $\langle \overline{\psi}\psi\rangle 
\sim - (300~MeV)^3$.  In the chiral matrix model, 
computation shows that the analogous quantity
is much smaller, $\langle \overline{\psi}\psi\rangle 
\sim - m_\pi^3 \sim - (140~MeV)^3$.  This difference is consistent
with chiral symmetry: in QCD the condensate only enters multiplied
by the current quark mass.  In the chiral matrix model, the pion
mass is related to the background field $h_u$, and has no direct
relation to the chiral condensate 
$\langle \overline{\psi}\psi\rangle$.

However, what matters for 
the associated chiral susceptibilities are the form factors, and so
$\langle \overline{\psi}\psi\rangle$.  These are too small by
an order of magnitude, and so cannot be used to constrain $c_A$.

\section{Flavor susceptibilities}
\label{sec:flavor_susp}

Besides the computation of bulk thermodynamic properties, most useful
insight is gained by computing derivatives with respect to quark chemical
potentials.

In principle this is straightforward, simply the derivative of the
effective potential with respect to the relevant $\mu$, evaluated
at $\mu = 0$.  For example, the baryon number
susceptibility is given by
\begin{equation}
\chi_n^B = T^{n-4} \;
\left. 
\frac{\partial^n P}{\partial \mu_B^n} 
\right|_{\mu = 0}
\; .
\label{Eq:defchiB}
\end{equation}
Particularly in our model, it is trivial to take derivatives with
respect to a given flavor, to compute the corresponding susceptibility.

At the outset we should note that because we treat the mesons in mean
field approximation, implicitly we neglect fluctuations from pions.
Pion fluctuations are not important in computing susceptibilities
with respect to baryon number and strangeness, but do matter in
computing those with respect to other chemical potentials,
including those for up and down flavor number, isospin, and charge.  

There is one
point which must be treated with care, as was discussed in Sec.
(\ref{sec:quark_potential}).  
Most quantities are even under charge conjugation, ${\cal C}$.  This
includes the effective potential, and the stationary points for the chiral
condensates, $\Sigma_u$ and $\Sigma_d$, and for the Polyakov loop, $q$.
The latter is not obvious: while the gauge potential $A_0 \rightarrow
- A_0^*$ under ${\cal C}$, because we assume that the stationary point
for the Polyakov loop is real, we always sum over $q$ and $-q$.  
That these quantities are even under ${\cal C}$ greatly simplifies
how they can enter into quark number susceptibilities.

Previously, however, we argued that when $\mu \neq 0$, that the
stationary point involves imaginary values of $r = i {\cal R}$,
Eq. (\ref{eq:imag_rR}).  This means that we can compute quark number
susceptibilities using a type of Furry's theorem: loops with
insertions of $\mu$ correspond to a type of coupling to an Abelian 
gauge field.  There must be an {\it even} number of insertions, where
{\it both} insertions of $\mu$ or $r$ can enter. Since we
work in mean field approximation, only one field can be exchanged.

\subsection{Second order susceptibilities}
\label{sec:second_sus}

\begin{figure}
\includegraphics[height=0.1\linewidth]{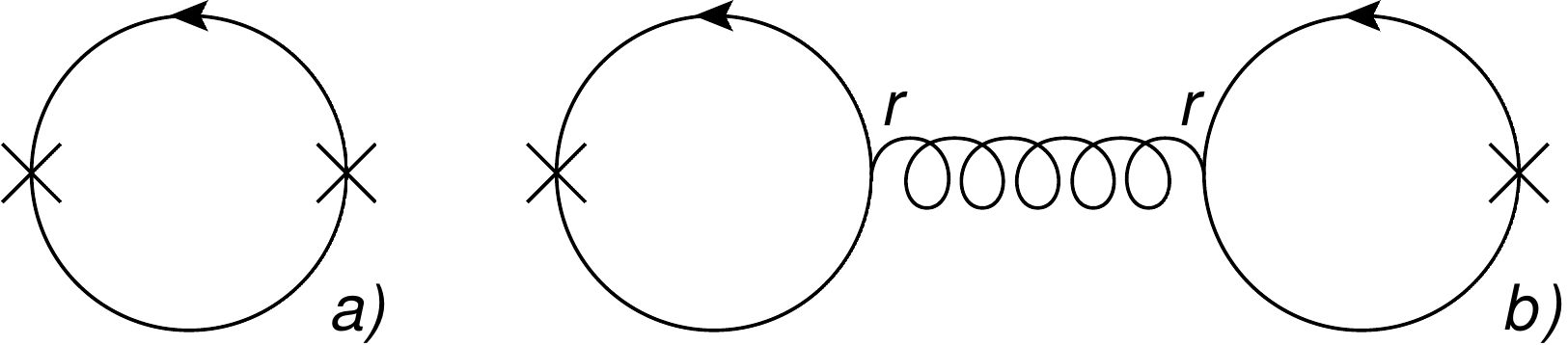}
\caption{
Contributions to the second order 
baryon number susceptibility  $\chi_2^B$: (a) the one particle irreducible 
and (b) the one particle reducible. 
The wiggly line denotes $r-r$ propagator.  
}
\label{fig:c2_d}
\end{figure}

Let us start with the simplest quantity, $\chi_2^B$.  The diagrams
which contribute are illustrated in Fig. (\ref{fig:c2_d}).  
The first diagram, on the left, is expected: two insertions of the
chemical potential into a quark loop.  What is unexpected is the
second diagram, where one has two quark loops, each with single insertions
of $\mu$ and $r$, coupled by a single propagator for $r$.  Since 
we are computing fluctuations, that the stationary point in $r$ is
imaginary is really secondary; what matters is that $r$, like $\mu$,
is ${\cal C}$ odd.  Thus both diagrams satisfy Furry's theorem.
Note that the second diagram is only nonzero when $q \neq 0$: otherwise,
as an insertion of $r$ brings in $\lambda_8$, Eq. (\ref{eq:def_cartan}),
the color trace vanishes.  

\begin{figure}
\includegraphics[width=0.45\linewidth]{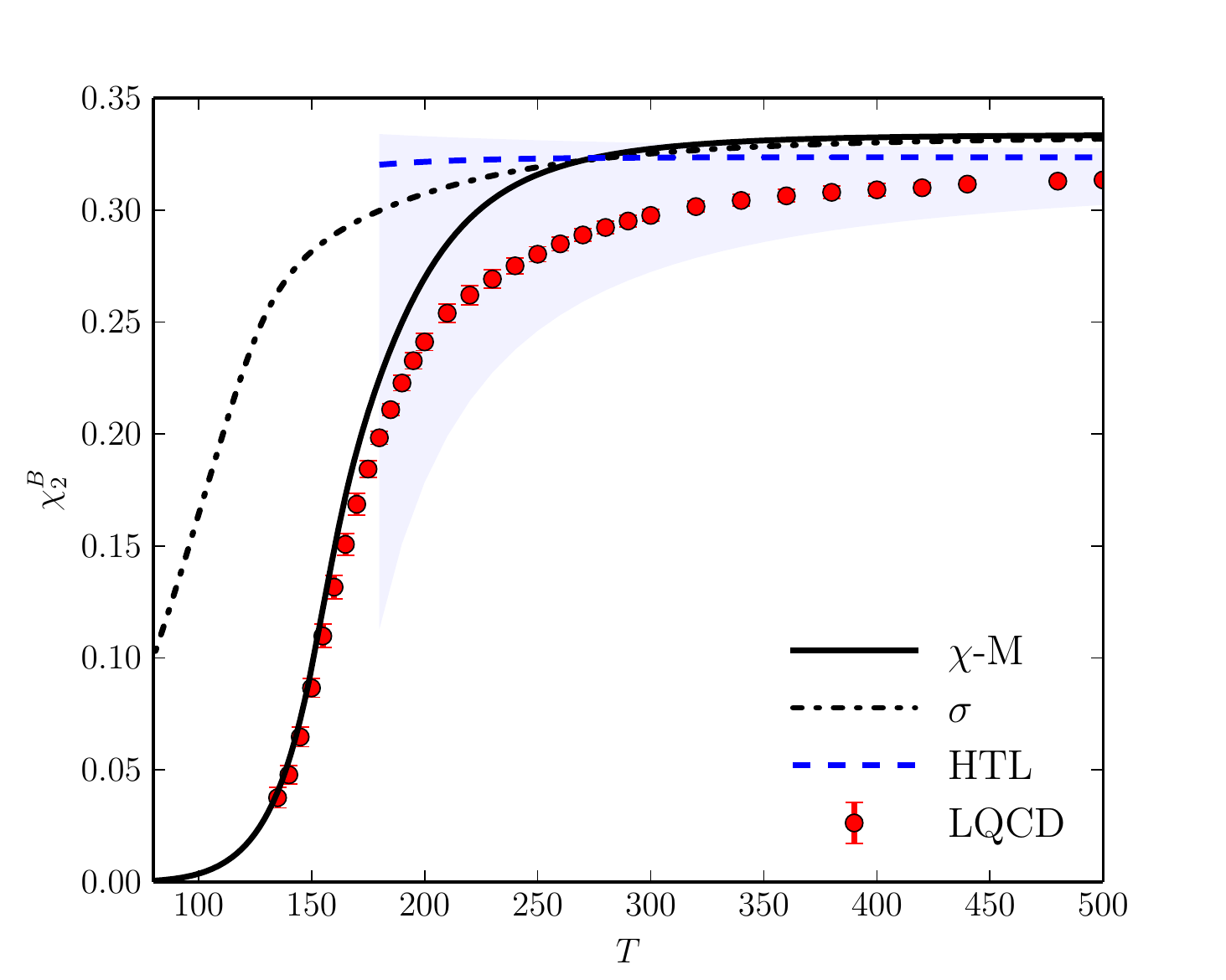}
\includegraphics[width=0.45\linewidth]{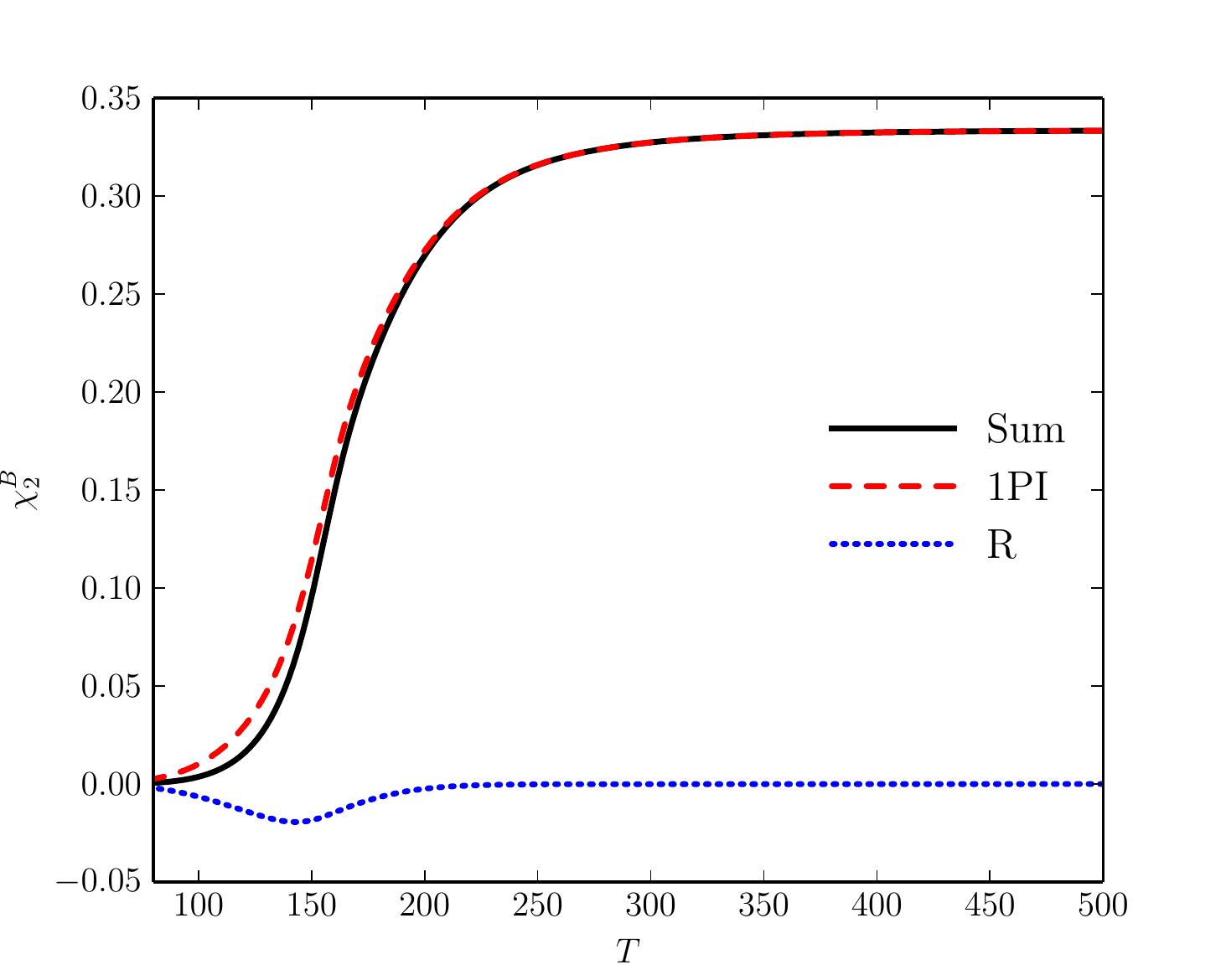}
\caption{The second order baryon number susceptibility  
as a function of the temperature.  
The left panel shows the results
in the chiral matrix ($\chi-M$) model, compared to a $\sigma$ model,
HTL resummation
\cite{andersen_three-loop_2010, *andersen_gluon_2010, *andersen_nnlo_2011, *andersen_three-loop_2011,*haque_two-loop_2013, *mogliacci_equation_2013, *haque_three-loop_2014}, 
and numerical simulations on the lattice.  The
right panel shows contributions to the chiral matrix model:
it is dominated by the contribution of the one particle irreducible
contribution, Fig.~(\ref{fig:c2_d}a), over $r$-exchange in
Fig.~(\ref{fig:c2_d}b).
}
\label{fig:c2_on_T}
\end{figure}

The results for $\chi_2$ are given in Fig. (\ref{fig:c2_on_T}).  
It is completely dominated over all temperatures by the one
particle irreducible contribution in Fig.~(\ref{fig:c2_d}a).
The second diagram, from the exchange of a $r$ gluon, is present,
but numerically small over all temperatures.

The results of the chiral matrix ($\chi-M$) model approach the asymptotic
value of $1/3$ faster than the lattice data.  However, the gross behavior
agrees with the lattice.  The chiral matrix model certainly agrees
much better with the lattice data than Hard Thermal Loop resummation,
which stays near $1/3$.  More surprisingly, it also agrees much better
than a sigma model, which incorporates chiral symmetry breaking, but
not the nontrivial holonomy of the Polyakov loop, $q \neq 0$.  We shall
see this is  true for higher susceptibilities as well.

\begin{figure}
\includegraphics[width=0.4\linewidth]{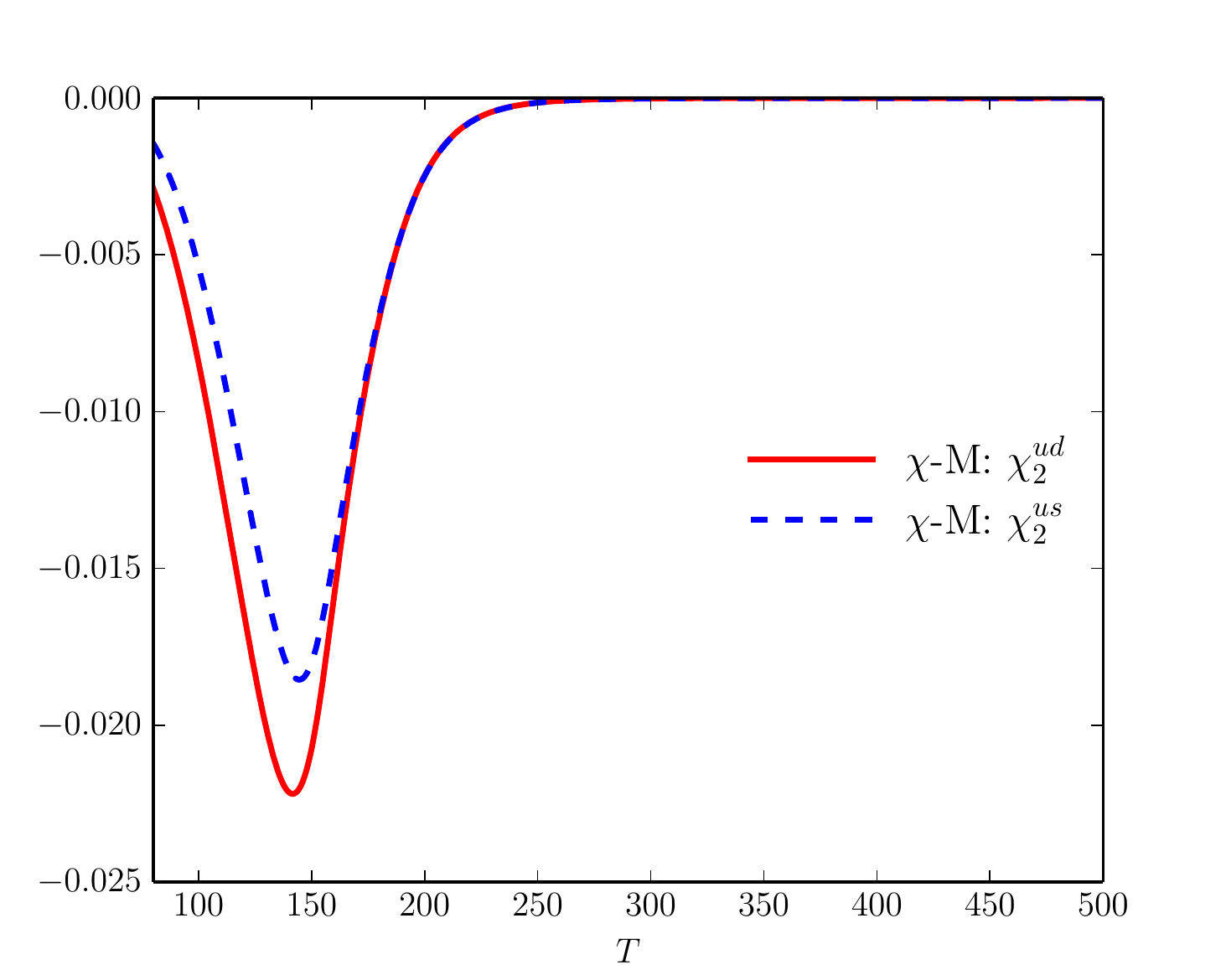}
\caption{The second order off-diagonal susceptibilities
as a function of the temperature.  
}
\label{fig:c2off}
\end{figure}

One can also compute the off-diagonal susceptibilities.
Those for light-light, $ud$,  and heavy-light, $us$, are illustrated
in Fig. (\ref{fig:c2off}).  
This is a very interesting quantity to compute, because on the lattice,
it is due to disconnected diagrams.  In our model, the off-diagonal
susceptibilities are due {\it entirely} not to the connected diagram,
Fig. (\ref{fig:c2_d}.a), but to the diagram from the exchange of
an $r$ gluon, Fig. (\ref{fig:c2_d}.b).

In our model, we find that the off-diagonal susceptibilities 
for $ud$ and $us$ are nearly equal.  This is easy to understand, because
the difference is only one of form factors: generating an $r$ gluon
from an up loop is about as probable  as from a strange loop.

The results of our model for the off-diagonal susceptibilities $us$
are in reasonable agreement with lattice simulations.  On the other
hand, the results for $ud$ are about an order of magnitude {\it smaller}
than measured on the lattice.  
This is because we do not include dynamical hadrons, in particular
pions, in our model.  The most direct way of including dynamical pions
would be to use the Functional Renormalization Group
\cite{skokov_non-perturbative_2012}.

\subsection{Fourth order susceptibilities}
\label{sec:fourth_sus}

\begin{figure}
\includegraphics[height=0.1\linewidth]{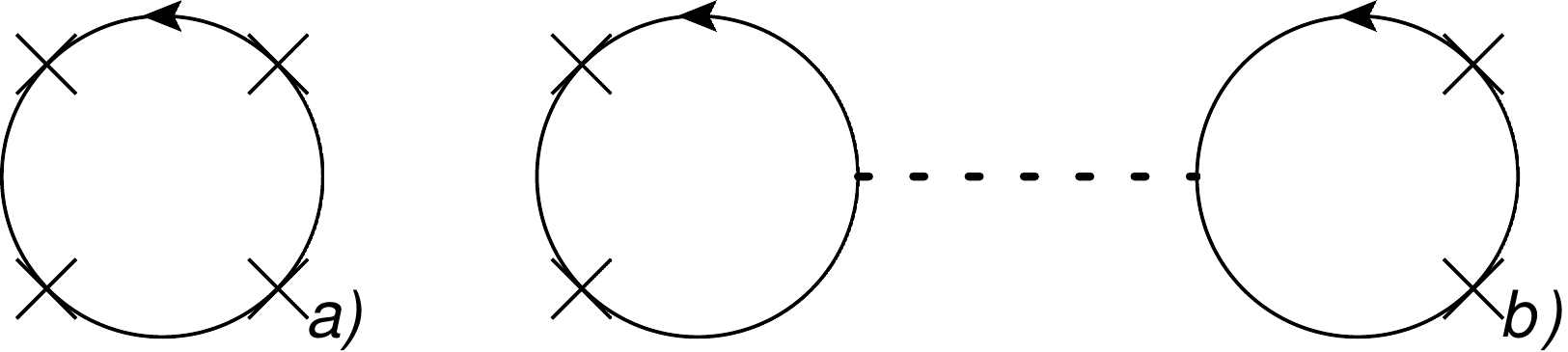} 
\includegraphics[height=0.1\linewidth]{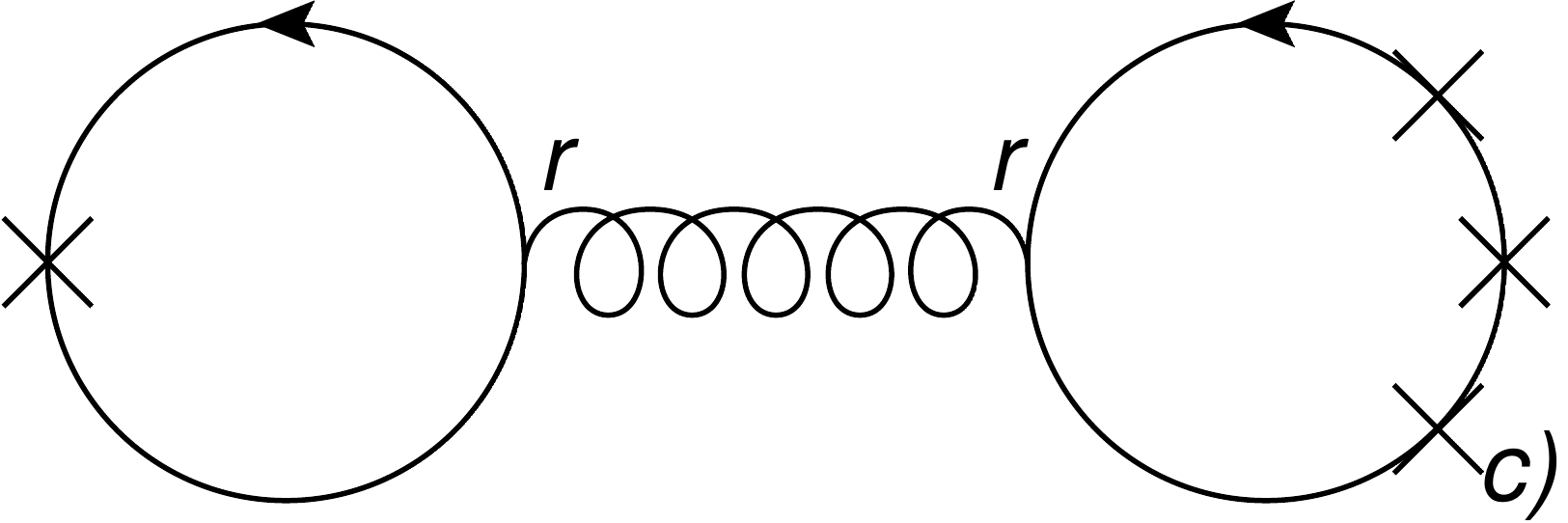}
\caption{Contributions to the fourth order baryon number susceptibility
$\chi_4^B$:  (a) the one particle irreducible 
and (b-c) the one particle reducible. 
  Only diagrams to two quark loop order are shown, which
are not inclusive.
The dashed line denotes the propagator for ${\cal C}$ even fields, 
$q$, $\Sigma_u$, and $\Sigma_s$; the gluon line, for $r$.
}
\label{fig:c4_d}
\end{figure}

Turning to the fourth order susceptibility, the diagrams which contribute
are those of Fig. (\ref{fig:c4_d}).  The diagrams include
four insertions of the chemical potential into a quark loop,
Fig. (\ref{fig:c4_d}.a).  Then there are two 
insertions of the chemical potential
into two different loops, connected by the exchange of ${\cal C}$ even
fields, either $q$, $\Sigma_u$, or $\Sigma_s$,
Fig. (\ref{fig:c4_d}.b).  Lastly, there is diagram from one quark
loop, with a single insertion of $\mu$, and another quark loop, with
three insertions of $\mu$, connected by the exchange of an $r$ gluon.

\begin{figure}
\includegraphics[width=0.45\linewidth]{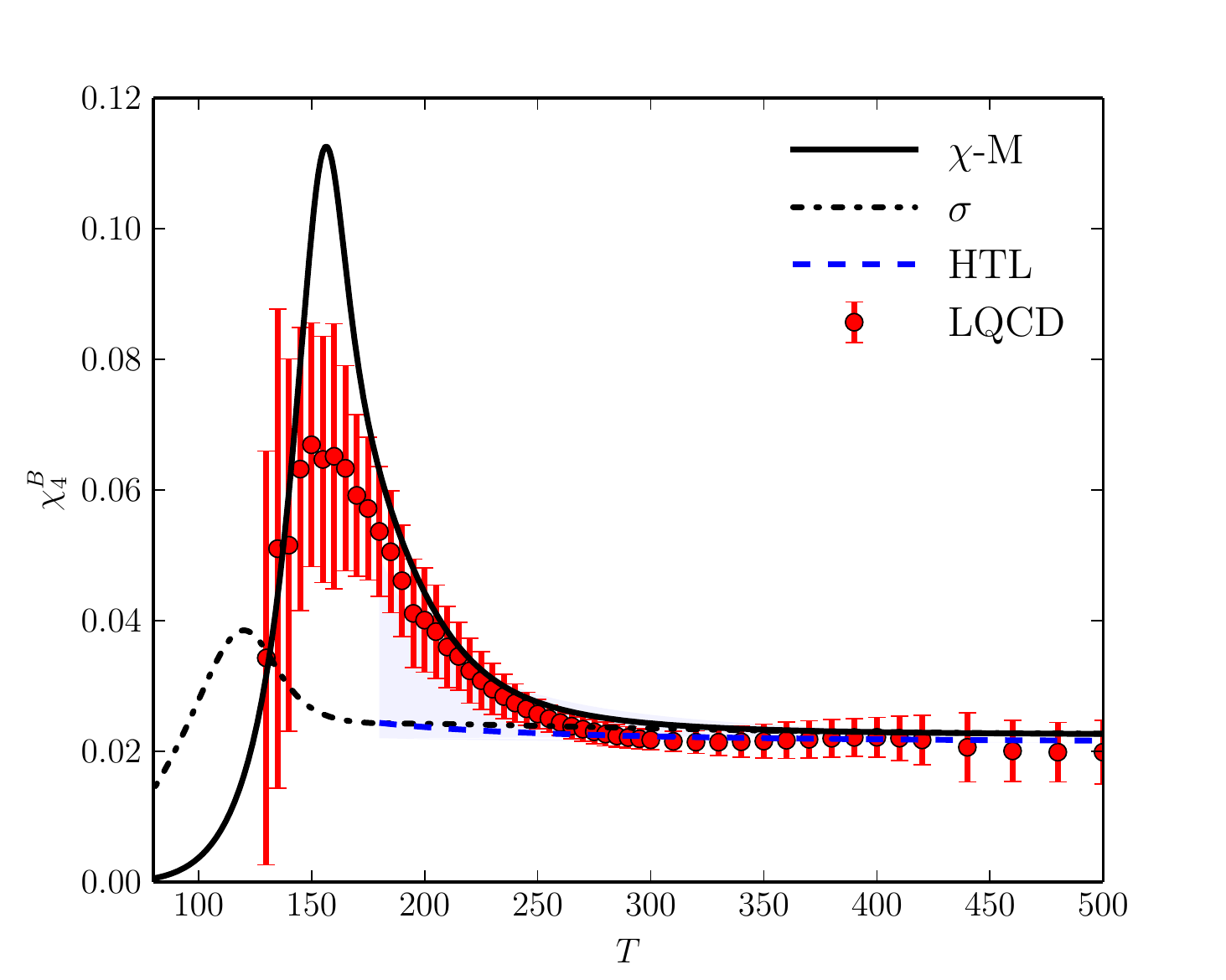}
\includegraphics[width=0.45\linewidth]{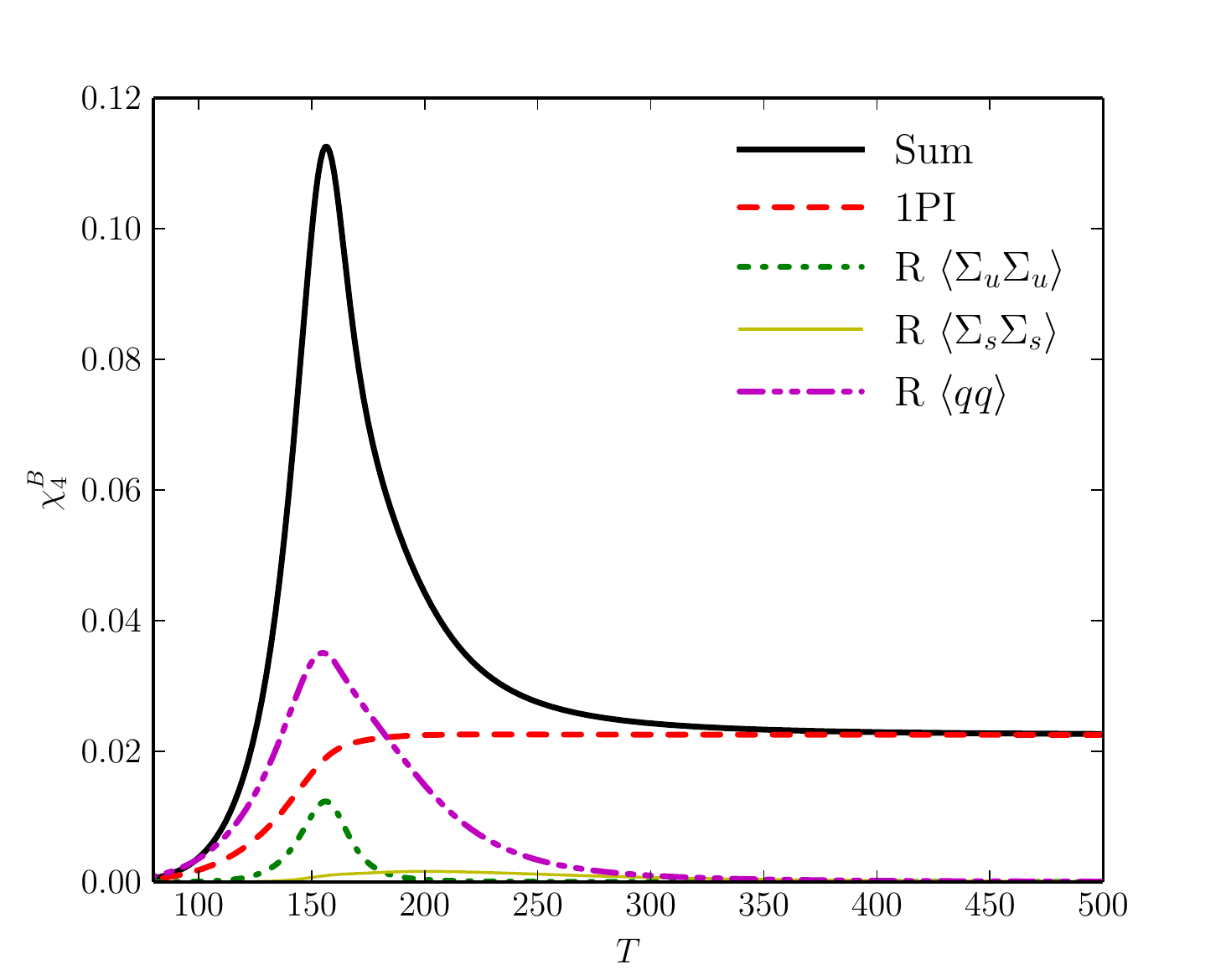}
\caption{The fourth order baryon number susceptibility 
as a function of the temperature.  The left panel shows the results
in the chiral matrix ($\chi-M$) model, compared to a $\sigma$ model,
HTL resummation
\cite{andersen_three-loop_2010, *andersen_gluon_2010, *andersen_nnlo_2011, *andersen_three-loop_2011,*haque_two-loop_2013, *mogliacci_equation_2013, *haque_three-loop_2014}, 
and numerical simulations on the lattice.  The
right panel shows contributions to the chiral matrix model; only
those from Fig. (\ref{fig:c4_d}) are shown, which are not inclusive.
}
\label{fig:c4_on_T}
\end{figure}

The results for the fourth order baryon number susceptibility
are shown in Fig. (\ref{fig:c4_on_T}).  In this case, the one
particle irreducible contribution of Fig. (\ref{fig:c4_d}.a) 
gives a smooth contribution which is no longer dominant.  Instead,
the exchange of a $q$ gluon gives the largest contribution near $T_\chi$.
Indeed, this is larger than that of the $\Sigma_u$ field.  

The results of the chiral matrix model for $\chi_4^B$ appear to
overshoot the results of the lattice by a factor of two near $T_\chi$,
but the lattice results have large error bars.  More striking is that
the Hard Thermal Loop result is essentially constant with respect to
temperature, while a sigma model gives a result which is too small,
and peaked at a temperature significantly below that of the lattice
data.

It is also interesting to plot the difference of the second and
fourth order baryon susceptibilities.  

Consider the contribution of the quarks to the pressure for a single
flavor, 
\begin{equation}
p =  - \, 2 \, T \, \sum_{a=1}^{N_c} \int \frac{d^3 k}{(2\pi)^3} 
	\ln\left( 1 + e^{-E_f/T-\mu/T 
+ i \frac{2\pi}{3} q_a}\right)
+  	\ln\left( 1 + e^{-E_f/T+\mu/T 
- i \frac{2\pi}{3} q_a}\right). 
\label{Eq:pressure1f}
\end{equation}
Expanding in powers of the fugacity,
\begin{equation}
	p = \frac{m^2 T^2 N_c}{\pi^2} 
\sum_{n=1}^\infty  \frac{(-1)^{n+1}}{n^2} K_2(n m/T) 
\left( 
\ell_n e^{n \mu/T} + \ell^\dagger_n e^{- n \mu/T} \right) 
	\label{Eq:hatp}
\end{equation}
where $\ell_n$ is the Polyakov loop in the fundamental representation
which wraps around in imaginary time $n$ times,
\begin{equation}
\ell_n = \frac1{N_c} \sum_{a=1}^{N_c}  
\exp\left( i \frac{2\pi}{3} q_a\right). 
	\label{Eq:ln}
\end{equation}
The $\ell_n$ can be expressed in terms of loops in various irreducible
representations.  We shall not need the detailed form.  All that
matters here is that those which wrap around a multiple of $N_c$ times
include the identity representation.  At small temperatures, these
terms are nonzero, and so dominate.

To eliminate the contribution of such ``baryonic'' loops, we construct
a quantity for which $\ell_{N_c}$ cancels.  
Notice that to second and fourth order, the baryon number susceptibilities
are
\begin{eqnarray}
\chi_2^B &=&  \frac{2 \, N_c \, m^2}{9 \, \pi^2 \, T^2} 
\sum_{n=1}^\infty (-1)^{n+1} K_2
\left(
\frac{n m}{T}
\right) 
\; \ell_n \; , \\
\chi_4^B &=&  \frac{2 \, N_c \, m^2}{81 \, \pi^2 \, T^2} 
\sum_{n=1}^\infty (-1)^{n+1}  n^2 K_2
\left(
\frac{n m}{T}
\right) \; \ell_n \; .
	\label{Eq:chi2chi4}
\end{eqnarray}
For three colors the difference between the two is
\begin{equation}
\chi_2^B - \chi_4^B \approx  \frac{2 \, m^2}{27 \, \pi^2\, T^2} 
\left(
8 K_2
\left(
\frac{m}{T}
\right) \; \ell_1 - 5 K_2
\left(
\frac{2 m}{T}
\right) \ell_2  + \ldots \right) \; .
	\label{Eq:differencechi2chi4}
\end{equation}
The contribution from the loop $\ell_3$ cancels in the difference.
One can show that $\ell_2 = \ell_1(3 \, \ell_1-2)$, so 
at small temperature, $\chi_2^B - \chi_4^B$ 
is proportional to the loop, $\ell_1$.  There are also terms
$\sim \ell_4, \ell_5$ and so on in Eq. (\ref{Eq:differencechi2chi4}),
but these are numerically small.

\begin{figure}
\includegraphics[width=0.45\linewidth]{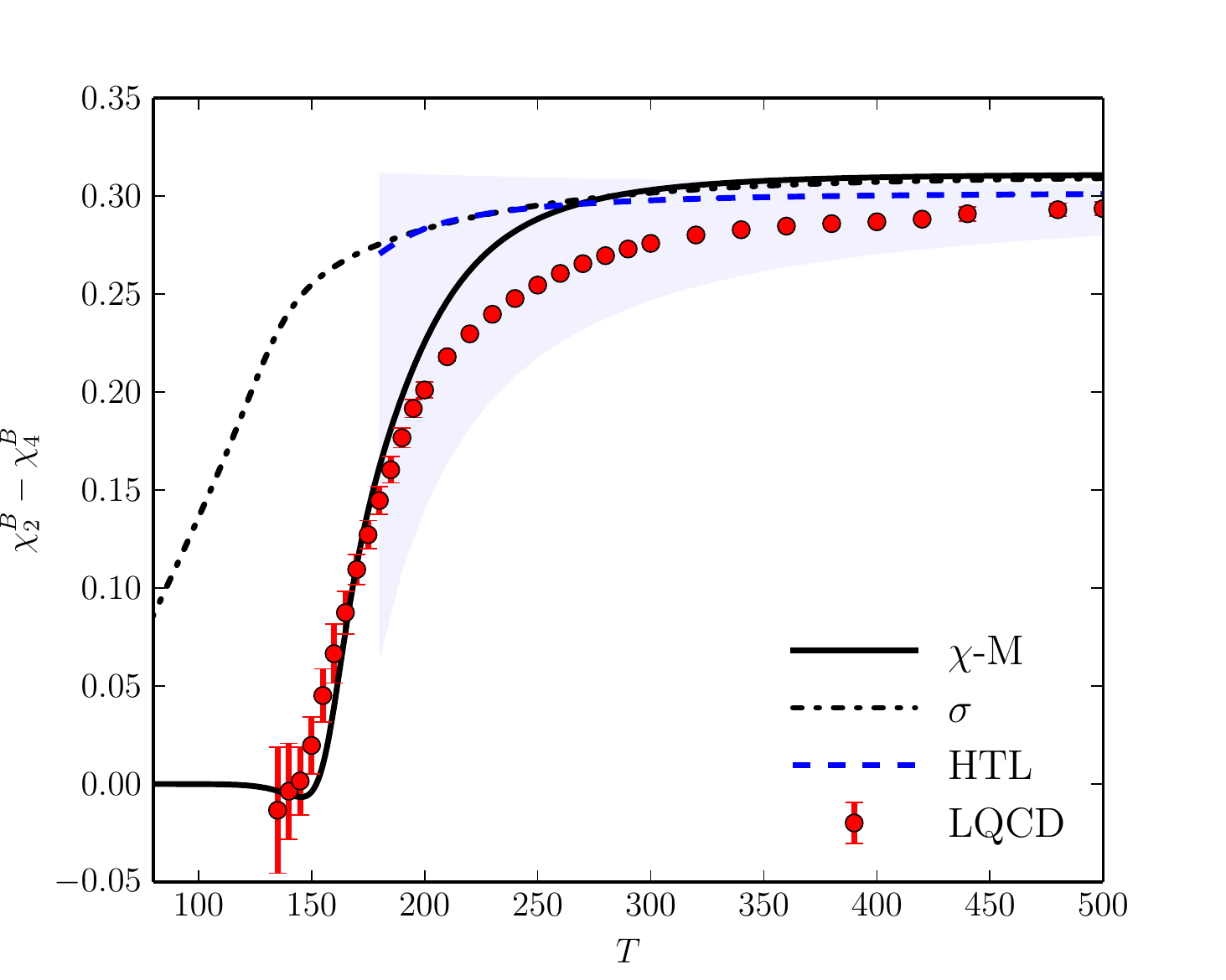}
\includegraphics[width=0.45\linewidth]{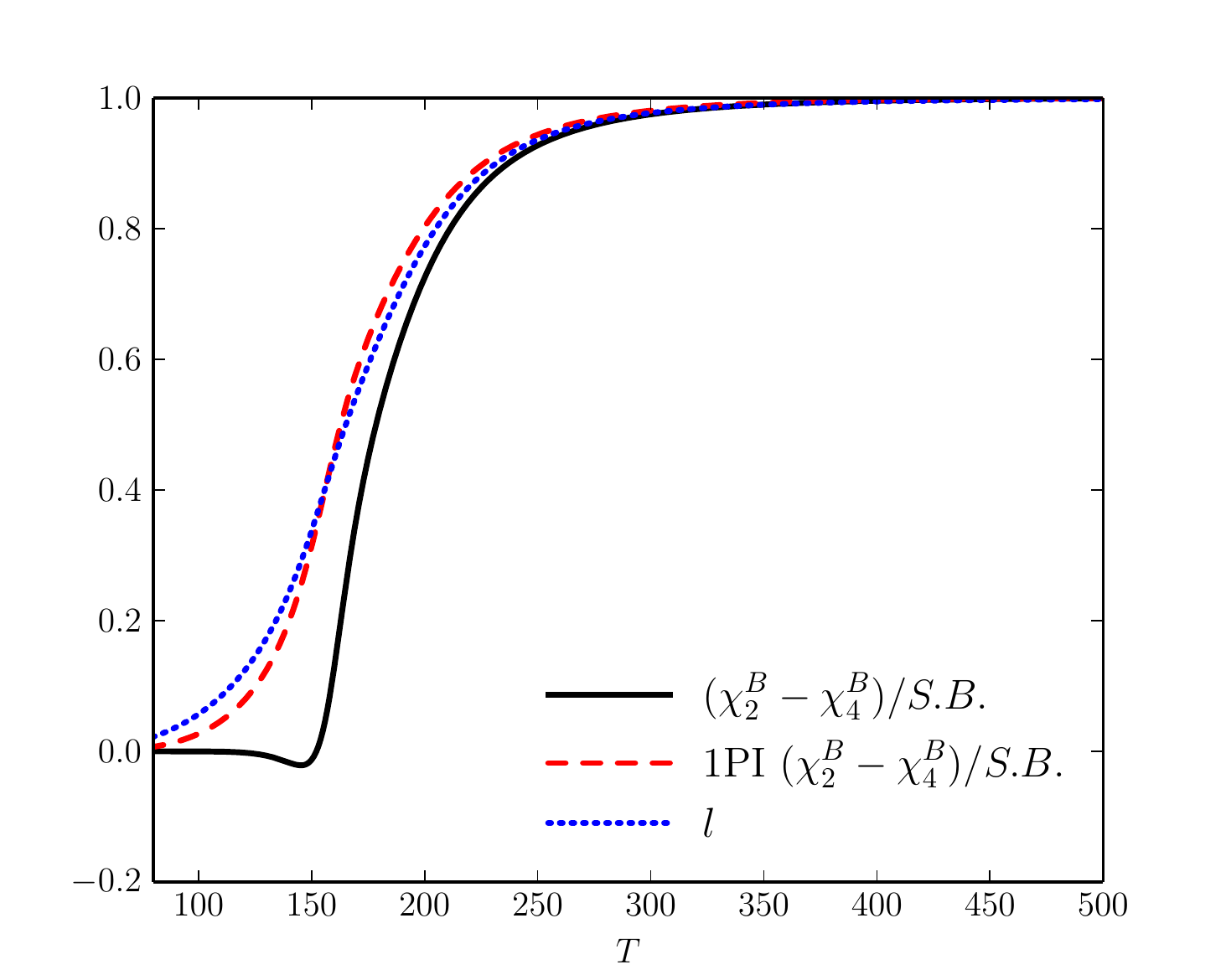}
\caption{The difference between the second and forth order baryon number 
susceptibilities  as a function of the temperature. 
}
\label{fig:c2_minus_c4_on_T}
\end{figure}

In Fig. (\ref{fig:c2_minus_c4_on_T}) we plot this difference
as a function of the temperature.  The chiral matrix model agrees
very well with the lattice results up to temperatures of $\sim 200$~MeV,
and then goes more quickly to a constant value than the lattice data.
In contrast, HTL resummation gives essentially a constant value
\cite{andersen_three-loop_2010, *andersen_gluon_2010, *andersen_nnlo_2011, *andersen_three-loop_2011,*haque_two-loop_2013, *mogliacci_equation_2013, *haque_three-loop_2014}.
More surprising, a sigma model, which includes chiral symmetry restoration
but not the change in the Polyakov loop, is much higher than the lattice
data.  As can be seen from the panel on the right-hand side, 
this difference of susceptibilities is nearly proportional to the
Polyakov loop.

\subsection{Sixth order susceptibilities}
\label{sec:sixth_sus}

\begin{figure}
\includegraphics[height=0.1\linewidth]{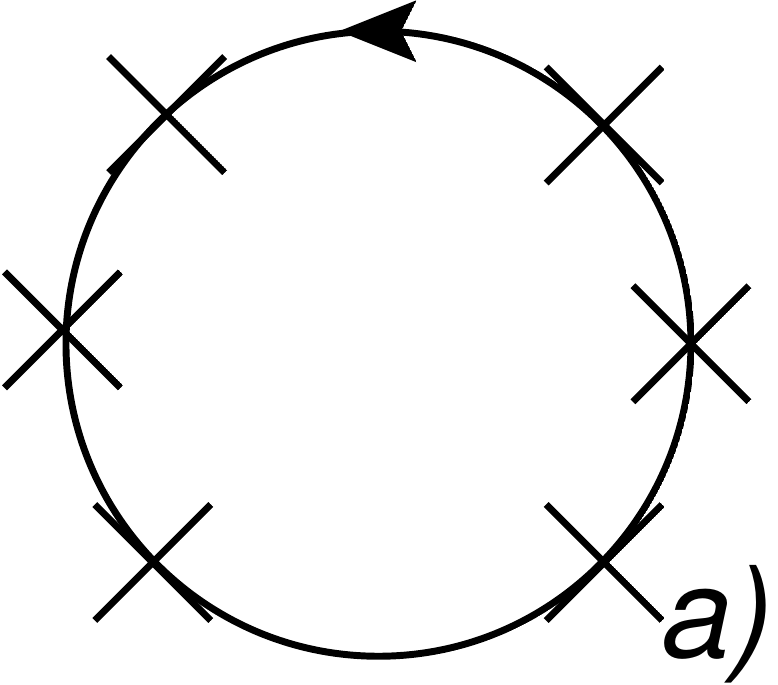} 
\includegraphics[height=0.1\linewidth]{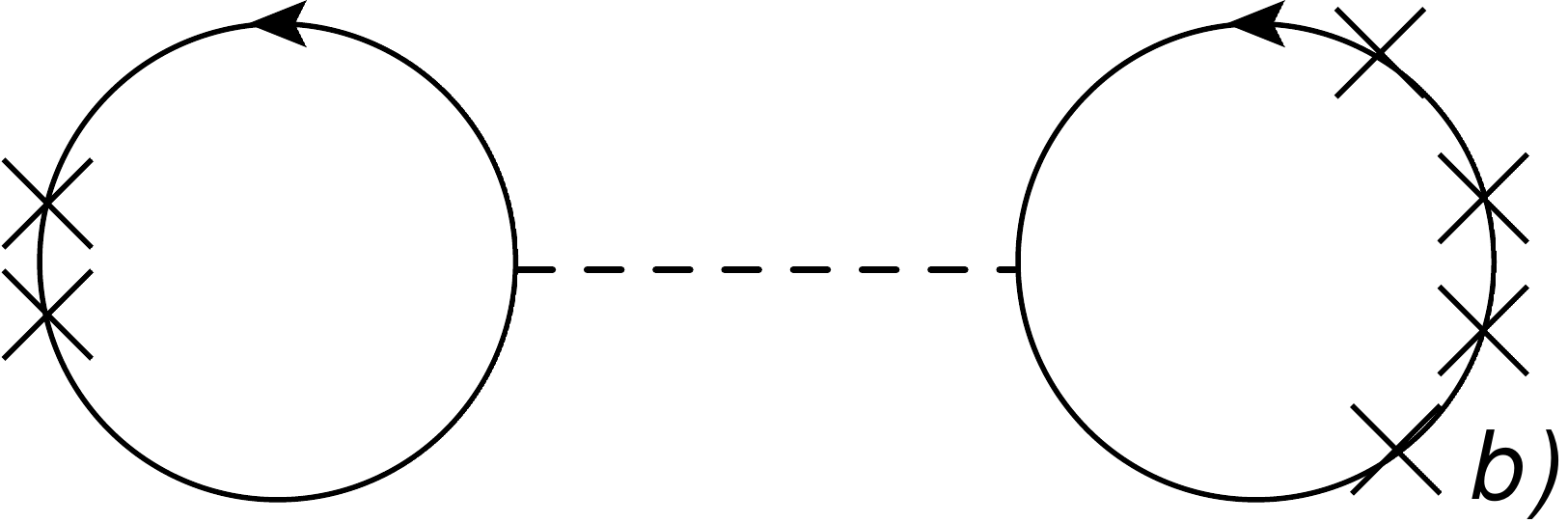}
\includegraphics[height=0.1\linewidth]{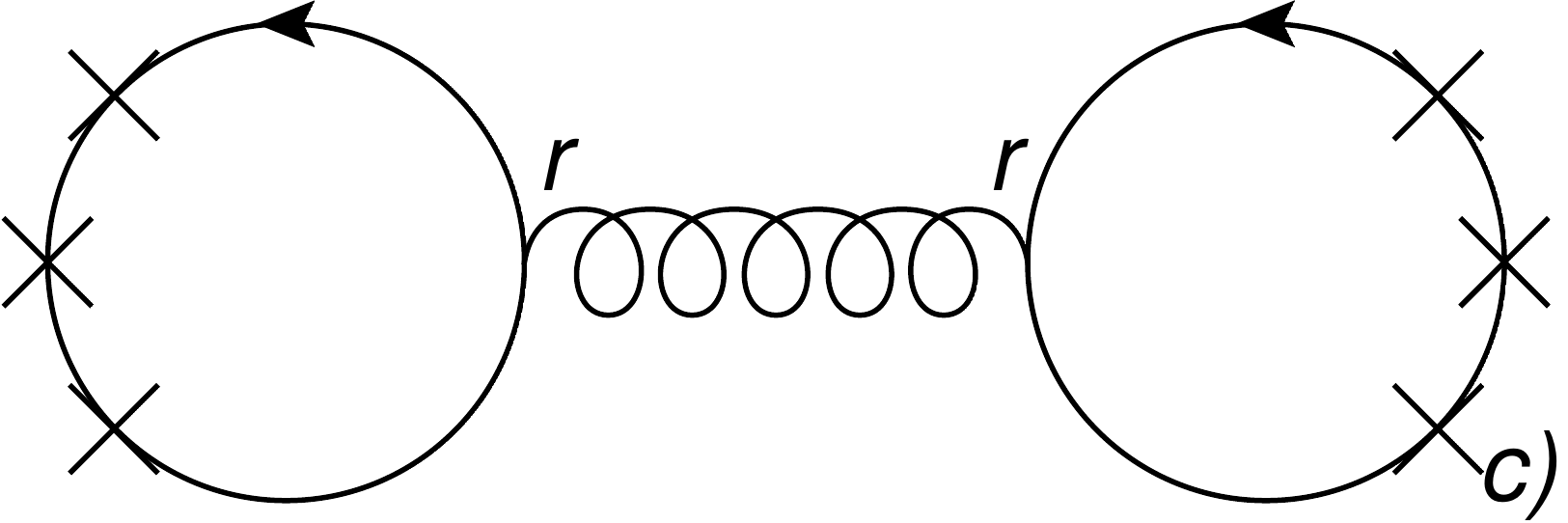}
\includegraphics[height=0.1\linewidth]{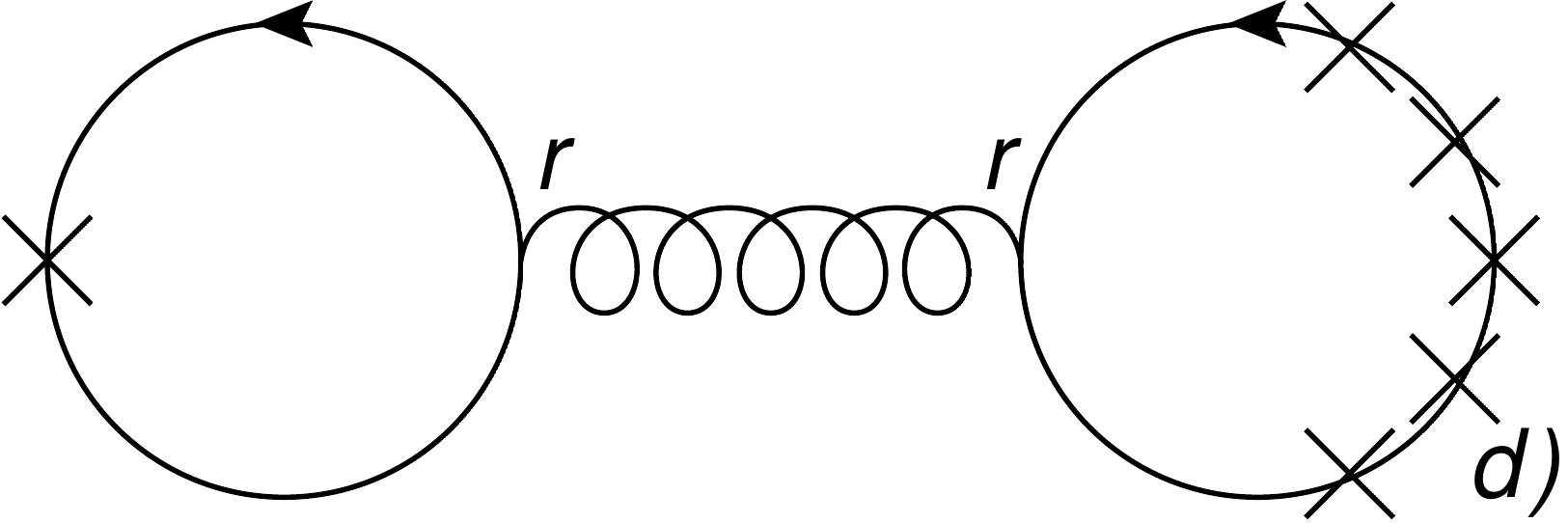}
\caption{
Contributions to the sixth order baryon number 
susceptibility  $\chi_6^B$:  (a) the one particle irreducible 
and (b-d) the one particle reducible. 
Only diagrams with two quark loops and less are shown. 
Diagrams  up to two quark loops are shown only. The dashed 
line denotes the propagator
of $\vec{\phi} = (q, \Sigma_u, \Sigma_s)$. 
The propagator has off-diagonal elements.    
}
\label{fig:c6_d}
\end{figure}

We conclude with results for the sixth order baryon susceptibility.
Some of the diagrams which contribute are illustrated in Fig.
(\ref{fig:c6_d}).  We only show the diagrams with up to two quark
loops.  We note, however, that the diagrammatic method is not
particularly useful for computing the susceptibilities.  Instead,
direct numerical evaluation was used.

\begin{figure}
\includegraphics[width=0.45\linewidth]{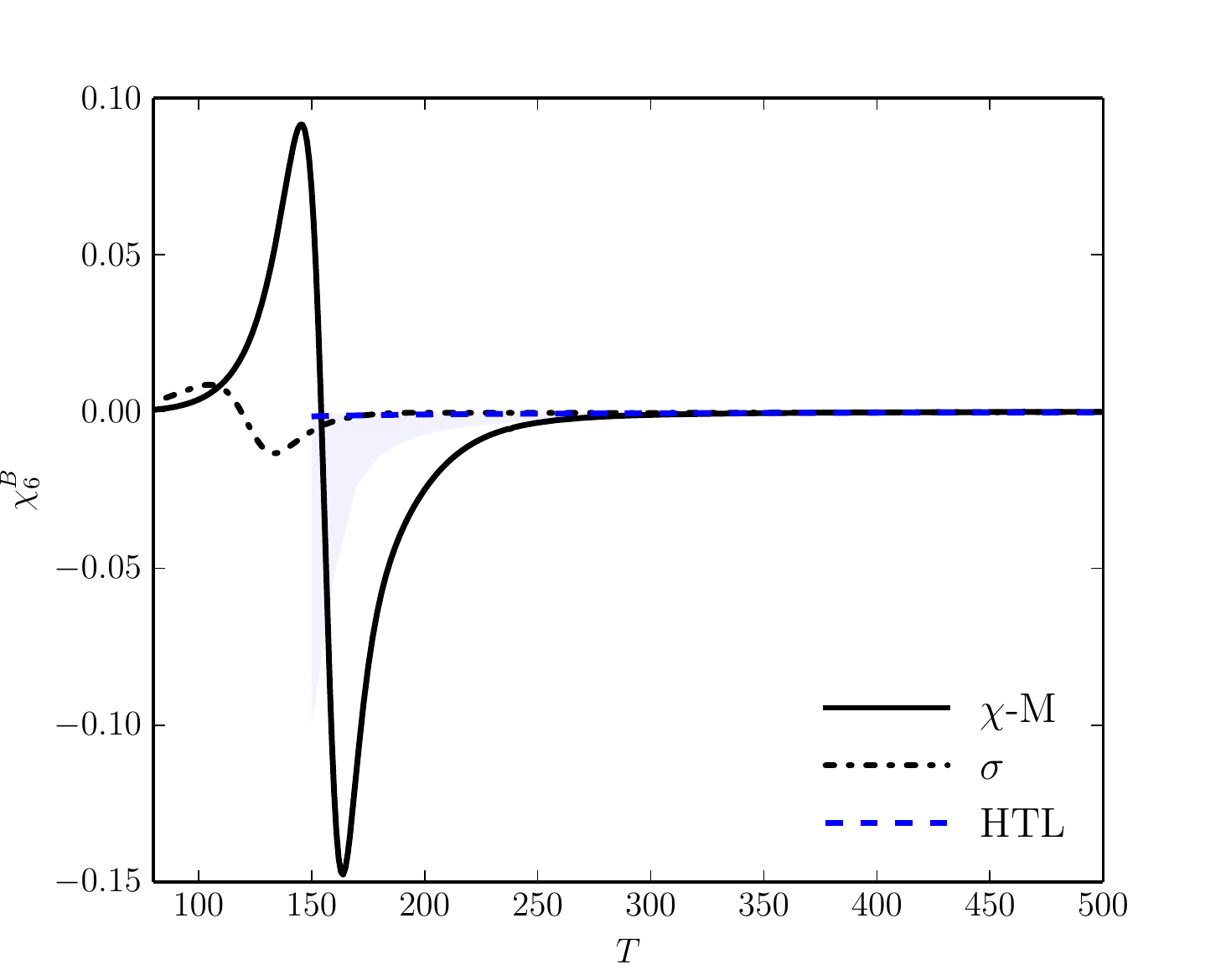}
\includegraphics[width=0.45\linewidth]{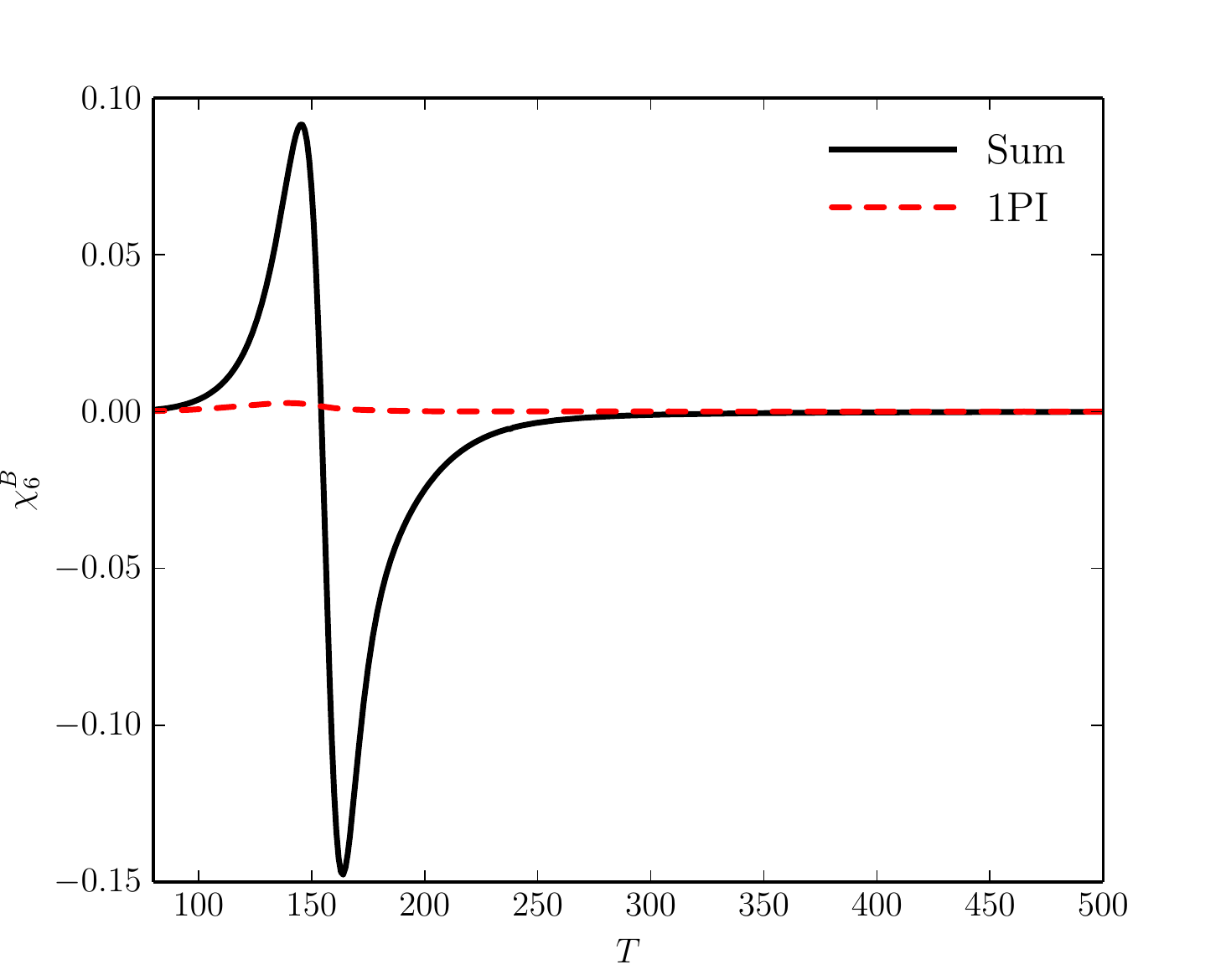}
\caption{The sixth order baryon number susceptibility 
as a function of the temperature, in the chiral matrix model
($\chi-M$), a sigma model, and HTL resummation
\cite{andersen_three-loop_2010, *andersen_gluon_2010, *andersen_nnlo_2011, *andersen_three-loop_2011,*haque_two-loop_2013, *mogliacci_equation_2013, *haque_three-loop_2014}.
}
\label{fig:c6_on_T}
\end{figure}

The results are shown in Fig. (\ref{fig:c6_on_T}). There are preliminary
results available on the lattice, but none are continuum extrapolated,
and so we do not show these.  The results of HTL resummation are
very small
\cite{andersen_three-loop_2010, *andersen_gluon_2010, *andersen_nnlo_2011, *andersen_three-loop_2011,*haque_two-loop_2013, *mogliacci_equation_2013, *haque_three-loop_2014}.
This is expected: in perturbation theory the pressure
is $\mu^4$ times a power series in the coupling constant.  Thus
contributions to $\chi_6^B$ are suppressed at least by powers of
$g^2$.  

What is not evident is not in contrast to a $\sigma$ model, the chiral
matrix model shows a {\it strong} nonmonotonic behavior, with a
large amplitude of oscillation.  The $\sigma$ model behaves similarly,
but occurs below $T_\chi$, and is almost an order of magnitude
smaller than the chiral matrix model.  

\section{Alternate models}
\label{sec:alternate}

The principle problem with the chiral matrix model is that while most
quantities agree well with lattice results,
that for the Polyakov loop, Fig. (\ref{fig:loop_compare}),
does not.  

Consequently, in this section we consider alternate models, where
we fix the value of the Polyakov loop to agree with the results from
numerical simulations.  We then compute various quantities, and 
consider if the agreement is better or worse  than with our
original model.  In all cases, we find that the agreement is worse.
We discuss this further in the Conclusions, Sec. (\ref{sec:conclusions}).

\subsection{Pure gauge theory}

We start with the theory without dynamical quarks.  In Sec. 
(\ref{sec:matrix_model_massless_quarks}) we took the nonperturbative
gluon potential from Refs.
\cite{dumitru_how_2011, dumitru_effective_2012}, where it is assumed
that the only terms are even powers of temperature,
$\sim T^4$, $T^2$, and $T^0$.  
The simplest generalization is then to assume arbitrary powers
of temperature.

In order to fit the Polyakov loop, we
use an observation of Megias, Ruiz Arriola, and Salcedo 
\cite{megias_quark-antiquark_2007}.  They showed that except close to $T_d$,
the expectation value of the Polyakov loop is close to an exponential
in $1/T^2$, $ \langle \ell \rangle \sim {\rm e}^{- \# T_d^2/T^2}$. 
Numerical simulations by Gupta, H\"ubner, and Kaczmarek
show this holds for 
both three colors \cite{gupta_renormalized_2008};
Mykkanen, Panero, and Rummukainen show it is valid for
two to six colors \cite{mykkanen_casimir_2012}.
At large $T$, then, $\langle \ell \rangle - 1 \sim 1/T^2$.
While a matrix model will not give an exponential behavior of the Polyakov
loop in any natural way, at least at large $T$
this parametrization indicates that $\langle q \rangle \sim 1/T$.

The perturbative gluon potential is
fixed by perturbation theory to be that of Eq. 
(\ref{perturbative_gluon_potential}).  For $r=0$, this potential
involves
\begin{equation}
{\cal V}_4(q,0) \equiv {\cal V}_4(q)
= \frac{2}{3} \, q^2 \, 
\left( 1 - \frac{10}{9} \, q + \frac{1}{3} \, q^2 \right) \; .
\label{v4_r0}
\end{equation}
We assume that we use the same kind of functions as before, Eq.
(\ref{nonpert_gluon_pot}).  Thus we also need
\begin{equation}
{\cal V}_2(q,0) \equiv {\cal V}_2(q)
= \frac{2}{3} \, q \, 
\left( 2 - q \right) \; .
\label{v2_r0}
\end{equation}

We then generalize the potential
of Eq. (\ref{nonpert_gluon_pot}) by assuming that the coefficients
of these functions involves not just $T^2$, 
but arbitrary powers of temperature, $T^3$, $T^2$, and $T$:
\begin{equation}
{\cal V}_{non}^{gl}(q)
= \frac{4 \pi^2}{3} \; T_d^4
\left(
\left( 
\alpha \, t^3 +  \beta \, t^2 + \gamma \, t 
\right)
{\cal V}_2(q)
+ 
\left( 
\alpha' \, t^3 +  \beta' \, t^2 + \gamma' \, t 
\right)
{\cal V}_4(q)
+ \frac{2}{15} c_3 \, t^2
\right) 
\; ,
\label{non_pert_alt}
\end{equation}
where 
\begin{equation}
t = \frac{T}{T_d} \; .
\end{equation}

Consider the behavior of this model at high temperature, where $q$ is small.
The dominant behavior is given by balancing the perturbative
potential, $\sim T^4 {\cal V}_4 \sim T^4 q^2$, against the nonperturbative
term, $\sim T^3 {\cal V}_2 \sim T^3 q$.
This gives $\langle q \rangle \sim 1/T$ at large $T$, which as we
discussed above is suggested
by measurements of the renormalized Polyakov loop.

Following Refs.
\cite{dumitru_how_2011, dumitru_effective_2012}, 
we impose two conditions. 
The first is that the pressure (approximately) vanishes at
the critical temperature.  This can be used to determine
the constant term, $\sim c_3$:
\begin{equation}
c_3 = \frac{1}{27}
\left( 
47 - 20 \alpha' - 20 \beta' - 20 \gamma'
\right) 
\; .
\label{cond_c3}
\end{equation}
The second condition is given by requiring that the transition
occurs at $T_d$.  

The previous potential, Eq. (\ref{nonpert_gluon_pot}), 
starts with three coefficients, which then reduce
to one free parameter.  The new model begins
with seven parameters, which reduce to five free parameters.
By some trial and error, we are led to the values
$$
\alpha = -0.403376 \; ; \;
\alpha' = -1.00819 \; ; \;
\beta  = -2.58495  \; ; \;
$$
\begin{equation}
\beta' = -8.6023 \; ; \;
\gamma = 1.12179\; ; \;
\gamma' = 5.57084 \; .
\label{values_glue_alt}
\end{equation}

\begin{figure}
\includegraphics[width=0.45\linewidth]{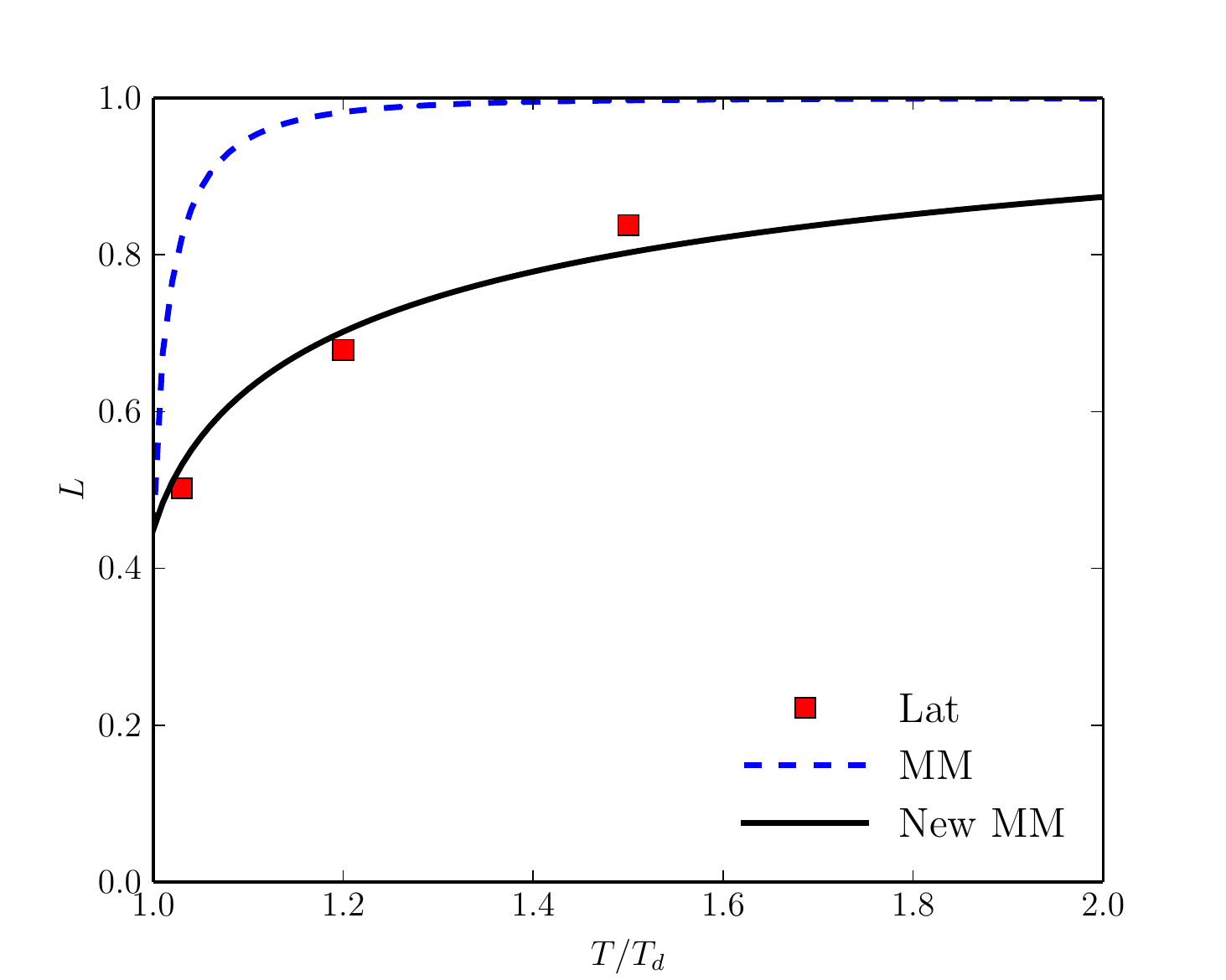}
\includegraphics[width=0.45\linewidth]{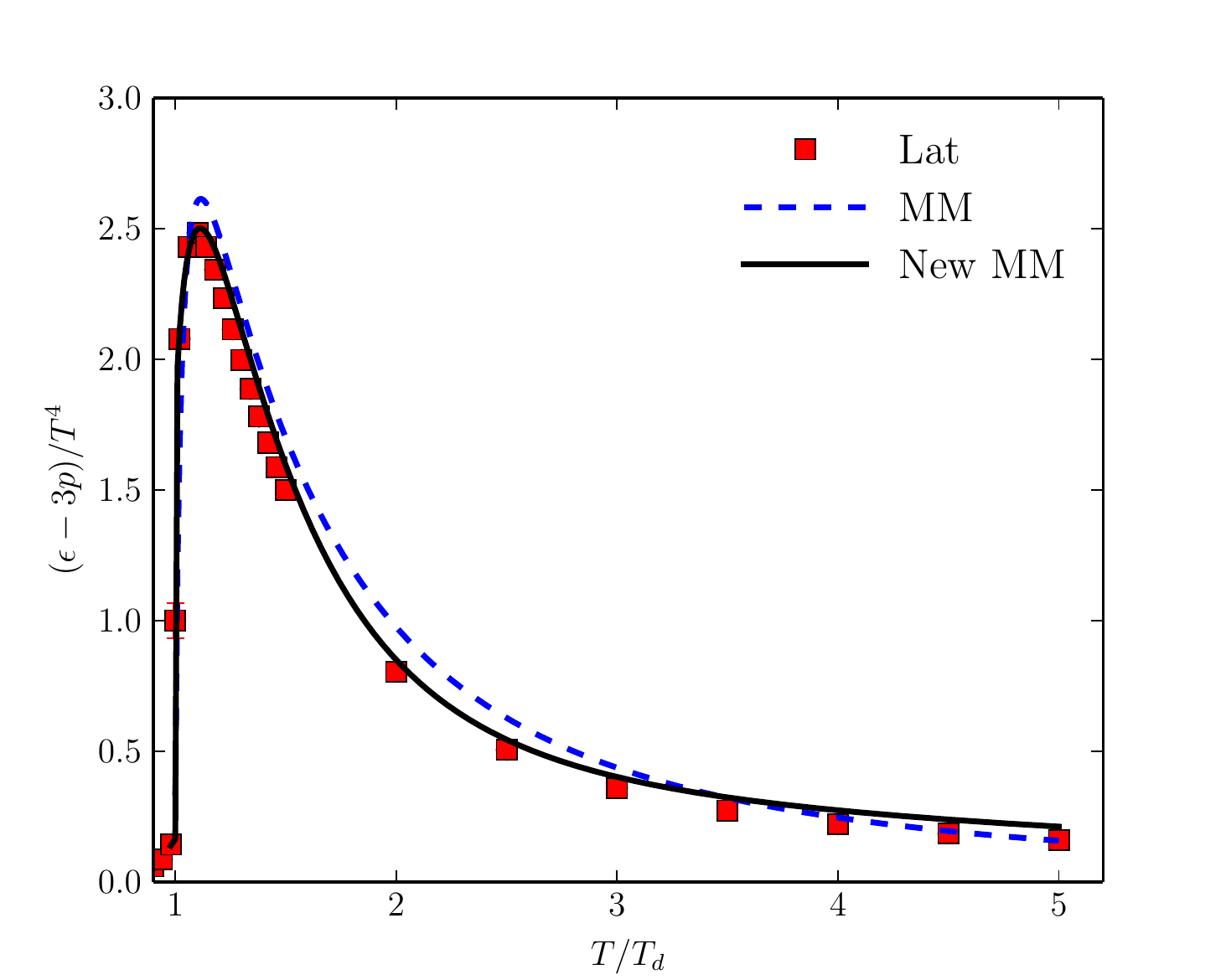}
\caption{
Results from the previous matrix model (MM), Eq. (\ref{nonpert_gluon_pot}), 
and in the new matrix model (New MM), Eqs. 
(\ref{non_pert_alt}) and (\ref{values_glue_alt}).  With five free parameters
in the New MM, as opposed to one in the MM, good fits for both the
Polyakov loop, in the left panel, and for the interaction measure, 
$(e-3p)/T^4$, in the right panel, can be obtained.
}
\label{fig:new_MM}
\end{figure}

\begin{figure}
\includegraphics[width=0.6\linewidth]{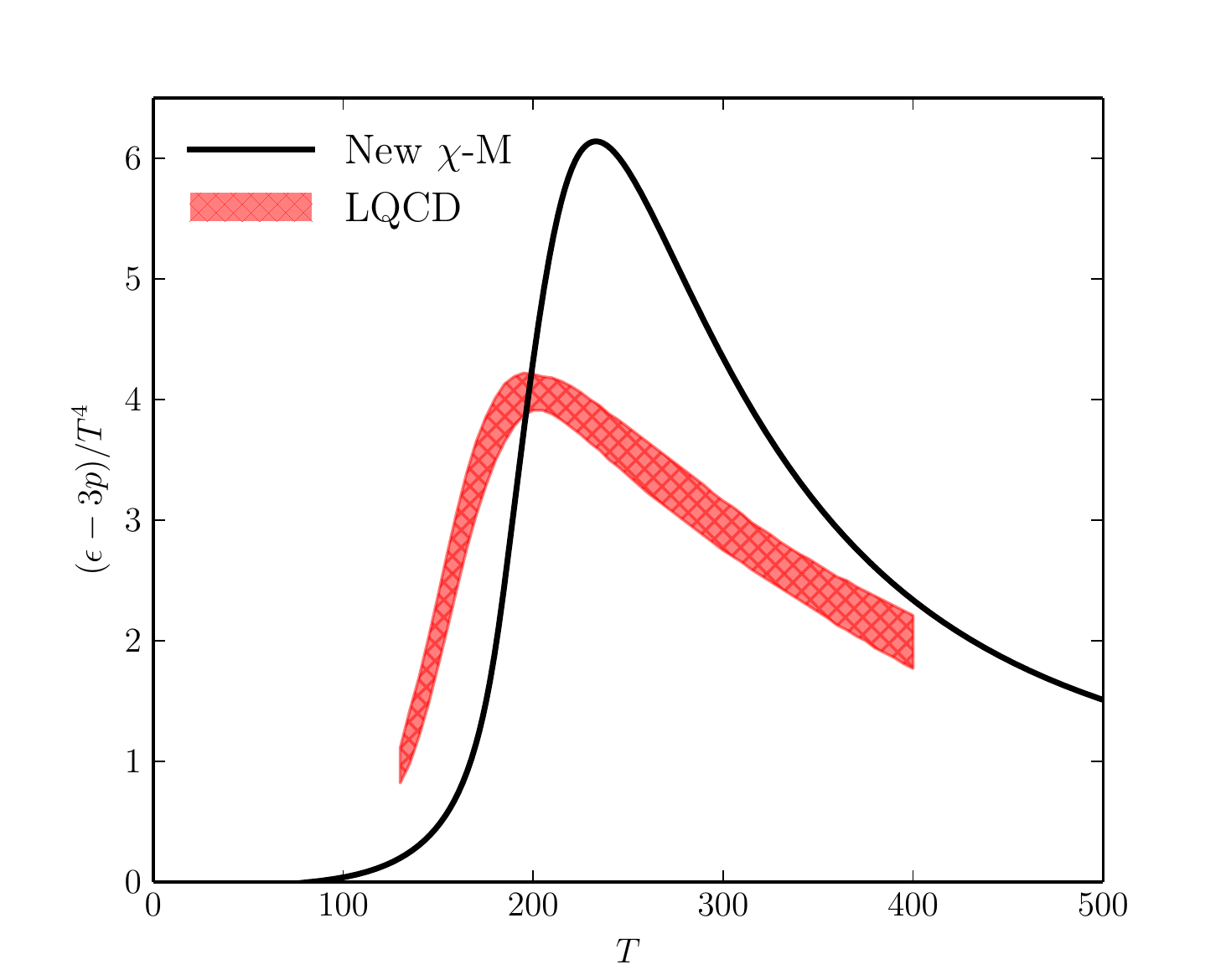}
\caption{
Results for $(e-3p)/T^4$ for the new matrix model, Eqs.
(\ref{non_pert_alt}) and (\ref{values_glue_alt}), 
with dynamical quarks.  The result is farther from the 
lattice values than our original model, see 
Fig. (\ref{fig:int_on_T}).  
}
\label{fig:new_MM_qks}
\end{figure}

The results for the Polyakov loop and the interaction measure,
$(e-3p)/T^4$, are shown in Fig. (\ref{fig:new_MM_qks}).  
Given the plethora of parameters, 
it is hardly surprising 
that we can fit both the Polyakov loop and the pressure at all
temperatures above $T_d$.

We then adopt the same approach as before to include dynamical
quarks.  The results are shown in Fig. (\ref{fig:new_MM}).
The results are not close to those of the lattice, with the
peak in the interaction measure in the new matrix
model at a much higher value, $\sim 6$ instead of $\sim 4$,
and at a significantly larger temperature, $\sim 250$~MeV instead
of $\sim 200$~MeV. 
This should be compared to the 
results in our original matrix model, Fig. (\ref{fig:int_on_T});
while these are not perfect, they are far closer than those
in the new matrix model of Fig. (\ref{fig:new_MM_qks}).

As before, the value of the Yukawa coupling is $y=5$, with
little sensitivity to varying the Yukawa coupling by $\sim 10\%$.
We have also computed the chiral order parameter and the susceptibility
for light quarks.  This shows that the temperature for the chiral
crossover in the new matrix model is 
$T_\chi \sim 186$ MeV, which is significantly higher than the lattice value of
$T^{lattice}_\chi \sim 155$~MeV.  

This shows that for the pressure and the transition temperature,
that assuming a model which fits
the Polyakov loop in the pure gauge theory gives a worst
fit to these quantities in QCD.  Needless to say, this is under
the assumption that there are no new nonperturbative
terms in the gluon potential.  We could certainly fit both
the pressure and loop 
in QCD by allowing new nonperturbative terms in the gluon potential
which are dependent upon presence of quarks.  Since we already have a model
with five parameters, fitting the pressure and loop in QCD with
even more parameters does not seem particularly noteworthy.

\subsection{The Polyakov loop and baryon susceptibilities}

To emphasize the physics, then, in this section 
we assume that the value of the Polyakov loop
is given by the value from the lattice.  We show in this section that doing
so, there is a large and persistent disagreement with the baryon
susceptibilities.

Taking the Polyakov loop from the lattice and computing with
our chiral model, we find that the chiral crossover temperature is like
that in the previous section, and is too large, $T_\chi \sim 191$~MeV.  
For the time being we ignore this to compute
the second order baryon susceptibility, $\chi_2^B$.  Our computation
is not complete, because we cannot compute the diagram including
``r'' exchange, which is the diagram on the right-hand side of 
Fig. (\ref{fig:c2_d}).  Nevertheless, as seen from the diagram 
on the right-hand side of Fig. (\ref{fig:c2_on_T}), this contribution
is generally small, and so we assume it can be neglected.

\begin{figure}
\includegraphics[width=0.6\linewidth]{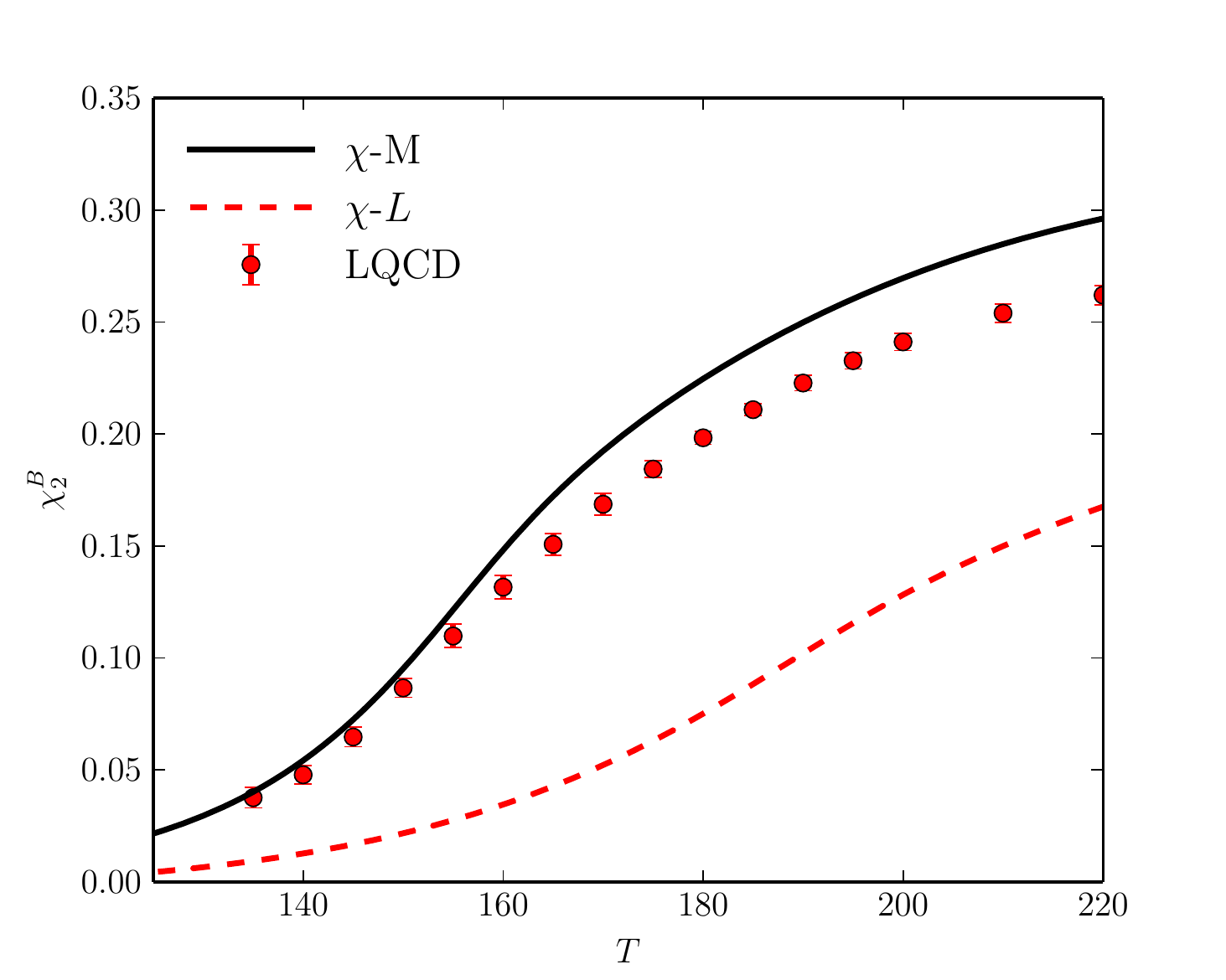}
\caption{
The second order baryon number susceptibility, $\chi_2^B$,
in a model where the Polyakov loop is fitted directly from the lattice,
$\chi-L$.  This model gives $T_\chi \sim 191$~MeV, instead of
$T_\chi^{lattice} = 155$~MeV.
Results in our chiral matrix model are shown in $\chi-M$;
these are much closer to the lattice data, LQCD.
}
\label{fig:c2_fp}
\end{figure}

We present the results in Fig. (\ref{fig:c2_fp}).  The overall
trend of the results is easy to understand.  
Because of confinement, $\chi_2^B$ vanishes in the confined phase,
and equals $1/3$ for ideal quarks.  Thus being deeper in the confined
phase decreases $\chi_2^B$.  
This is exactly what is shown in Fig. (\ref{fig:c2_fp}):
while the results in our chiral matrix model are
slightly higher than those from the lattice, the results with
a chiral model which fits the Polyakov loop from the lattice are
much lower than the results from $\chi_2^B$ on the lattice.
For example, at $T = 200$~MeV, our chiral matrix model is too high
by about $\sim 10\%$; in contrast, the value computed from the lattice
Polyakov loop is smaller than the lattice $\chi_2^B$ by about
half.

Thus perhaps the problem is that $T_\chi$ is too high.  
Motivated by including the pion degrees of freedom, 
by hand we adjust the mass squared in the linear sigma model
to {\it fit} $T_\chi$ to be $155$~MeV, as on the lattice.
We find that a fit
\begin{equation}
m^2 \rightarrow m^2 \left( 1 + 0.1 \left( 
\frac{T}{f_\pi} \right)^2 \right) \; ,
\label{eq:change_m2}
\end{equation}
suffices: the coefficient of $0.1$ is chosen to obtain
$T_\chi = 155$~MeV.

\begin{figure}
\includegraphics[width=0.6\linewidth]{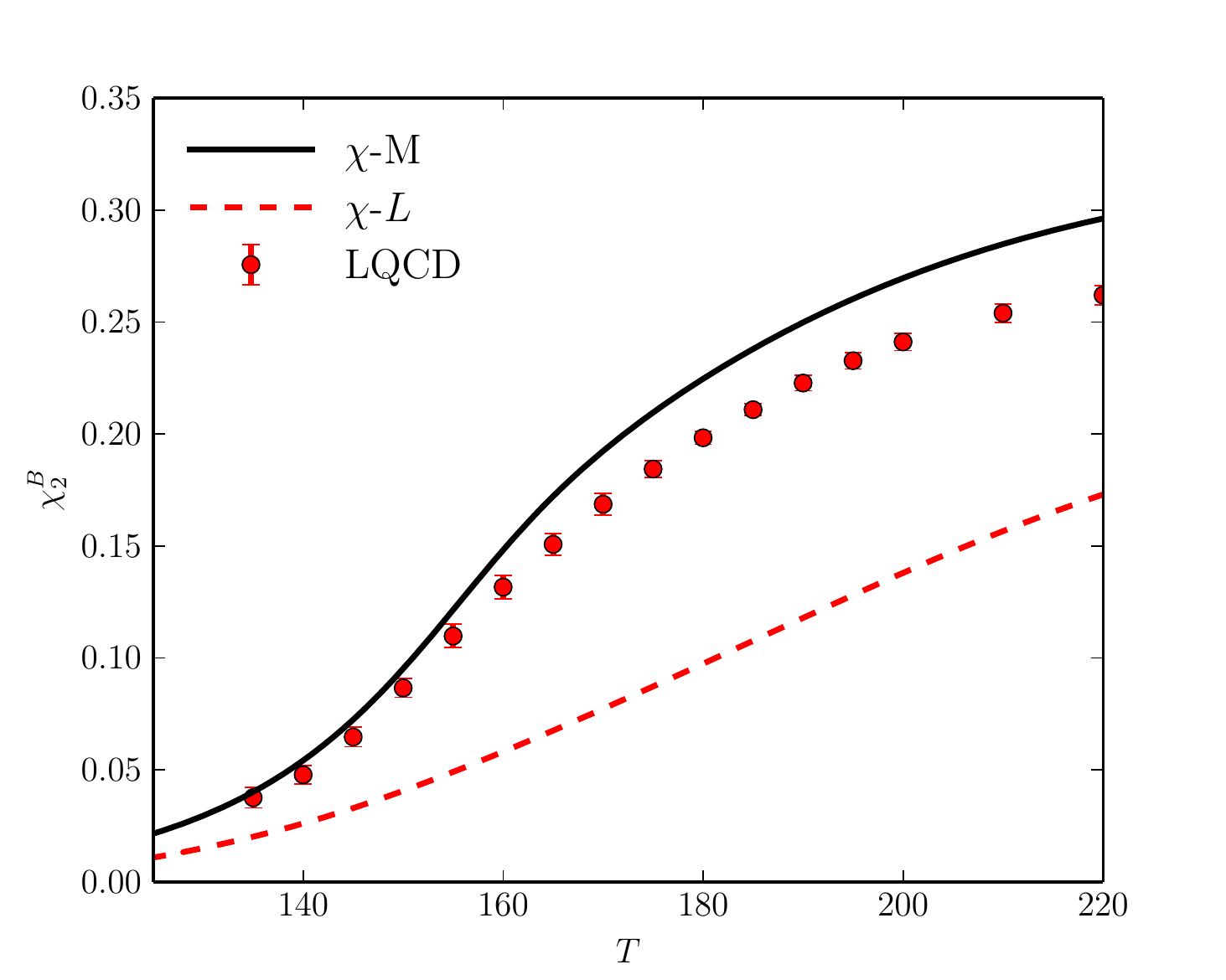}
\caption{
The second order baryon number susceptibility, $\chi_2^B$,
in a model where the Polyakov loop is fitted directly from the lattice,
$\chi-L$, and where
the mass parameter in the sigma model is tuned by
hand, Eq. (\ref{eq:change_m2}), so that $T_\chi = 155$~MeV.  
Results in our chiral matrix model are shown in $\chi-M$;
these are much closer to the lattice data, LQCD.
}
\label{fig:c2_fp_mod_m}
\end{figure}

The results in Fig. (\ref{fig:c2_fp_mod_m})
show that while this approach moves $\chi_2^B$ upward,
closer to the lattice results, it is not by
much.  As in Fig. (\ref{fig:c2_fp}), fitting to the lattice Polyakov 
loop gives a result in which $\chi_2^B$ is rather far from the
lattice results.

We have also computed higher order baryon susceptibilities,
and find similar results.  We also computed using the model
of the previous section, and find that the baryon susceptibilities
are uniformly farther from the lattice results than in our original
chiral matrix model.

We compare our analysis with those of the
Functional Renormalization Group (FRG)
\cite{braun_nature_2010, herbst_phase_2011, *fister_confinement_2013, *herbst_phase_2013, *haas_gluon_2014, *herbst_thermodynamics_2014, *mitter_chiral_2015, herbst_confinement_2015, fu_relevance_2015, fu_correlating_2015}.
In the FRG, the loop approaches unity quickly, as in the chiral
matrix model; for the pure gauge theory, see Fig. (1) of
Ref. \cite{braun_nature_2010}.  Herbst, Luecker, and Pawlowski 
argued that corrections to the FRG modify this so that
the loop is much closer to the lattice, 
Fig. (6) of Ref. \cite{herbst_confinement_2015}.
In QCD, though, 
Fu and Pawloski computed the baryon susceptibilities
\cite{fu_relevance_2015, fu_correlating_2015}, and find good agreement
with the lattice.  This requires, however, that the loop is relatively
large at $T_\chi$: from Fig. (7) of Ref. \cite{fu_correlating_2015},
$\langle \ell \rangle \sim 0.4$ at $T_\chi$.  This agrees with
our conclusions in this section.

\section{Conclusions}
\label{sec:conclusions}

The analysis in the previous section shows that the baryon 
susceptibilities can be a sensitive test of how quickly QCD deconfines.

\begin{figure}
\includegraphics[width=0.45\linewidth]{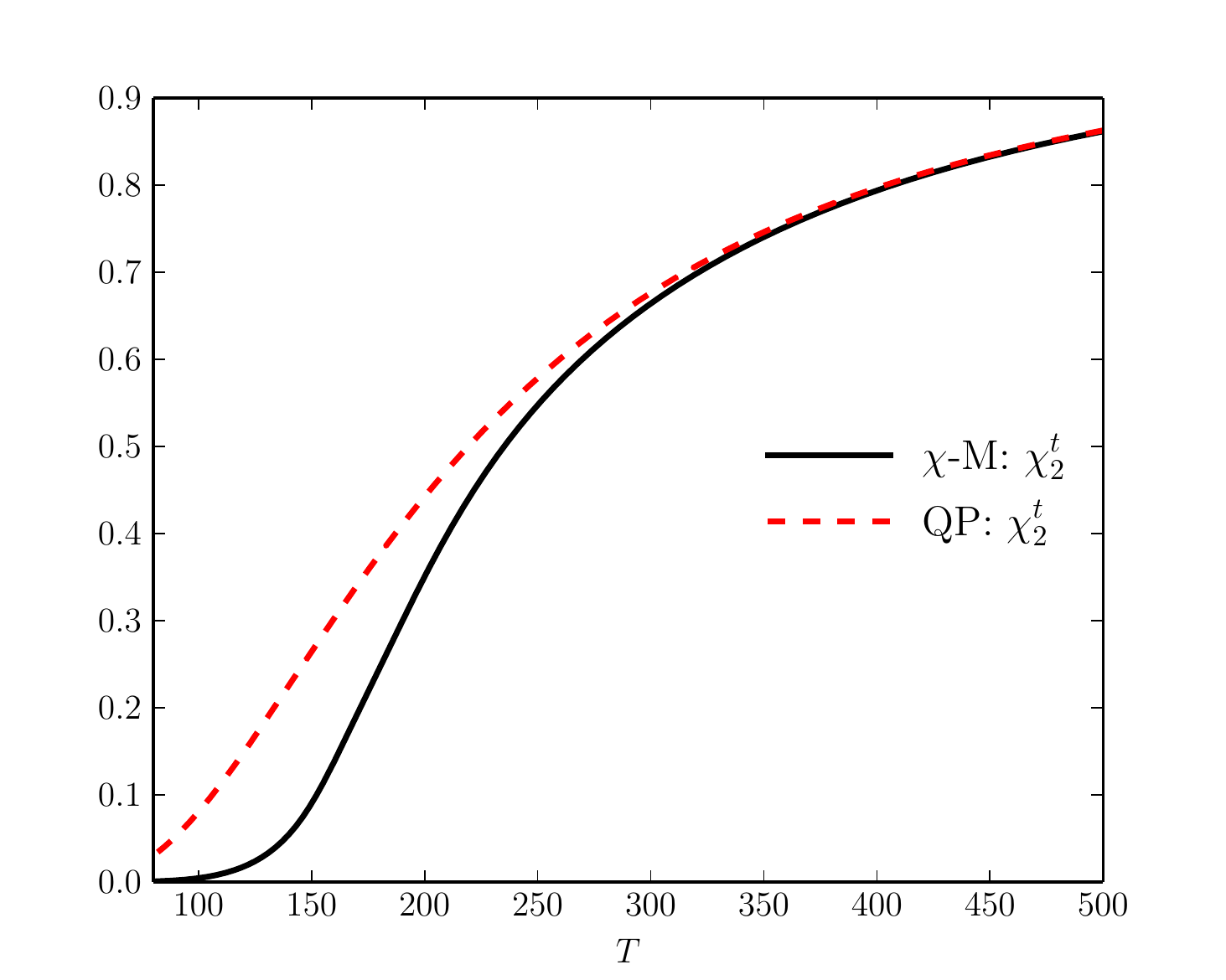}
\caption{
The baryon susceptibility to second order for a heavy test quark,
$m = 500$~MeV.  The solid black line is in our chiral matrix model,
the dotted line, for free quarks.
}
\label{fig:c2_test}
\end{figure}

For light quarks, the baryon susceptibilities are clearly tied to
the restoration of chiral symmetry.  This suggests that a sensitive
test of deconfinement would be to measure the second order
baryon susceptibility for a relatively heavy quark.  The quark
cannot be too heavy, or the entire signal is Boltzmann suppressed.
To illustrate this, we show in Fig. (\ref{fig:c2_test}) 
$\chi_2^B$ in our chiral matrix model, versus the results for
free, deconfined quarks, with $q=0$.  For such a heavy quark,
the approximate restoration of chiral symmetry at $T_\chi = 155$~MeV
should not be of relevance.  Nevertheless, at this temperature
there is a large difference between the two curves, by more than 
a factor of two.

Thus we suggest that it may be useful to measure $\chi_2^B$ for
a heavy quark in QCD.  In the lattice, this heavy quark can be
treated in the valence approximation, which should simplify the
analysis.  

In this vein, we comment on the difference between the second
order chiral susceptibilities for light and strange quarks.
Bellwied, Szabolcs, Fodor, Katz, and Ratti 
\cite{bellwied_is_2013} have described this difference as
due to a ``flavor hierarchy'' between light and strange quarks.
In our chiral matrix model, the difference between the two is
simply a consequence that because the strange quark is heavier,
$\chi_2$ tends to lag behind that for a heavier quark.  In any
case, measuring the susceptibility for a test quark should
enable one to disentangle the effects of chiral symmetry restoration
versus deconfinement.

Indeed, one can view the discrepancy between the chiral matrix model
and the Polyakov loop more generally.  In our model, the approximate
restoration of chiral symmetry for light quarks is closely
coincident with deconfinement, Fig. (\ref{fig:op_on_T}).
Notably, at the chiral crossover temperature, the Polyakov loop is large,
$\sim 0.5$.  Similarly, the peaks in the susceptibilities for
the chiral order parameter for light quarks coincide with the peak
for the loop susceptibilities, Fig. (\ref{fig:op_on_T}).  Moreover,
as demonstrated in Secs. (\ref{sec:flavor_susp}) and
(\ref{sec:alternate}), this is also consistent with the
baryon susceptibilities, which approach their ideal
values rather quickly, certainly by temperatures of $\sim 300$~MeV.

In contrast, the value of the renormalized Polyakov loop 
\cite{bazavov_polyakov_2013, bazavov_polyakov_2016} from the lattice
is {\it extremely} small at $T_\chi$, $\langle \ell \rangle \sim 0.1$.
That is, chiral symmetry is restored in a phase which is nearly
confined, not deconfined.  If taken at face value, this indicates
that even for $\mu = 0$, chiral symmetry is restored in a quarkyonic
phase
\cite{mclerran_phases_2007, *andronic_hadron_2010, *kojo_quarkyonic_2010, *kojo_interweaving_2012}.  
Even by temperatures of $200$~MeV, the Polyakov
loop is still very small, $\langle \ell \rangle \sim 0.3$.  
It is hard to understand why the quark susceptibilities are close
to their ideal values at relatively low temperatures, $\sim 300$~MeV,
if the renormalized Polyakov loop indicates the  theory is still
close to confining.

We have not settled this question here, but it demonstrates that the
thermodynamic behavior of QCD is more involved than naive prejudice 
might suggest.

\begin{acknowledgements}
We thank F. Karsch, S. Mukherjee, P. Petreczky, 
S. Rechenberger, 
D. Rischke, J. Schaffner-Bielich, and S. Sharma for
discussions, and S. Borsanyi and C. Ratti for sharing their
data. 
R.D.P. would like to thank P. Kovacs and Gy. Wolf for discussions about their model, 
and P. Levai for his hospitality at the Wigner Research Center for Physics in Budapest in July, 2016.
R.D.P. thanks the U.S. Department of Energy for support
under contract DE-SC0012704.
\end{acknowledgements}

{\em Note added.} --  There is significant overlap between our analysis and that
of Kovacs et al.~\cite{kovacs_existence_2016},
who use a Polyakov loop model coupled to both scalars and vector mesons. While
the details of our analyses differ, they also find that the value of the
Polyakov loop in their model is much larger than that measured by lattice
simulations.

\newpage

\appendix

\section{Integrals in the semi-QGP}

In this appendix we collect some useful integrals.

The trace at zero temperature is defined in Eq. (\ref{eq:trace_zeroT}).
The basic integral for a single massive field is
given in Eq. (\ref{zeroTselfsame}).  For two fields whose masses differ,
the corresponding integral is
$$
\left. {\rm tr} \; \frac{1}{(K^2 + m_1^2)(K^2 + m_2^2)} \right|_{T=0}
$$
\begin{equation}
= \; + \; \frac{1}{16 \pi^2}
\left( \frac{1}{\epsilon} 
+\frac{1}{m_1^2-m_2^2}
\left( m_1^2 \log\left(\frac{\mu^2}{m_1^2}\right)
- m_2^2 \log\left(\frac{\mu^2}{m_2^2}\right)\right)
- 1
+ \log(4 \pi) - \gamma \right) \; .
\label{zero_T_self_diff}
\end{equation}
Taking $m_1 \rightarrow m_2$, this reduces to Eq. (\ref{zeroTselfsame}).

At nonzero temperature the trace is defined in Eq. (\ref{eq:trace_T}).
In this case we need to compute for $Q \neq 0$ as well.
To compute the integrals, it is useful to Fourier transform the propagator
in $k_0$ space to that in imaginary time, $\tau$:
\begin{equation}
\frac{1}{K_c^2 + m^2}
= \int^{1/T}_0 d\tau \; \frac{{\rm e}^{i k_0^c \tau} }{2 E}
\left( (1 - \widetilde{n}_q(E)) {\rm e}^{- E \tau}
- \widetilde{n}_{-q}(E) {\rm e}^{+ E \tau} \right) \; ,
\end{equation}
where $E$ is the energy,
\begin{equation}
E = \sqrt{k^2 + m^2} \; ,
\end{equation}
and $\widetilde{n}_q(E)$ is the Fermi-Dirac statistical distribution function
with an (imaginary) chemical potential $2 \pi i q$,
\begin{equation}
\widetilde{n}_q(E)
= \frac{1}{ {\rm e}^{E/T - 2 \pi i q} + 1 } \; ,
\label{saclay_ferm}
\end{equation}
The term at zero temperature is obviously 
due to the piece independent of the $\widetilde{n}$'s, the $1$ in
$1 - \widetilde{n}_q(E)$.  For future reference, 
$\widetilde{n}(E)$ is just the usual Fermi-Dirac function.

The advantage of this method is that the sum over $k_0$ is trivial:
it gives
a delta function in $\tau$, leaving an integral over the spatial
momentum.  For example, the equation of motion for $\sigma$, and
the pion self-energy, involves
\begin{equation}
{\rm tr} \; \frac{1}{K_c^2 + m^2} = 
\left. {\rm tr} \; \frac{1}{K^2 + m^2}\right|_{T=0}
- \frac{1}{4 \pi^2} \int_0^\infty dk \; \frac{k^2}{E}\;
\left( \widetilde{n}_q(E) + \widetilde{n}_{-q}(E) \right) \; .
\label{quad_therm}
\end{equation}
We note that by using Eq. (\ref{saclay_ferm}), to regularize
the integral we need to
continue the spatial integral to $3 - 2 \epsilon$ dimensions,
\begin{equation}
\left. {\rm tr} \; \frac{1}{K^2 + m^2}\right|_{T=0}
= \mu^{2 \epsilon} 
\int \frac{d^{3 - 2 \epsilon}k}{(2 \pi)^{3 - 2 \epsilon}} 
\; \frac{1}{2 E}\; .
\end{equation}
The result is identical to that in $4-2 \epsilon$ dimensions.

The sum of Fermi-Dirac statistical distribution functions for $q$ and $-q$ is
\begin{equation}
\frac{1}{2} 
\left( \widetilde{n}_q(E) + \widetilde{n}_{-q}(E) \right)
= \frac{\cos(2 \pi q) \, {\rm e}^{E/T} \; 
+ 1}{{\rm e}^{2 E/T} + 2 \cos( 2 \pi q) \, {\rm e}^{E/T} + 1 } \; .
\label{nq_plus_nminusq}
\end{equation}

For the equation of motion of the $q$ field, the integral which enters is
\begin{equation}
{\rm tr} \; \frac{k_c^0}{K_c^2 + m^2} \; .
\end{equation}
To evaluate this, it is easiest to write 
\begin{equation}
k_c^0 = - \, i \; 
\frac{\partial}{\partial \tau} \; {\rm e}^{i k_c^0 \tau}  \; ,
\end{equation}
and then to integrate by parts in the $\tau$ integral.  In this way, we find
\begin{equation}
{\rm tr} \; \frac{k_c^0}{K_c^2 + m^2} \; = \;
\frac{1}{4 \pi^2} \int_0^\infty dk \; k^2 \;
(i) \left(\widetilde{n}_q(E) - \widetilde{n}_{-q}(E) \right) \; .
\label{int_q_tadpole}
\end{equation}
The term at zero temperature vanishes, because the integral is then odd
in $k^0$.  The difference of the Fermi-Dirac 
statistical distribution functions is
\begin{equation}
(-i) \; \left(\widetilde{n}_q(E) - \widetilde{n}_{-q}(E) \right)
= \;  \frac{2 \, \sin(2 \pi q) \, {\rm e}^{E/T} }{{\rm e}^{2 E/T} + 2 \cos( 2 \pi q) \, {\rm e}^{E/T} + 1 } \; .
\label{int_q_tadpole_ns}
\end{equation}

For a given $q$, each Fermi-Dirac statistical distribution function 
$\widetilde{n}_{q}(E)$ is complex.  However, 
we shall show that uniformly what enters is either a sum of
distribution functions, 
as $\widetilde{n}_{q} + \widetilde{n}_{-q}$
in Eq. (\ref{nq_plus_nminusq}), or $i$ times the difference
of distribution functions, as
$i(\widetilde{n}_{q} - \widetilde{n}_{-q})$ in Eq. (\ref{int_q_tadpole_ns}).
In all cases, in the end what enters is manifestly real, and so the
complexity of $\widetilde{n}_{q}$ does not cause any problems, at least
for the quantities which we compute herein.

For the self energies, there are several integrals which enter.  We start
with the simplest,
$$
{\rm tr} \; \frac{1}{(K_c^2 + m^2)^2}
= - \; \frac{\partial}{\partial m^2} \; {\rm tr} \; \frac{1}{K_c^2 + m^2}
$$
\begin{equation}
= \left. {\rm tr} \; \frac{1}{(K^2 + m^2)^2} \right|_{T=0}
- \; \frac{1}{8 \pi^2}\; \int^{\infty}_0 dk \; \frac{k^2}{E^3}
\left( \widetilde{n}_q(E) \left( 1 + 
\frac{E}{T} \left( 1 - \widetilde{n}_q(E) \right) \right)
+ (q \rightarrow -q) \right) \; .
\label{int_four_over_over}
\end{equation}
For this integral we also need the 
sum of the Fermi-Dirac statistical distribution functions
\begin{equation}
\frac{1}{2} \left( \widetilde{n}_q(E) \left( 1 - \widetilde{n}_q(E) \right) 
+ (q \rightarrow -q)  \right)
= \frac{ {\rm e}^{E/T} (\cos(2 \pi q) \, ( {\rm e}^{2E/T} +1 ) 
\; + 2 \, {\rm e}^{E/T})}{({\rm e}^{2 E/T} 
+ 2 \cos( 2 \pi q) \, {\rm e}^{E/T} + 1 )^2} \; .
\end{equation}

It is useful to make a comment about infrared divergences.  At zero
temperature, the integral ${\rm tr}1/(K^2+m^2)^2$ has a logarithm in
mass, $\sim m^4 \log(\mu/m)$, Eq. (\ref{zeroTselfsame}).  This is
evident, as $\sim \int d^4K/(K^2+m^2)^2$ has both ultraviolet and
infrared divergences.

The ultraviolet divergence is unchanged at nonzero temperature, but
the nature of the infrared divergence changes.  
To isolate the infrared divergence, for a Fermi-Dirac statistical
distribution function we can take the energy $E$ to vanish.  At
$q=0$, $\widetilde{n}(0) = 1/2$.  At $q \neq 0$, the sum of
$\widetilde{n}$'s satisfies the same identity,
\begin{equation}
\widetilde{n}_q(0) + \widetilde{n}_{-q}(0) = 1 \;\;\; ; \;\;\;
q \neq \frac{1}{2} \; .
\end{equation}
(The restriction that $q \neq 1/2$ is necessary because then the Fermi-Dirac
statistical distribution function becomes Bose-Einstein, with $n(E)
\sim T/E$ at small $E$. In practice, for three colors $q \leq 1/3$, so
this never presents a problem.)
From Eq. (\ref{int_four_over_over}) there is then an infrared divergence
from 
\begin{equation}
- \; \frac{1}{8 \pi^2}\; \int^{\infty}_0 dk \; \frac{k^2}{E^3} 
(\widetilde{n}_q(E) + \widetilde{n}_{-q}(E))
\sim - \; \frac{1}{8 \pi^2}\; \int^T_m \frac{dk}{k} 
\sim - \frac{1}{16 \pi^2} \log\left( \frac{T^2}{m^2} \right) \; .
\end{equation}
Comparing with Eq. (\ref{zeroTselfsame}), we see that the logarithm
in mass, $\sim m^4 \log(m)$, cancels identically, and is replaced by
a logarithm in temperature, $\sim m^4 \log(T)$
\cite{skokov_vacuum_2010}.  In all, when $m \ll T$,
\begin{equation}
\left. {\rm tr} \; \frac{1}{(K^2 + m^2)^2}\right|_{m \ll T}
= \; + \; \frac{1}{16 \pi^2}
\left( 
\frac{1}{\epsilon} 
+ \log \left( \frac{\mu^2}{T^2} \right)
+ \log \left( \frac{4}{\pi}     \right) 
+ \gamma 
\right) \; .
\label{nonzeroTselfsame}
\end{equation}
This expression is only valid for masses much less than the temperature.
It is easy to understand why a logarithm in mass 
is replaced by one in temperature.
At zero temperature the only infrared cutoff is the mass.
At nonzero temperature, for fermions with $q \neq 1/2$ the temperature
acts as an infrared cutoff, so that one can smoothly take the limit
of $m \rightarrow 0$ without effect.  

We shall not need Eq. (\ref{nonzeroTselfsame}), as in general the masses we
consider are on the order of the temperature.
In this case, it is more useful to compute
the part at zero temperature analytically, and the part at nonzero
temperature numerically.  However, this expression 
illustrates the necessity of including
terms at zero temperature. Otherwise we would include terms with a
a logarithm of the mass which properly are not there.
 
The susceptibility with respect to a real quark chemical potential,
and the $q$ self-energy involves the integrals
\begin{equation}
- \; {\rm tr} \; \frac{1}{K_c^2 + m^2}
+ \; {\rm tr} \; \frac{2 \; E^2}{(K_c^2 + m^2)^2} \; .
\end{equation}
At zero temperature this term vanishes,
\begin{equation}
\int \frac{d^{3-2\epsilon} k}{(2 \pi)^{3-2 \epsilon}}
\left( - \; \frac{1}{2E} 
+ 2 \; E^2 \; \left(- \; \frac{\partial}{\partial m^2} \right)
\; \frac{1}{2 E}
\right) = 0 \; .
\label{int_q_self_zeroT}
\end{equation}
This is most reasonable, since we do not expect any
ultraviolet divergence for the quark susceptibility, or
from fluctuations in $A_0 \sim q$.
Thus the only contribution is at nonzero temperature, 
\begin{equation}
- \; {\rm tr} \; \frac{1}{K_c^2 + m^2}
+ \; {\rm tr} \; \frac{2 \; E^2}{(K_c^2 + m^2)^2}
= - \frac{1}{4 \pi^2}  \int^\infty_0 dk \; k^2
\left( \widetilde{n}_q(E) \left( 1 - \widetilde{n}_q(E) \right) 
+ (q \rightarrow -q)  \right) \; .
\label{int_q_self}
\end{equation}

There is also a mixing between the $\sigma$ channel and $q$,
\begin{equation}
{\rm tr} \; \frac{k_c^0}{(K_c^2 + m^2)^2} \; = \;
\frac{1}{4 \pi^2 T} \int_0^\infty dk \; \frac{k^2}{E} \;
(i) \left(\widetilde{n}_q(E) \left( 1 - \widetilde{n}_q(E) \right)
 - (q \rightarrow  -q) \right) \; ,
\label{int_mix_sigma_q}
\end{equation}
where 
\begin{equation}
(i) \left(\widetilde{n}_q(E) \left( 1 - \widetilde{n}_q(E) \right)
 - (q \rightarrow  -q) \right) = 
- \; \frac{ \sin(2 \pi q)\; {\rm e}^{E/T}\; ({\rm e}^{2E/T} - 1 )}{({\rm e}^{2 E/T} + 2 \cos( 2 \pi q) \, {\rm e}^{E/T} + 1 )^2} \; .
\end{equation}

For mesons such as kaons, with one strange and one light quark, we require
integrals such as
$$
{\rm tr} \; \frac{1}{(K_c^2 + m_1^2)(K_c^2 + m_2^2)} \; = \;
\left. {\rm tr} \; \frac{1}{(K^2 + m_1^2)(K^2 + m_2^2)}\right|_{T=0}
$$
\begin{equation}
- \; \frac{1}{8 \pi^2}
\int^\infty_0 dk \; \frac{k^2}{E_1 \, E_2}
\left( \frac{1}{E_1 + E_2}
\left(\widetilde{n}_q(E_1) + \widetilde{n}_q(E_2) \right)
- \;  \frac{1}{E_1 - E_2} 
\left(\widetilde{n}_q(E_1) - \widetilde{n}_q(E_2) \right)
+ ( q \rightarrow - q) \right) \; ,
\end{equation}
where
\begin{equation}
E_1 = \sqrt{k^2 + m_1^2} \;\;\; , \;\;\; 
E_2 = \sqrt{k^2 + m_2^2} \; .
\end{equation}
Naturally one can check that this reduces to
Eq. (\ref{int_four_over_over}) as $m_1 \rightarrow m_2$.

\section{Meson masses at finite temperature}
In this appendix,  we list the results for the thermal meson masses 
used in computing Fig.~(\ref{fig:masses}).
The pion mass is given by 
\begin{equation}
m_{\pi}^2 =  \frac{\partial  {\cal V}_u^{qk}} {\partial \Sigma_u^2} 
- \, c_A \, \Sigma_s+2 \, \lambda  \, \Sigma_u^2+m^2 = 
\frac{\hat{h}_u    }{\Sigma_u} \; .
\label{Eq:mpi_fT}
\end{equation}
As at zero temperature, the equations of motion were used to obtain
the final expression, $m_\pi^2 = \hat{h}_u/\Sigma_u$, and do this
consistently in what follows.
To ease the notation, we also redefine the symmetry breaking field as
\begin{equation}
\hat{h}_u = h_u + \Sigma^0_u \; \frac{\partial^2}{\partial \Sigma_u^2} 
\; {\cal V}^{qk, T}_u \, .
	\label{Eq:hat_h_u}
\end{equation}
The last term is due to our symmetry breaking term at nonzero temperature.
The corresponding expression for $\hat{h}_s$ is
\begin{equation}
\hat{h}_s = h_s + \Sigma^0_s \; \frac{\partial^2}{\partial \Sigma_s^2} 
\; {\cal V}^{qk, T}_s \; .
	\label{Eq:hat_h_s}
\end{equation}

The kaon mass is 
\begin{equation}
	m_{K}^2  = \frac{\hat{h}_u+\hat{h}_s}{\Sigma_u + \Sigma_s}.
	\label{Eq:K_fT}
\end{equation}

The masses of the $K^*_0$ and  $a_0$ are
\begin{eqnarray}
	m_{K^*_0}^2 &=& \frac{\hat{h}_s-\hat{h}_u}{\Sigma_s - \Sigma_u}\,,\\ 
	m_{a_0}^2 &=&  \frac12 \; \frac{\partial^2}{\partial \Sigma_u^2 } 
	\hat{\cal V}^{qk}_u  
	+c_A \, \Sigma _s+6 \, \lambda  \, \Sigma _u^2+\, m^2\,,
	\label{Eq:Kappa_a0_fT}
\end{eqnarray}
where 
\begin{equation}
	\hat{\cal V}^{qk}_f = 
	{\cal V}^{qk}_f
	-  \Sigma^0_f \frac{\partial}{\partial \Sigma_f} {\cal V}^{qk, T}_f\,. 
	\label{hatV}
\end{equation}

The sigma and $f_0$ masses are given by
\begin{eqnarray}
m_\sigma^2  = \frac{1}{2}\left(   m_{\sigma_{00}}^2 + m_{\sigma_{88}}^2 
+ \sqrt{(m_{\sigma_{00}}^2-m_{\sigma_{88}}^2)^2+4m_{\sigma_{08}}^4}
 \right)\,,\\
 m_{f_0}^2  = \frac{1}{2}\left(   m_{\sigma_{00}}^2 + m_{\sigma_{88}}^2 
-\sqrt{(m_{\sigma_{00}}^2-m_{\sigma_{88}}^2)^2+4m_{\sigma_{08}}^4}
 \right)\,,
	\label{Eq:sigma_fo_fT}
\end{eqnarray}
where 
\begin{eqnarray}
	m_{\sigma_{00}}^2 
= 
\frac{1}{3}  \; \frac{\partial^2}  {\partial \Sigma_u^2 } \hat{\cal V}^{qk}_u  
+ \frac{1}{6} \;\frac{\partial^2}  {\partial \Sigma_s^2 } \hat{\cal V}^{qk}_s  
- \frac{2}{3} \, c_A \, \left(2 \, \Sigma_u+\Sigma_s\right)
+ 2 \, \lambda  \left(2 \, \Sigma_u^2+\Sigma_s^2\right)+ m^2\,,\\
m_{\sigma_{08}}^2 = 
\frac{1}{3 \sqrt2} \; \frac{\partial^2}{\partial \Sigma_u^2 } 
\hat{\cal V}^{qk}_u  
- \frac{1}{3 \sqrt2}\;  \frac{\partial^2}{\partial \Sigma_s^2 } 
\hat{\cal V}^{qk}_s  
+  \frac{\sqrt{2}}{3} \; c_A \left(\Sigma _u-\Sigma _s\right)
-2 \, \sqrt{2} \, \lambda  \, \left(\Sigma_s^2-\, \Sigma_u^2\right)\,,\\  
m_{\sigma_{88}}^2 = 
\frac{1}{6} \;   \frac{\partial^2}{\partial \Sigma_u^2 } \hat{\cal V}^{qk}_u  
+ \frac{1}{3} \;\frac{\partial^2}  {\partial \Sigma_s^2 } \hat{\cal V}^{qk}_s  
+ \frac{1}{3} \; c_A \; \left(4 \, \Sigma_u-\, \Sigma_s\right)
+ 2 \, \lambda  \left(2 \,\Sigma_s^2+\Sigma_u^2\right)+m^2\,.
	\label{Eq:sigmas_fT}
\end{eqnarray}

Finally, the $\eta$ and $\eta'$ meson masses are obtained from
the expressions at zero temperature, Eqs.~\eqref{eq:sum_eta_etap_masses}
and~\eqref{eq:diff_masses_eta_etap} by replacing $h_f \to \hat{h}_f$.

\bibliography{qks}

\newpage

U.S. Department of Energy Office of Nuclear Physics or High Energy Physics

{\it Notice:} 
This manuscript has been co-authored by employees of Brookhaven 
Science Associates, LLC under Contract No. DE-SC0012704 with 
the U.S. Department of Energy. The publisher by accepting the manuscript for 
publication acknowledges that the United States Government retains a 
non-exclusive, paid-up, irrevocable, world-wide license to publish or 
reproduce the published form of this manuscript, or allow others to do so, 
for United States Government purposes.
This preprint is intended for publication in a journal or proceedings.  
Since changes may be made before publication, it may not be cited or 
reproduced without the author’s permission.
{\it DISCLAIMER}:
This report was prepared as an account of work sponsored by an agency of the 
United States Government.  Neither the United States Government nor any 
agency thereof, nor any of their employees, nor any of their contractors, 
subcontractors, or their employees, makes any warranty, express or implied, 
or assumes any legal liability or responsibility for the accuracy, 
completeness, or any third party’s use or the results of such use of any 
information, apparatus, product, or process disclosed, or represents that 
its use would not infringe privately owned rights. Reference herein to any 
specific commercial product, process, or service by trade name, trademark, 
manufacturer, or otherwise, does not necessarily constitute or imply its 
endorsement, recommendation, or favoring by the United States Government or 
any agency thereof or its contractors or subcontractors.  The views and 
opinions of authors expressed herein do not necessarily state or reflect 
those of the United States Government or any agency thereof.

\end{document}